\documentclass[usenatbib]{mn2e}

\usepackage{epsfig,times,amsmath,supertabular,colortbl,amsfonts,amssymb,multirow,array,fleqn,longtable}

\title[Flexion measurement in simulations]{Flexion measurement in simulations of \emph{Hubble Space Telescope} data}

\author[Rowe et al.]{Barnaby~Rowe,$^{1,2,3,4}$\thanks{E-mail: browe@star.ucl.ac.uk}
David~Bacon,$^5$
Richard~Massey,$^6$
Catherine~Heymans,$^7$ 
\newauthor
Boris H\"{a}u\ss{}ler,$^8$
Andy~Taylor,$^7$ 
Jason Rhodes,$^{2,3}$
Yannick Mellier$^4$ \\  % CHECK IF YANNICK STILL WANTS TO COAUTHOR
$^1$Department of Physics and Astronomy, University College London,
Gower Place, London WC1E 6BT, UK  \\
$^2$Jet Propulsion Laboratory, California Institute of Technology,
4800 Oak Grove Drive, Pasadena, CA 91109, USA \\
$^3$California Institute of Technology, 1200 East California
Boulevard, Pasadena CA 91106, USA \\
$^4$Institut d'Astrophysique de Paris, UMR7095 CNRS, Universit\'{e}
Pierre et Marie Curie -- Paris 6, 98 bis, Boulevard Arago, 75014
Paris, France \\
$^5$Institute of Cosmology and Gravitation, University of Portsmouth,
Mercantile House, Hampshire Terrace, Portsmouth, Hants PO1 2EG, UK \\
$^6$Institute for Computational Cosmology, Durham University, South
Road, Durham DH1 3LE, UK\\
$^7$SUPA\thanks{Scottish Universities Physics Alliance}, 
Institute for Astronomy,
University of Edinburgh, The Royal Observatory, Blackford Hill,
Edinburgh EH9 3HJ, UK\\
$^8$School of Physics and Astronomy, The University of Nottingham,
University Park, Nottingham NG7 2RD, UK}

\begin{document}
\maketitle

\newcommand{\expect}[1]{\left\langle #1 \right\rangle} % expectation value

\newcommand{\fflex}{\mbox{$\mathcal{F}$}}
\newcommand{\fflexsmall}{\mbox{\tiny $\mathcal{F}$}}
\newcommand{\gflex}{\mbox{$\mathcal{G}$}}
\newcommand{\gflexsmall}{\mbox{\tiny $\mathcal{G}$}}
\newcommand{\fflext}{\mbox{$\mathcal{F_T}$}}
\newcommand{\gflext}{\mbox{$\mathcal{G_T}$}}
\newcommand{\stild}{\tilde{\gamma}}
\newcommand{\ftild}{\tilde{\mbox{$\mathcal{F}$}}}
\newcommand{\gtild}{\tilde{\mbox{$\mathcal{G}$}}}
\newcommand{\mR}{\mbox{$\mathcal{R}$}}
\newcommand{\trans}{\mbox{$\mathcal{T}$}}
\newcommand{\tri}{\mathcal{T}}
\newcommand{\dif}{\mbox{$\mathrm{d}$}}
\newcommand{\x}{\mbox{\boldmath$x$}}
\newcommand{\y}{\mbox{\boldmath$y$}}
\newcommand{\thetab}{\mbox{\boldmath$\theta$}}
\newcommand{\Delthetab}{\mbox{\boldmath$\Delta\theta$}}
\newcommand{\alphab}{\mbox{\boldmath$\alpha$}}
\newcommand{\betab}{\mbox{\boldmath$\beta$}}
\newcommand{\xib}{\mbox{\boldmath$\xi$}}
\newcommand{\nablab}{\mbox{\boldmath$\nabla$}}
\newcommand{\me}{\mbox{$\mathrm{e}$}}
\newcommand{\mi}{\mbox{$\mathrm{i}$}}
\newcommand{\mes}{\mbox{\scriptsize$\mathrm{e}$}}
\newcommand{\mos}{\mbox{\scriptsize$\mathrm{o}$}}
\newcommand{\mess}{\mbox{\tiny$\mathrm{e}$}}
\newcommand{\moss}{\mbox{\tiny$\mathrm{o}$}}
\newcommand{\msep}{\mbox{\scriptsize$\mathrm{sep}$}}
\newcommand{\mis}{\mbox{\scriptsize$\mathrm{i}$}}
\newcommand{\ml}{\mbox{\scriptsize$\mathrm{l}$}}
\newcommand{\mls}{\mbox{\scriptsize$\mathrm{ls}$}}
\newcommand{\ms}{\mbox{\scriptsize$\mathrm{s}$}}
\newcommand{\mE}{\mbox{\scriptsize$\mathrm{E}$}}
\newcommand{\mB}{\mbox{\scriptsize$\mathrm{B}$}}
\newcommand{\mc}{\mbox{\scriptsize$\mathrm{c}$}}
\newcommand{\mH}{\mbox{$\mathrm{H}$}}
\newcommand{\mkms}{\mbox{$\mathrm{kms}$}}
\newcommand{\mkg}{\mbox{$\mathrm{kg}$}}
\newcommand{\mMpc}{\mbox{$\mathrm{Mpc}$}}
\newcommand{\mcov}{\mbox{$\mathrm{cov}$}}
\newcommand{\mN}{\mbox{\scriptsize$\mathrm{N}$}}
\newcommand{\mint}{\mbox{\scriptsize$\mathrm{int}$}}
\newcommand{\mcrit}{\mbox{\scriptsize$\mathrm{crit}$}}
\newcommand{\mtot}{\mbox{\scriptsize$\mathrm{tot}$}}
\newcommand{\mrot}{\mbox{\scriptsize$\mathrm{rot}$}}
\newcommand{\mprop}{\mbox{\scriptsize$\mathrm{prop}$}}
\newcommand{\mcom}{\mbox{\scriptsize$\mathrm{com}$}}
\newcommand{\mang}{\mbox{\scriptsize$\mathrm{ang}$}}
\newcommand{\mlum}{\mbox{\scriptsize$\mathrm{lum}$}}
\newcommand{\mvir}{\mbox{\scriptsize$\mathrm{vir}$}}
\newcommand{\mstar}{\mbox{\scriptsize$\mathrm{star}$}}
\newcommand{\mobs}{\mbox{\scriptsize$\mathrm{obs}$}}
\newcommand{\mMC}{\mbox{\scriptsize$\mathrm{MC}$}}
\newcommand{\mMCs}{\mbox{\tiny$\mathrm{MC}$}}
\newcommand{\atanh}{\mbox{$\mathrm{tanh}$}}
\newcommand{\nn}{\nonumber \\}
\newcommand{\sextractor}{\textsc{SExtractor}}
\newcommand{\mmin}{\mbox{\scriptsize$\mathrm{min}$}}
\newcommand{\mmax}{\mbox{\scriptsize$\mathrm{max}$}}
\newcommand{\munweighted}{\mbox{\scriptsize$\mathrm{unweighted}$}}
\newcommand{\mGaussian}{\mbox{\scriptsize$\mathrm{Gauss}$}}
\newcommand{\mdiag}{\mbox{\scriptsize$\mathrm{diag}$}}
\newcommand{\mmag}{\mbox{\scriptsize$\mathrm{unweighted}$}}
\newcommand{\mav}{\mbox{\scriptsize$\mathrm{av}$}}
\newcommand{\mshot}{\mbox{\scriptsize$\mathrm{shot}$}}
\newcommand{\mlin}{\mbox{\scriptsize$\mathrm{lin}$}}
\newcommand{\mback}{\mbox{\scriptsize$\mathrm{back}$}}
\newcommand{\mcoll}{\mbox{\scriptsize$\mathrm{coll}$}}
\newcommand{\mf}{\mbox{\scriptsize$\mathrm{f}$}}
\newcommand{\mhalos}{\mbox{\scriptsize$\mathrm{halos}$}}
\newcommand{\mm}{\mbox{\scriptsize$\mathrm{m}$}}
\newcommand{\lcdm}{$\Lambda$CDM}
\newcommand{\ard}{\hat{a}^{\dagger}_r}
\newcommand{\ald}{\hat{a}^{\dagger}_l}
\newcommand{\ards}{\hat{a}^{\dagger 2}_r}
\newcommand{\alds}{\hat{a}^{\dagger 2}_l}
\newcommand{\ar}{\hat{a}_r}
\newcommand{\al}{\hat{a}_l}
\newcommand{\ars}{\hat{a}^2_r}
\newcommand{\als}{\hat{a}^2_l}
\newcommand{\vsos}{\mbox{$V_{606}$}}

\newcommand{\nat}{Nat}
\newcommand{\mnras}{MNRAS}
\newcommand{\pasp}{PASP}
\newcommand{\apj}{ApJ}
\newcommand{\apjl}{ApJL}
\newcommand{\apjs}{ApJS}
\newcommand{\physrep}{Phys.~Rep.}
\newcommand{\aap}{A\&A}
\newcommand{\aaps}{A\&AS}
\newcommand{\aj}{AJ}

\begin{abstract}
  We present a simulation analysis of weak gravitational lensing
  flexion and shear measurement using shapelet decomposition, and
  identify differences between flexion and shear measurement noise in
  deep survey data.  Taking models of galaxies from the \emph{Hubble
    Space Telescope} Ultra Deep Field (HUDF) and applying a correction
  for the HUDF point spread function we generate lensed simulations of deep, optical
  imaging data from \emph{Hubble's} Advanced Camera for Surveys (ACS),
  with realistic galaxy morphologies.  We find that flexion and shear
  estimates differ in our measurement pipeline: whereas intrinsic
  galaxy shape is typically the dominant contribution to noise in
  shear estimates, pixel noise due to finite photon counts and
  detector read noise is a major contributor to uncertainty in flexion
  estimates, across a broad range of galaxy signal-to-noise.  This
  pixel noise also increases more rapidly as galaxy signal-to-noise
  decreases than is found for shear estimates.  We provide simple
  power law fitting functions for this behaviour, for both flexion and
  shear, allowing the effect to be properly accounted for in future
  forecasts for flexion measurement.  Using the simulations we also
  quantify the systematic biases of our shapelet flexion and shear
  measurement pipeline for deep \emph{Hubble} data sets such as Galaxy
  Evolution from Morphology and SEDs,
  Space Telescope A901/902 Galaxy Evolution Survey or the Cosmic Evolution
  Survey.  Flexion measurement biases are found to be
  significant but consistent with previous studies.
% that used galaxy models with simpler morphological  properties.  
 %The simulated images are made freely available on request.
\end{abstract}


\section{Introduction}
The study of weak gravitational lensing has progressed to become one
of the most important techniques in observational cosmology (see
e.g. \citealp{schneider06,hoekstrajain08} for a review).  As it does
not depend upon the microscopic composition of the mass by which it is
caused, gravitational lensing requires no assumptions to be made
regarding baryon-dark matter physics.  It can thus be used to make
direct observations of the matter distribution on large scales (e.g.\
\citealp{hoekstraetal06,masseyetal07nature,benjaminetal07,fuetal08,
schrabbacketal10,huffetal11,heymansetal12cfhtlens,kilbingeretal13,benjaminetal13,
heymansetal13,jeeetal13,vanwaerbekeetal13}).

Weak lensing studies have typically measured the small but coherent
distortions in the ellipticities of distant galaxies, due to the shear
$\gamma$, and have used these measurements to constrain the
distribution of the intervening matter field. The weak lensing
description has been extended to higher order via the flexion
formalism (see \citealp{goldbergbacon05,irwinshmakova05,baconetal06},
hereafter B06; \citealp*{irwinshmakova06,irwinetal07}),
which describes the slight arcing of galaxy shapes.  Despite being a
weaker effect than shear, it has been hoped that the flexion signal
may yet be profitably measured -- galaxies typically display less
intrinsic curvature than intrinsic ellipticity, and so the
contribution to noise from intrinsic shape is reduced for measurements
of flexion.  Studies have suggested (e.g.\ B06;
\citealp*{okuraetal07}; \citealp{leonardetal07};
\citealp*{okuraetal08,leonardetal11}) that flexion measurements can
provide extra constraints upon galaxy-galaxy lensing results and
cluster mass reconstructions.

For all weak lensing analyses, the correct treatment of systematic
errors is vital: image distortions and shape bias due to convolution
with an anisotropic Point Spread Function (PSF), as possessed by all
optical instruments, are commonly an order of magnitude greater than
the gravitational signal we wish to measure. The optimal, unbiased
estimation of weak lensing signals from real data is the subject of
much ongoing research, involving a variety of different approaches
towards the accurate inference about galaxy shapes, accounting for
telescope optics, detector effects and noise. Many of the current
methods used to correct for the effects of the PSF are based on the
scheme proposed by \citet*{kaiseretal95}, \citet*{luppinokaiser97} and
\citet{hoekstraetal98}, commonly referred to as KSB or KSB+. The use
of these techniques has proved to be both successful and widespread,
to date.

Despite its practical success, there are certain elements of the KSB
treatment that are unsatisfactory: \citet{kaiser00} provides one
compelling discussion of these potential limitations. This has
prompted efforts to develop alternative weak lensing methods (e.g.\
\citealp*{kaiser00,rhodesetal00,bernsteinjarvis02,refregier03,refregierbacon03};
\citealp{hirataetal04};
\citealp*{masseyrefregier05,kuijken06,melchioretal07,nakajimabernstein07};
\citealp{milleretal07,kitchingetal08};
\citealp*{bernstein10,violaetal11};
\citealp{melchioretal11,kacprzaketal12,milleretal13,zuntzetal13}).  The
Shear TEsting Program (STEP: see \citealp{heymansetal06step};
\citealp{masseyetal07step}) and GREAT challenge series
\citep{bridleetal09,bridleetal10,kitchingetal11,kitchingetal12} have
compared a wide range of current weak shear estimation methods, using
blind-tests on simulated lensing data.

The \emph{shapelet} approach, proposed by \citet{bernsteinjarvis02}
and \citet{refregier03}, is one such alternative to KSB
methods. Shapelets expresses galaxy images as a sum of simple basis
functions --- Gauss-Laguerre or Gauss-Hermite polynomials --- that
behave well under deconvolution with a modelled PSF. In addition, the
first method for the practical estimation of the flexion signal was
built within the shapelets framework (\citealp{goldbergbacon05}; B06).
Further work by \citet{masseyetal07polar}, hereafter referred to as
M07, investigated shear and flexion measurement within the
\emph{polar} shapelets formalism of \citet{masseyrefregier05}; results
suggested that polar shapelets provided an apparently natural
framework for estimating both quantities. \citet*{velanderetal11} used
shapelets (in an implementation closely related to that of
\citealp{kuijken06}) to constrain flexion in \emph{Hubble Space
  Telescope} (\emph{HST}) data, although following a somewhat
different modelling strategy to that suggested in M07.  An alternative
method for measuring flexion, referred to as Higher Order Lensing
Image Characteristics (HOLICS) has also been suggested, and indeed
employed in cluster modelling from real data
\citep{okuraetal07,okuraetal08,goldbergleonard07,leonardetal07,leonardetal11}.
Showing signs that it may provide less noisy measurements than
shapelets \citep{leonardetal07,leonardetal11}, HOLICS is conceptually a
generalization of KSB methods to higher order image moments.  The
correction for an anisotropic PSF within HOLICS is, however, of
significant complexity \citep{okuraetal08}.

%In this paper, we present new shapelet measurements of the
%gravitational shear and flexion signal from galaxy images in the
%\emph{Hubble Space Telescope} Galaxy Evolution from Morphologies and
%SEDs (\emph{HST} GEMS) survey \citep{rixetal04}.
%These measurements require a weak lensing pipeline that is
%significantly different from those of previous KSB style analyses,
%and we describe this pipeline in detail, and calibrate it using
%simulated data designed to closely imitate galaxy shear and flexion in
%\emph{HST} survey data.  In a forthcoming paper (Rowe et al., in
%prep.) we will examine the
%constraints on GEMS galaxy haloes available from these shear and flexion
%catalogues by performing a full, combined shear-flexion 
%galaxy-galaxy lensing analysis.   

In this paper we present an analysis of simulations of space-based
lensing data, such as that taken using the \emph{HST} Advanced Camera
for Surveys (ACS: see, e.g., \citealp{hartigetal03,rhodesetal07}).
Several wide-area imaging surveys that may be used for weak lensing
exist for this instrument, including the Galaxy Evolution from
Morphology and S\textsc{ED}s survey (GEMS: see, e.g.,
\citealp{rixetal04,heymansetal05}), the Cosmic Evolution Survey
(COSMOS: see, e.g., \citealp{scovilleetal07,leauthaudetal07}) and the
Space Telescope A901/902 Galaxy Evolution Survey (STAGES: see, e.g.,
\citealp{grayetal09}).  The imaging data from these surveys share
important characteristics as regards lensing measurement, having a
small but non-Gaussian PSF and significant correlation in the noise
between pixels due to the dithering and drizzling strategies employed
(see e.g., \citealp{leauthaudetal07}).  In addition, there exist a
wealth of galaxy cluster imaging data in the ACS
archive that are also of interest for shear and flexion lensing
analysis.  Mindful of these factors, we construct realistic simulations
of ACS lensing data using real sky noise taken from blank areas in the
GEMS survey data, and with a PSF that matches the radial profile of
the GEMS PSF.  Within these simulations we apply known input shears
and flexions, and use the resulting measurements to calculate the
necessary calibration for the shapelet measurement of shear and
flexion from space.

Our paper is organized as follows.
Sections~\ref{sect:wlform} and \ref{sect:shapelets} begin with a brief
description of the flexion formalism and a summary of our adopted
shapelet measurement method.  In Section~\ref{sect:udf} we describe
how we make shapelet models of \emph{Hubble} Ultra Deep Field data, including
both stars and galaxies. These are used to generate simulations of
lensed ACS data, which we describe in Section~\ref{sect:sims}. In
Section~\ref{sect:flippipe} we test how well we can measure flexion on
this simulated data. Finally, we discuss our findings in Section
\ref{sect:conc}.

\section{Weak shear and flexion formalism}\label{sect:wlform}
To begin, we review the flexion formalism developed by
\citet{goldbergbacon05} and B06, examining briefly how weak flexion is
defined in relation to weak shear.  We restrict the discussion to
lensing measurements in the weak regime, so we do not consider the
$\emph{reduced}$ shear or flexion (see
\citealp{schneiderseitz95,schneiderer08}).

Gravitational lensing conserves surface brightness, so the effects of
lensing may be described in terms of coordinate transformations
between the lensed and unlensed sky coordinate plane. In general, the
relationship between these coordinates is non-linear and is described
by the lens equation (e.g., \citealp{bartelmannschneider01,schneider06}).

However, if we may assume that changes in the lens properties of a
system occur only on angular scales that are large compared to the
angular size of the lensed light source, then the lens equation may be
locally linearized. In what follows, lensed and unlensed coordinates
are denoted by $\thetab$ and $\thetab'$ respectively, and we define
position relative to a galaxy centroid as $\Delthetab = \thetab -
\thetab_c$ and $\Delthetab' = \thetab' - \thetab'_c$, where
$\thetab_c$ and $\thetab_c'$ are the coordinate centroids of the
galaxy in the image and source plane, respectively.  Approximately
linearizing the lens equation around this centroid, we write
\begin{equation}\label{eq:lintran}
\Delta\theta'_i \simeq A_{ij} \Delta\theta_j ,
\end{equation}
where $A_{ij}$ is the Jacobian matrix of the transformation given by
the lens equation.  This matrix may be written as
\begin{eqnarray}\label{eq:Adef}
A_{ij}({\thetab})&\equiv&
\frac{\partial{\thetab_i'}}{\partial {\thetab_j}} ~ = ~
\delta_{ij}-\partial_i \partial_j \psi({\thetab}) \\ \nonumber
~ &=&
\left(
\begin{array}{cc}
1-\kappa-\gamma_1 & -\gamma_2 \\
-\gamma_2 & 1-\kappa+\gamma_1
\end{array}
\right), 
\end{eqnarray}
where $\psi(\theta)$ is the lensing potential, a two-dimensional
projection of the gravitational potential along the line of
sight (e.g.\ \citealp{schneider06}). These equations define the two components of weak shear
$\gamma_1$ and $\gamma_2$, and the convergence $\kappa$, which is a
measure of the projected matter density.

However, the assumption of gradual variation in lens properties across
the sky is not always justified, especially in very dense regions or
those with significant dark matter substructure along the line of sight.  In these situations
the lens transformation is more accurately described by
\begin{equation}\label{eq:flextran}
\Delta\theta'_i \simeq A_{ij}\Delta\theta_j +\frac{1}{2}D_{ijk} 
\Delta\theta_j \Delta\theta_k,
\end{equation}
which is simply the expansion of \eqref{eq:lintran} to second order, where 
\begin{equation}\label{eq:dijk}
D_{ijk}\equiv \frac{\partial^2 \theta'_i}{\partial \theta_j \partial \theta_k}
=\partial_k A_{ij}=-\partial_i \partial_j \partial_k \psi.
\end{equation}

Equation \eqref{eq:flextran} describes the lensing distortions known
as \emph{flexion}, which skew galaxy light distributions and lead to
weak lensing arcs, and which may be described upon an image $I (
  \thetab )$ using the conservation of surface brightness under lensing
\begin{equation}\label{eq:surf}
I(\thetab) = I^{(s)}(\thetab') = I^{(s)}(\thetab_c' + \Delthetab' )  .
\end{equation}
Employing equation \eqref{eq:flextran} in this surface brightness
transformation takes weak lensing one order closer to the fully
generalized non-linear treatment.  By expanding the surface
brightness as a Taylor series and substituting \eqref{eq:flextran},
\citet{goldbergbacon05} showed that the lensed surface brightness of a
galaxy may be approximated as
\begin{equation}
I(\thetab) \simeq I^{(s)}(\thetab) +
	\left[
	(A-I)_{ij}\Delta\theta_j+\frac{1}{2}D_{ijk}\Delta\theta_j
       \Delta\theta_k\right]
	\partial_i I^{(s)}(\thetab),
\label{eq:flextransb}
\end{equation}
an expression which is useful in the construction of weak shear and flexion
estimators using shapelets
\citep{refregier03,baconetal06,masseyetal07polar}.

For its clarity and convenience we will often employ the complex
notation introduced in B06.  The complex gradient operator on the sky
plane is defined as:
\begin{equation}\label{eq:compgrad}
\partial \equiv \partial_1 + \mi \partial_2.
\end{equation}
It is shown in B06 that using this notation the convergence and
complex shear $\gamma= \gamma_1 + \mi \gamma_2$ may be written as:
\begin{eqnarray}
\kappa & = & \frac{1}{2}\partial^* \partial \psi, \\
\gamma & = & \frac{1}{2}\partial \partial \psi = |\gamma|\me^{2 \mis \phi},
\end{eqnarray}
which neatly encapsulates the spin-2 rotational symmetry properties of
the shear.  Taking a further complex gradient, we may define two more
complex fields:
\begin{eqnarray}\label{eq:flexdef}
\fflex & = & \frac{1}{2}\partial^* \partial \partial \psi =
|\fflex|\me^{\mis \phi} , \\
\gflex & = & \frac{1}{2} \partial \partial \partial \psi =
|\gflex|\me^{3 \mis \phi},
\end{eqnarray}
referred to as the first flexion (spin-1) and second flexion (spin-3)
respectively. Using equation \eqref{eq:dijk} we may write $D_{ijk}$
entirely in terms of the components of $\fflex$ and $\gflex$, as
follows:
\begin{eqnarray} \label{eq:dijkflex}
D_{ij1} & = & -\frac{1}{2} \left(
\begin{array}{cc}
 3\fflex_1 + \gflex_1 & \fflex_2 + \gflex_2  \\
  \fflex_2 + \gflex_2  &  \fflex_1 - \gflex_1 
\end{array}
\right), \\ \nonumber
D_{ij2}&=& -\frac{1}{2} \left(
\begin{array}{cc}
 \fflex_2 + \gflex_2 &  \fflex_1 - \gflex_1 \\
 \fflex_1 - \gflex_1 & 3\fflex_2 - \gflex_2
\end{array}
\right).
\end{eqnarray}
Equations \eqref{eq:Adef}, \eqref{eq:flextransb} and
\eqref{eq:dijkflex} allow the formulation of practical estimators for
$\gamma$, $\fflex$ and $\gflex$.  In the following Section we outline
how shapelets may be used to make estimators of shear and flexion.

\section{Shapelet Modelling}\label{sect:shapelets}
\subsection{Shapelet basis sets}
The underlying concept of the shapelet approach, as introduced by
\citet{refregier03} and \citet{bernsteinjarvis02}, is the expression
of an object's surface brightness as a sum of orthonormal,
two-dimensional basis functions:
\begin{equation}\label{eq:f}
f(\thetab) = \sum_{n_1 = 0}^{\infty} \sum_{n_2=0}^{\infty} f_{n_1,n_2}
B_{n_1,n_2}(\thetab;\beta).
\end{equation}
The choice of basis functions is free in general, but the Cartesian
shapelet basis set is defined by the basis function
\begin{equation}\label{eq:cartbasis}
B_{n_1,n_2}(\thetab;\beta) =
\frac{H_{n_1}(\theta_1/\beta)H_{n_2}(\theta_2/\beta) 
\me^{-|\theta|^2/2\beta^2}}
{2^{(n_1n_2)} \beta \sqrt{\pi n_1\! n_2\!}} ,
\end{equation}
where $H_{n_i}(x)$ is a Hermite polynomial of order $n_i$, and the
important free quantity $\beta$ is the scale size of the shapelet
basis set. We refer to the sum of the two parameters $n_1$ and $n_2$
as the order of the shapelet basis function, and will generally
truncate shapelet models to some limiting order $n_{\mmax}$ such that
$n_1 + n_2 \le n_{\mmax}$.

The formalism of polar shapelets, introduced by
\citet{masseyrefregier05}, is closely related to that of Cartesian
shapelets. Instead of the basis set defined by Equations \eqref{eq:f}
and \eqref{eq:cartbasis}, polar shapelets express the object surface
brightness $f(\thetab)$ as
\begin{equation}\label{eq:fpolar}
f(\thetab) = f(\theta,\phi) = \sum_{n=0}^{\infty} \sum_{m=-n}^{n} 
f_{n,m} P_{n,m} (\theta,\phi;\beta),
\end{equation}
where $\theta$ is the modulus of the complex sky position vector
$\theta_1 + \mi \theta_2$, and $\phi = \arctan{(\theta_2/\theta_1)}$.
The polar shapelet basis functions, which we label $P_{n,m}$, are
defined by \citet{masseyrefregier05} as
\begin{eqnarray}\label{eq:polarbasis}
P_{n,m}(\theta,\phi;\beta) &=& \frac{(-1)^{(n-|m|)/2}}{\beta^{|m|+1}}
\left\{ \frac{ [(n-|m|)/2]! }{ \pi [(n-|m|)/2]!} \right\}^{1/2}
\\
&\times& \theta^{|m|} L^{|m|}_{(n-|m|)/2} \left( \frac{\theta^2}{\beta^2}
\right) {\me}^{-\theta^2/2\beta^2} {\me}^{- {\rm i} m \phi} \nonumber
\end{eqnarray}
using the following definition of the associated Laguerre polynomials
(see, e.g., \citealt{arfkenweber05}):
\begin{equation}
L^q_p(x) \equiv \frac{x^{-q} {\me}^x }{ p!} \frac{{\dif}^p}{\dif x^p}
\left( x^{p+q} {\me}^{-x} \right). 
\end{equation}
Each separate member of the basis set is uniquely described using the
two integers $n$ and $m$, with $n > 0$ and $|m| \le n$.

Both the Cartesian and polar shapelet basis set have relatively simple
behaviour under convolution \citep{refregier03} and deconvolution
\citep{refregierbacon03}; this makes them particularly suited for
correcting images for the effects of an instrumental point spread
function (PSF). We now describe the methodology required for this
correction.

\subsection{Image deconvolution using shapelets}\label{sect:shapedeconv}
Within the shapelet framework, there are two possible methods with
which to correct galaxy images for the effects of the optical system.
Both approaches begin with the construction of a shapelet model of the
PSF $g(\thetab)$ at the location of each galaxy: this model should be
as accurate as possible and will include any variation of the PSF
across the image plane of the instrument (e.g.\ \citealp{jarvisjain04,hoekstra04,rowe10,heymansetal12atmos,kitchingetal13star}).
The model may also need to include some treatment of time dependent
effects (see, e.g.,
\citealp{heymansetal05,schrabbacketal07,rhodesetal07}).

The deconvolution method used in this work is that proposed by
\citet{masseyrefregier05}, which is implemented within the shapelet
software package made available by these
authors\footnote{http://www.astro.caltech.edu/$\sim$rjm/shapelets/}. In
this approach, the deconvolved shapelet coefficients $f_{n_1,n_2}$ are
estimated by ``forward'' convolving the shapelet basis functions with
the PSF model in advance, creating a new basis set which we label
\begin{equation}
D_{n_1,n_2}(\thetab;\beta) = g(\thetab) * B_{n_1,n_2}(\thetab; \beta), 
\end{equation}
with an equivalent expression for the case of the polar shapelet basis
functions $P_{n,m}(\thetab;\beta)$. Fitting the data $h(\thetab)$ with
this new basis set $D_{n_1,n_2}$, one returns a deconvolved shapelet
model as follows:
\begin{eqnarray}\label{eq:fconv}
h(\thetab) &=& g(\thetab)*f(\thetab) \\ 
 & = & 
g(\thetab) * \left[ \sum_{n_1 = 0}^{\infty} \sum_{n_2=0}^{\infty} f_{n_1,n_2} 
B_{n_1,n_2}(\thetab;\beta) \right]  \nonumber \\
& = & \sum_{n_1 = 0}^{\infty} \sum_{n_2=0}^{\infty} f_{n_1,n_2} \left[
g(\thetab)*B_{n_1,n_2}(\thetab;\beta) \right] \nonumber \\
 & = & \sum_{n_1 = 0}^{\infty} \sum_{n_2=0}^{\infty} f_{n_1,n_2}
 D_{n_1,n_2}(\thetab;\beta) \nonumber .
\end{eqnarray}
As can be seen by comparison with Equation \eqref{eq:f}, the returned
shapelet coefficients $f_{n_1,n_2}$ will reconstruct the deconvolved
image when they are used with the original basis set
$B_{n_1,n_2}(\thetab;\beta)$.

There are obvious caveats to this approach, particularly that the
convolved basis set $D_{n_1,n_2}(\thetab;\beta)$ will in general no
longer be orthogonal.  However, errors due to this fact are small
wherever the scale size of the galaxy image is larger than that of the
PSF \citep{masseyrefregier05}. The alternative shapelet deconvolution
approach is that described by \citet{refregierbacon03} and developed
in some detail by \citet{melchioretal09}, involving the inversion of a
`PSF matrix' that describes the transformation between the shapelet
model coefficients of $f$ and $h$. This PSF matrix is large and may be
sparse, even despite efficient truncation
\citep{refregier03,refregierbacon03}, causing the latter authors and
\citet{masseyrefregier05} to argue against its inversion as a slow and
potentially unstable process. However, \citet{melchioretal09} find
this not to be necessarily the case for sufficiently small and
lightly-structured PSF models, and so argue in favour of a modified
inversion scheme for computational efficiency.

For ACS data we find (see Section \ref{sect:modudfpsf}) that
a simple, low $n_{\textrm{max}}$ model is insufficient to describe the
PSF.  This makes matrix inversion schemes problematic, and thus for
this \emph{HST}-focused analysis we will test a shape measurement
pipeline based upon the software of \citet{masseyrefregier05}, using
the forward deconvolution method described above.

The estimation of deconvolved galaxy images using shapelet modelling
is important at two stages in this Paper. In Section
\ref{sect:flippipe}, shapelet models will provide estimates of the
weak lensing signal in samples of simulated images.  First however,
the technique is used to estimate the deconvolved shapes of a sample
of galaxies from which simulated images will be created.  Shapelet
models are convenient for creating such simulations: they are
trivially rotated and inverted, and as they provide an estimate for
the full surface brightness profile of a galaxy they allow a full
range of distortions (including flexion) to be simulated. In the
following Section we describe the construction of a shapelet galaxy
weak lensing simulation using a set of deep ACS galaxy images.

\section{Shapelet models of the \emph{Hubble} Ultra Deep Field}\label{sect:udf}

In order to simulate lensing data, we require a set of real galaxy
images which are used in turn to create a set of galaxy shapelet
models which we will refer to as the `starter set'. The real images
are obtained from the publicly-available \emph{Hubble} Ultra Deep
Field (HUDF: see \citealp{beckwithetal06} for a detailed description),
a multicolour galaxy survey image composed of a single ACS pointing,
with a $10^6$s total integration time over four broad spectral bands:
F435W ($B_{435}$), V606W ($V_{606}$), F775W ($i_{775}$) and F850LP
($z_{850}$). The highest redshift objects are only visible in the
$i_{775}$ and $z_{850}$ bands, but the \vsos$~$ filter provides good
sensitivity to objects at redshifts lower than $\simeq 4$
\citep{beckwithetal06}. As this is also the ACS filter used in the
GEMS and STAGES lensing studies (due to its rich source number
density, see \citealp{heymansetal05,heymansetal08}), we choose the
\vsos$~$ filter image for the construction of the shapelet starter
set, as representative of typical lensing source galaxies. 

\begin{figure}
\begin{center}
\psfig{figure=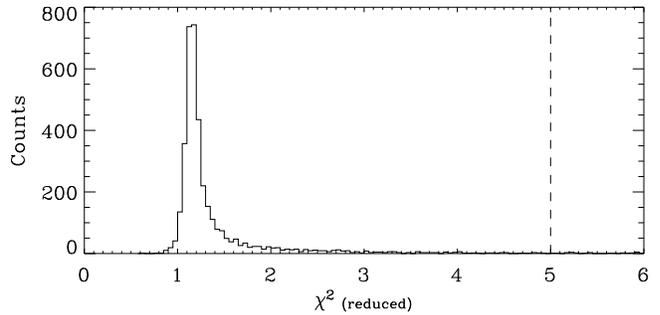, width=0.5\textwidth}
{\caption[Reduced $\chi^2$ for the HUDF galaxy models.]{Histogram of
    reduced $\chi^2$ values for the HUDF deconvolved galaxy models;
    the dashed line is the cut introduced to exclude very poorly-modelled galaxies.
\label{fig:chisqhist}}}
\end{center}
\end{figure}
A detailed description of the modelling of galaxies in the HUDF can be
found in Appendix \ref{app:udf}.  Given the input parameters, postage
stamps, and model of the HUDF PSF described in this Appendix, the
shapelets software was then used to find a best-fitting model of each
HUDF galaxy.  A total of 3867 of the 4128 input HUDF galaxies (94\%)
were successfully modelled.

On examination of these 6\% of catastrophic modelling failures, it was
found that they were caused either by:
\begin{enumerate}
\item Galaxy images lying too close to the edge of the HUDF or other
  objects in the field (being thus automatically rejected and flagged
  in the output shapelet catalogue); or
\item A population at faint magnitudes which appear to suffer from
  either star-galaxy confusion or general modelling degeneracy.
\end{enumerate}
The great majority of this second type of catastrophic failures were
for objects with $V_{606} > 27$, which is currently beyond the realm
of plausible lensing measurement with even deep imaging surveys such
as GEMS or COSMOS.  Without a clear means of further reducing this
failure rate, it is tolerated given the fact that it predominantly
affects objects beyond the detection limit of the simulations (See
Section~\ref{sect:noise}).

An additional cut was imposed on the sample, based upon the value of
the reduced $\chi^2$ output for each galaxy model. A histogram of this
output statistic is shown in Figure \ref{fig:chisqhist} for the HUDF
galaxies.  A cut of reduced $\chi^2 < 5$, beyond which the chances of a
good fit are vanishingly small for the high degree-of-freedom shapelet
models used, removed a further 153 very
poorly-modelled galaxies not identified by shapelet flags or other
indicators of total modelling failure. 

\begin{figure}
\begin{center}
  \psfig{figure=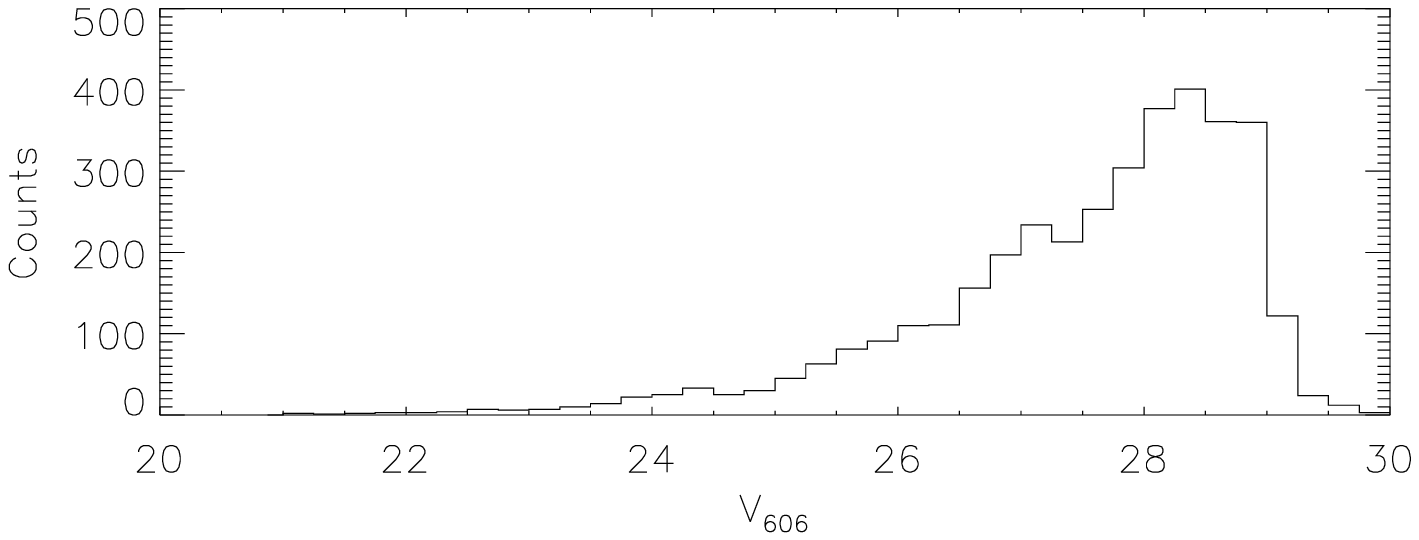, width=0.48\textwidth}
  \psfig{figure=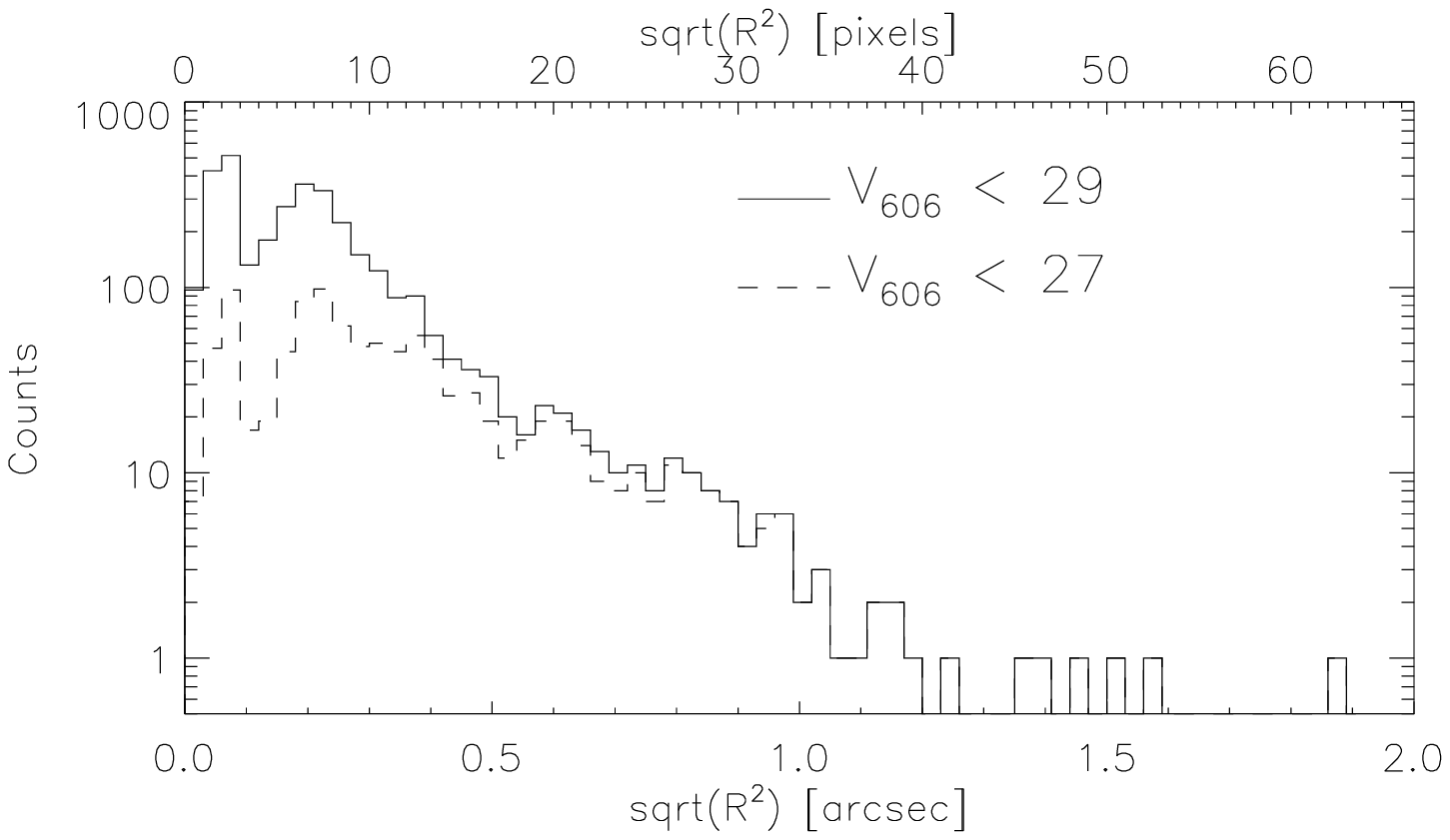, width=0.48\textwidth}
  {\caption{Upper panel: Distribution of $V_{606}$ magnitudes of
      shapelet galaxy models in the HUDF starter set.   Lower panel:
      Distribution of $\sqrt{R^2}$ (see equation \ref{eq:r2} for a
      definition)  in
      the HUDF starter set for two different magnitude cuts.  The
      $V_{606} < 27$ cut approximately corresponds to the limiting
      magnitude at which weak lensing measurements may be made in
      ACS surveys such as GEMS \citep{heymansetal05} and COSMOS
      \citep{leauthaudetal07}.
\label{fig:starterprops}}}
\end{center}
\end{figure}
The remaining 3714 PSF-corrected galaxy models form the galaxy starter
set that are now used to construct simulated lensing survey images
with realistic galaxy morphologies.  This corresponds to a high
density of approximately 410~galaxies~arcmin$^{-2}$, giving
considerable freedom to realistically simulate shallower \emph{HST} data than
the HUDF (an important caveat is the limited size of the sample
itself, which we discuss in the following Section).  Figure
\ref{fig:starterprops} shows the distribution of $V_{606}$ magnitudes
and an rms size measure in the model starter set (this size is based
on the shapelet-derived quantity $R^2$: see equation \ref{eq:r2} in
the Appendices).  The distribution of sizes illustrates the compact
nature of galaxies in deep ACS images, also illustrated in Figure
\ref{fig:sizemag}.

\section{Simulations of ACS data}\label{sect:sims}
\subsection{Source galaxy images}\label{sect:starter}

The galaxy models in the starter set described in the previous Section
are the fundamental data used to generate our simulations. However,
there is additional processing required before models may be used to
accurately simulate flexion and shear measurement.

A first problem is that the starter set itself contains the signature
of gravitational shear and flexion due to the matter structure along
the line of sight of the HUDF, along with any residual, uncorrected
HUDF PSF anisotropy.  These signals are largely eradicated by the
random rotation and inversion of galaxy models. These two
transformations can be performed by simple analytic manipulations of
the galaxy shapelet models in the polar shapelets formalism of \cite
{masseyrefregier05}, and so this process is fast and accurate.

A second problem is that the starter galaxy set represents a limited
sample of galaxy morphologies.  This may be alleviated to an extent by
introducing small random perturbations to the shapelet models of the
galaxy images (e.g., \citealp{masseyetal04,meneghettietal08}).  We
follow \citet{meneghettietal08} and add an independent Gaussian random variable
$N(0, \sigma_{n_1,n_2})$ to the real and imaginary part of
each shapelet coefficient $f_{n_1,n_2}$. We choose $\sigma_{n_1,n_2} =
0.1 \times |f_{n_1,n_2}|$, which we found ensured that galaxy shapes
are randomized in a way that did not introduce a high incidence of
noticeably unphysical features such as negative flux (see
\citealp{masseyetal04} for a discussion of over-randomized galaxy
models).

Randomly rotated, inverted and perturbed galaxy models are then a
suitable population of source galaxies, and can be assigned to
locations in the final output images.  Unlike the GREAT08 or GREAT10
simulations \citep{bridleetal10,kitchingetal12}, we choose to assign
galaxy images to random locations upon the output science tile, rather
than on a regular grid.  Although this complicates the analysis it has
the benefit that the impact of object detection and deblending is
included in our pipeline tests.  These effects are important as they
likely contribute to noise on shear and flexion estimates, one of the
primary topics of investigation in this work, enhancing the value of
any noise forecasts based on the results.

\subsection{Applying shear and flexion: shapelet transformations
  versus raytracing}
\label{sect:raytrace}

In the previous Section we arrived at a set of shapelet models
suitable for use as source galaxies. The desired weak lensing
distortion, a coordinate transformation governed by the lens equation,
must now be applied to these galaxies. There are two approximate means
by which this coordinate transformation may be applied in practice:

\begin{enumerate}
\item In `shapelet-space': shear and flexion distortions may be
  represented by an infinite sum of transformations upon the shapelet
  coefficient values in the galaxy model itself
  \citep{refregier03,masseyrefregier05,masseyetal07polar}. Applying
  successive shapelet transformations therefore allows the lensing
  distortion to be represented with arbitrary precision. The lensed
  shapelet model may then be analytically integrated across square
  pixels as described by \citet{masseyrefregier05}.\\

\item In `real-space': the shear and flexion distortions may be
  represented directly using the lens equation $I(\thetab) =
  I^{(s)}(\thetab')$ to raytrace from the regular grid of image pixels
  back to an irregular sampling of points on the source plane. This is
  possible as each shapelet model encodes the surface brightness
  $I^{(s)}$ at \emph{all} points in the source plane.
\end{enumerate}
Previous shapelet-based simulation methods have preferred the former
\citep{masseyetal07step,meneghettietal08} for weak lensing
transformations. For our simulations, designed to mimic space-based
flexion and shear observations as closely as possible, we choose the
latter real-space method for reasons that we now describe.

In the shapelet-space approach it is necessary to apply multiple
shapelet transformations to each galaxy model, each operation
capturing successive terms in a series expansion representation of the
true distortion, to ensure that the final distortion approximates a
real shear or flexion accurately. Doing this is important if
subsequently we wish to test shapelet measurement methods fairly.  As
shown by equation (41) of \citep{masseyrefregier05}, performing a
shapelet shear transformation accurate to first order in $\gamma$
increases the order $n_{\mmax}$ of the model from $n_{\mmax}$ to
$n_{\mmax} + 2$, resulting in $2n_{\mmax} + 5$ extra coefficients
overall.  At third order in $\gamma$, the minimum that should be
considered for precision work where the applied shear $\gamma$ may
approach 0.1, the order of the model is increased to $n_{\mmax} + 6$,
an extra $6n_{\mmax} + 27$ coefficients in total.

The situation is worse for flexion. \citet{masseyetal07polar} showed
that a first-order approximation to gravitational flexion increases
the order of a shapelet model to $n_{\mmax} + 3$.  Performing a
shapelet flexion transformation that is accurate to third order
therefore increases $n_{\mmax}$ by 9, meaning that a modest $n_{\mmax}
= 12$ shapelet galaxy requires a far more complex $n_{\mmax} = 21$
model once flexed.  The shapelet software suffers extreme memory
demands and reductions in speed beyond $n_{\mmax} = 20$, and so this
method of introducing accurate distortions becomes prohibitively slow
(see Appendix \ref{sect:modudfgals}). Unfortunately, re-truncating the
shapelet model back to a more manageable $n_{\mmax}$ is not a simple
solution to these problems. This degrades the congruence between the
exact and shapelet-approximated lensing transformations in a way that
is difficult to describe, as it varies with both the distortion
applied and the surface brightness distribution of each individual
object.

Also of concern is the action of the lensing transformations upon the
$\beta$ scale parameter.  Lensing distortions magnify images, which
for the linear-order shear and convergence transformations can be
simply represented by a small change in $\beta$.  With flexion a
fundamental complication with $\beta$ arises: as shown by
\citet{schneiderer08}, the determinant of the Jacobian matrix varies
as a function of position for flexion-order lensing.  This varying
transformation in the area element cannot be reproduced via a single
rescaling of the $\beta$ scale parameter.

This potentially increases the difficulty of exactly reproducing
non-linear flexion transformations using the shapelet formalism. When
constructing simulations for flexion measurement we must be extremely
careful that we are accurately describing the distortion if we wish to
construct a fair test of current and future methods. The real-space,
raytracing option listed above is therefore adopted for applying both
flexion and shear in the simulations.  As this scheme requires
numerical approximation of the integration of model surface brightness
over pixels, it was necessary to test the degree of approximation
necessary for accurately describing shear and flexion.  Details of
these tests, and their results, can be found in Appendix
\ref{sect:upsample}.  It is found that shear and flexion can be added
to our simulation galaxies with sufficient accuracy for current
purposes.  We turn now to the problem of convolution with an
instrumental PSF.

% whereas its approximate shapelet counterpart as derived by
%\citet{refregier03} and \citet{masseyrefregier05} conserves flux.
%To account for this $\beta$ should increase by a factor $1 / \sqrt{\mu}$, %where $\mu$ is the magnification and given by the inverse of the determinant %of $A_{ij}$ as defined in equation \eqref{eq:Adef} (see %\citealp{bartelmannschneider01}). This is constant across the image and for %pure shear alone is given by $\sqrt{1 - |\gamma^2|}$. This corresponds to an %overall very slight increase in overall size and flux for the galaxy model, %and so for weak shear simulation studies is typically and safely ignored.

\subsection{Applying the PSF}\label{sect:applyingpsf}
\begin{figure*}
\begin{center}
  \psfig{figure=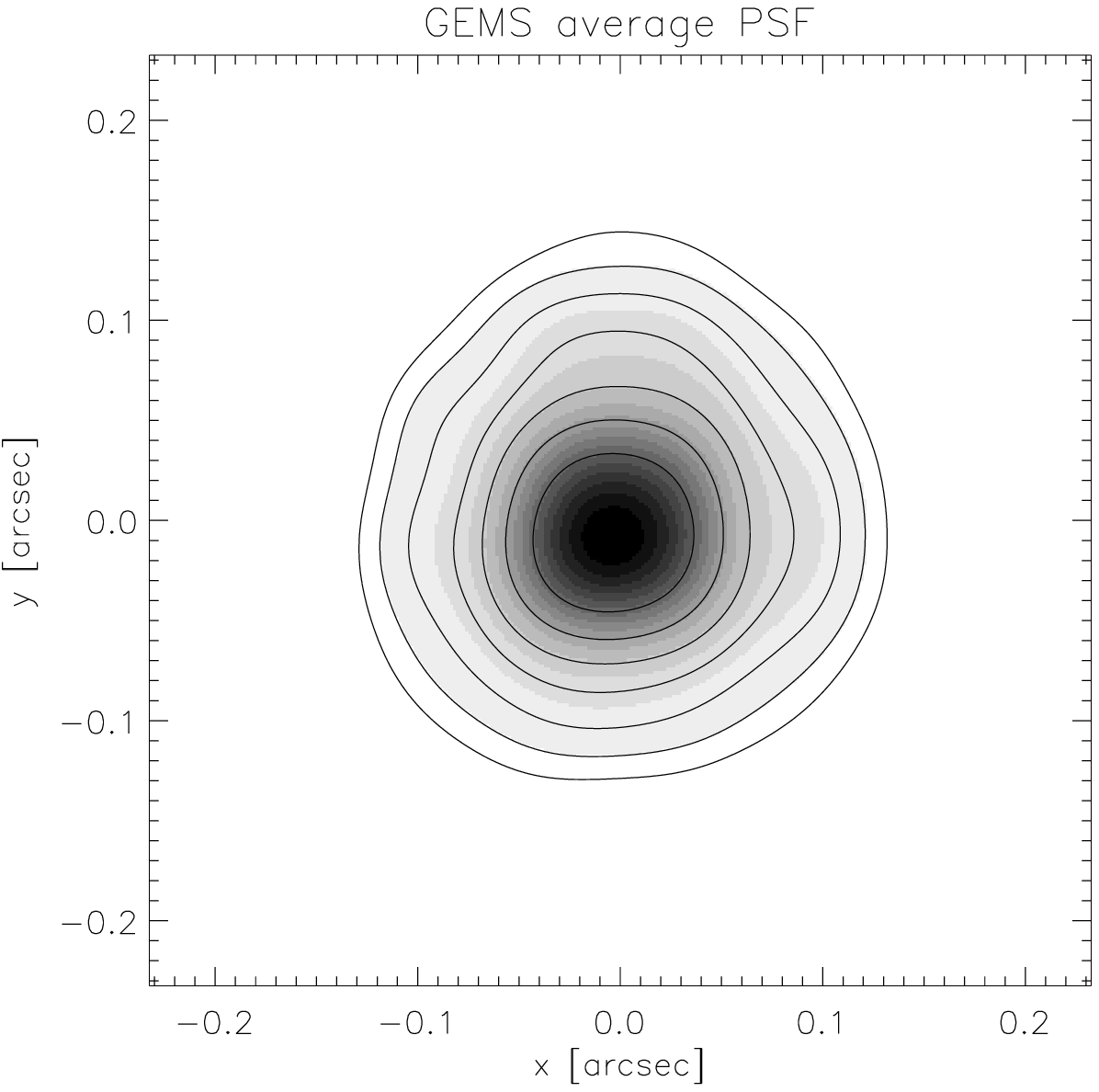, width=0.33\textwidth}
  \psfig{figure=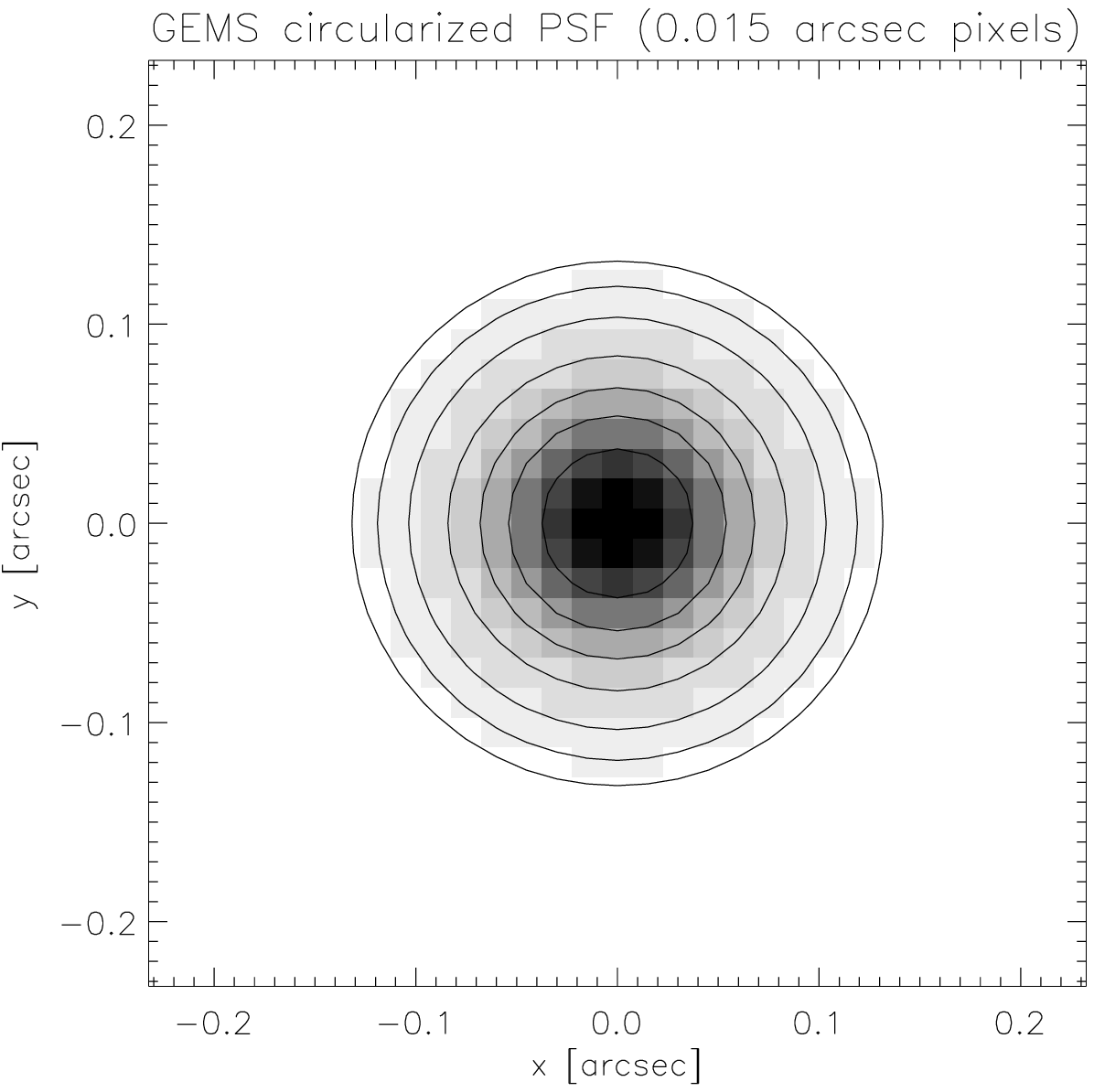, width=0.33\textwidth}
  \psfig{figure=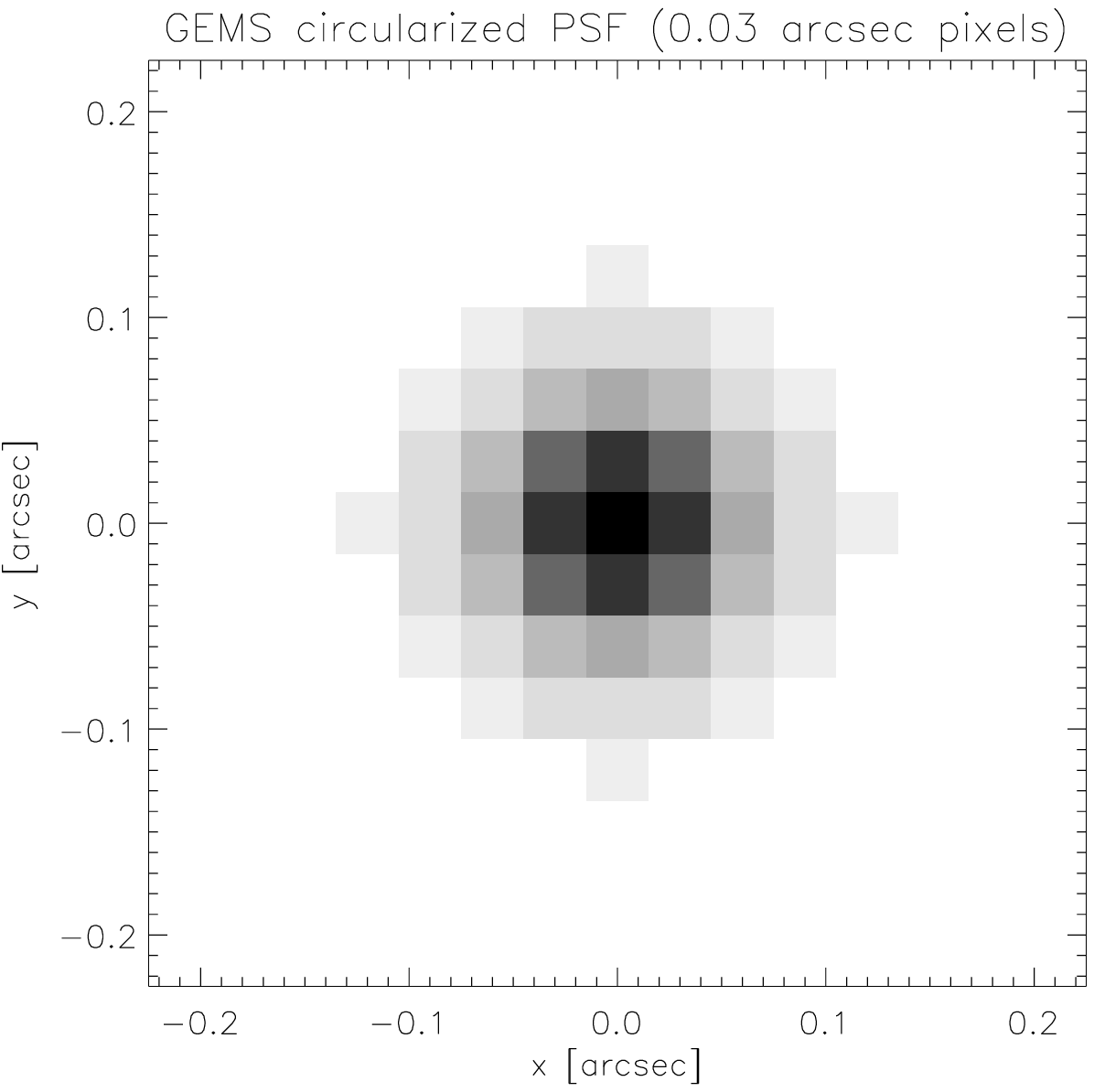, width=0.33\textwidth} {\caption{Left-hand panel: PSF
      pattern created from the weighted average of 909 shapelet models
      of selected stars in the GEMS $V_{606}$ science tiles. Central
      panel: circularized version of this same GEMS PSF made by
      setting to zero all the polar shapelet coefficients $f_{n,m}$
      with $m \ne 0$, and shown at the upsampled 0.015 arcsec
      pixel$^{-1}$ resolution used for performing the real-space
      convolution. Right-hand panel: the circularized GEMS PSF, shown at
      the final ACS 0.03 arcsec pixel$^{-1}$ resolution for reference.
      In all panels the grey-scale is linear in surface brightness
      whereas the contours are logarithmic. \label{fig:gemspsf}}}
\end{center}
\end{figure*}
Once flexion and shear has been applied to simulation galaxies, the
next stage is to convolve these correctly-lensed objects with a
simulation kernel that mimics a realistic target PSF.  The PSF we
choose for the simulations in this study is based on ACS observations
of stars from $V_{606}$ observations in the GEMS survey
\citep{rixetal04,caldwelletal08}, specifically the 909 stellar objects
selected as described by \citet{heymansetal05}. Shapelet
decompositions of these objects are created using $n_{\textrm{max}} =
20$, $\beta = 1.80$ pixels, and stacked to make an inverse-variance
weighted average GEMS PSF model (the same procedure as described in
Section~\ref{sect:modudfpsf} for modelling the HUDF PSF).  This model
can be seen as the leftmost panel in Figure~\ref{fig:gemspsf}.

\begin{figure}
\begin{center}
  \psfig{figure=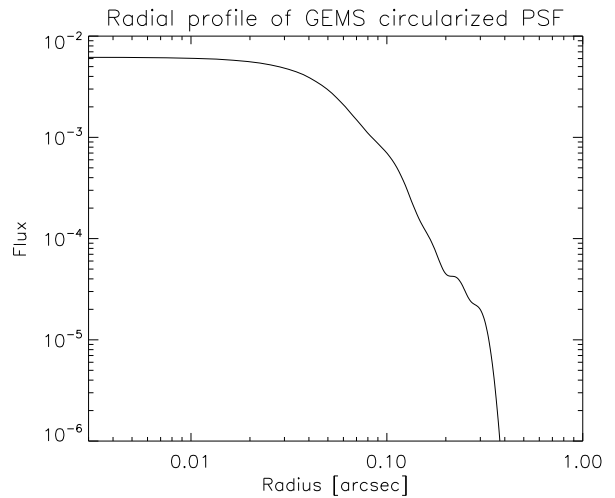, width=0.48\textwidth}
      {\caption{Radial profile of the shapelet model describing the 
         circularized GEMS PSF as depicted
          in the central and right-hand panels of Figure~\ref{fig:gemspsf}.
\label{fig:gemspsf_radial}}}
\end{center}
\end{figure}

As described by \citet{masseyrefregier05}, any image described using
polar shapelets may be easily `circularized' (i.e. circularly
averaged) by setting the model coefficients $f_{n,m} = 0$ for all $m
\ne 0$.  The circularized GEMS PSF generated in this way can be seen
in the central panel of Figure~\ref{fig:gemspsf} (imaged at a
resolution of 0.015 arcsec pixel$^{-1}$), and in the rightmost panel
(at the final output resolution of the simulations, 0.03 arcsec
pixel$^{-1}$). We choose to use the circularized GEMS PSF for the
simulations in this Paper as its symmetry simplifies the
interpretation of lensing measurement results, while still
incorporating a radial profile characteristic of space-based data such
as that from ACS.  Diffraction spikes, such as caused by support
struts for the secondary mirror, will be lacking from this
adopted PSF model, but the extent to which these are
successfully characterized by shapelets PSF models is unclear even
when not circularizing, as in Figure~\ref{fig:psfudf}.  The radial
profile of this circularized PSF model is shown in
Figure~\ref{fig:gemspsf_radial}.

Since shapelet models are no longer being used to describe galaxies
after shear or flexion is applied, the convolution must be performed
numerically using a pixelized image of this PSF.  As a shapelet model
PSF such as that in Figure~\ref{fig:gemspsf} is not \emph{formally}
band-limited (see, e.g., \citealp{marks09}), this therefore requires
another empirical investigation into the effects of finite sampling.

That this investigation is numerically feasible also illustrates a
further advantage of not performing the calculation using shapelet
transformations. As shown by \citet{melchioretal09}, an exact shapelet
treatment of convolution results in a convolved image of order
$n_{\textrm{max}}$ given by
\begin{equation}
n_{\textrm{max, convolved}} = n_{\textrm{max, galaxy}} + n_{\textrm{max, PSF}}.
\end{equation}
The PSF in ACS images requires $n_{\textrm{max, PSF}} \gtrsim 20$,
which makes describing perfect convolution on even modestly-sized
galaxy shapelet models extremely expensive. Re-truncating the
convolved model to the original $n_{\textrm{max}}$ spoils the
exactness of the shapelet convolution treatment in a way that varies
depending on the shape of each galaxy. In contrast convolutions on
images, performed using the fast Fourier Transform (FFT, a fast
algorithm for performing discrete Fourier Transforms: see e.g.\
\citealp{pressetal92}), can be performed to great accuracy in a
fraction of the time.  We test the accuracy that may be achieved using
this technique in Appendix \ref{sect:convol}, and find that, as in the
case of the raytracing representation of shear and flexion upon
shapelet galaxy models, a desired level of accuracy can be achieved
without great additional numerical cost.

\subsection{Design strategy for distorted, convolved images for the
  ACS simulations}\label{sect:strategy}

The results of Appendices \ref{sect:upsample} and \ref{sect:convol} now
allow informed decisions to be made regarding the construction of the
ACS simulations to test the measurement of flexion and shear with
shapelets.  Current shear measurement methods are able
to measure shear at percent-level accuracy
\citep{bridleetal10,kitchingetal12}, and it is unlikely that flexion
measurement will approach this capability within the near term. A
conservative requirement is therefore that the treatment of distortion
and convolution in the simulations should be accurate at the $10^{-3}$
level in terms of the impact on $\gamma$, $\fflex$ and $\gflex$ in
simulated galaxies.  This also matches stated requirements on the
estimation of shear for an all-sky, space-based survey mission
\citep{amararefregier08}.

We therefore adopt an upsampling ratio of $r=2$ when applying both the
lensing image distortions (shear and flexion) and for subsequent
convolution with the circularized PSF of Figure~\ref{fig:gemspsf},
corresponding to an absolute resolution of 0.015 arcsec pixel$^{-1}$.
It is computationally convenient that both distortion and convolution
occur at this same resolution.  The results of
Appendices~\ref{sect:upsample}~and~\ref{sect:convol} suggest that
errors in the pertinent moments of simulated galaxy images will be
thus controlled to better than 0.1$\%$ on average.

A remaining issue is the level of input shear and flexion distortions
which should be applied to the simulation images.  For shear we
require successful recovery of the signal due to correlated large
scale structure, but also for galaxy-galaxy lensing and the stronger
shears expected around cluster lenses.  For flexion, it is unlikely
that the cosmological signal is measurable in the near future, but the
galaxy-galaxy signal may be of interest (e.g.\ B06) and there are
certainly applications in the field of cluster reconstruction (e.g.,
\citealp{leonardetal07,leonardetal11}).

For the suite of simulations we choose to split our input signals into
three broad groups exploring a range of distortion signal strengths,
labelled as `high', `mid' and `low'. These designations refer to the
magnitude of the input gravitational signal applied, chosen to be
$|\gamma_{\textrm{input}}| = 0.1$, 0.05, 0.01 respectively for the
shear simulations.

For the flexion simulations it was decided to bring the `mid' and `low'
samples closer together (to concentrate on exploring sensitivity to
galaxy-galaxy flexion) and extend the high signal somewhat further to
investigate measurement of values which may be found in cluster
studies (B06; \citealp{leonardetal07,leonardetal11}).  We choose $|\fflex_{\textrm{input}}|,
|\gflex_{\textrm{input}}| = 0.05$, 0.01, 0.005 arcsec$^{-1}$ for `high',
`mid' and `low', respectively.

In order to explore any anisotropy in signal recovery (due, e.g., to
alignment with pixel axes) we split each of the three sets into a
further three subsets by the angle of orientation of the input signal
with respect to the image $x$-axis.  Orientations of $\phi =
0^{\circ}$, 30$^{\circ}$, 45$^{\circ}$ were chosen for the input
signals $\fflex_{\textrm{input}} = |\fflex_{\textrm{input}}|
{\me}^{\mis \phi}$, $\gamma_{\textrm{input}} =
|\gamma_{\textrm{input}}| {\me}^{2 \mis \phi} $ and
$\gflex_{\textrm{input}} = |\gflex_{\textrm{input}}| {\me}^{ 3 \mis
  \phi}$, giving a total of $3 \times 3 = 9$ subsets overall for each
of the three lensing distortions.  These values and choices are
summarized in Table \ref{tab:inputs}.  We note that these values span
a predominantly positive range of values in the components of
$\gamma$, $\fflex$ and $\gflex$: this asymmetry in the applied signal
is benign due to the circular symmetry of the adopted simulation PSF
(Figure~\ref{fig:gemspsf}).

\begin{table}
  \caption{Flexion and shear input signal values for the 270 simulated ACS tiles, where the first flexion input is given by $\fflex_{\textrm{input}} = |\fflex_{\textrm{input}}| {\me}^{\mis \phi}$, the shear by $\gamma_{\textrm{input}} = |\gamma_{\textrm{input}}| {\me}^{2 \mis \phi} $ and the third flexion input by $\gflex_{\textrm{input}} = |\gflex_{\textrm{input}}| {\me}^{ 3 \mis \phi}$ (see Section \ref{sect:strategy}).} 
\label{tab:inputs}
\begin{center}
\begin{tabular}{ccccrc}
\hline
 & Set & $|\fflex_{\textrm{input}}|$, $|\gflex_{\textrm{input}}|$  & $|\gamma_{\textrm{input}}|$  & $\phi$ & Tiles \\
 &     & arcsec$^{-1}$ &   &  & \\
\hline \hline
 &1 & 0.05 & 0.1 & 0$^{\circ}$ & 1-10\\
`high' &2 & 0.05 & 0.1 & 30$^{\circ}$ & 11-20 \\
 &3 & 0.05 & 0.1 &  45$^{\circ}$ & 21-30 \\
\hline 
 &4 & 0.01 & 0.05 & 0$^{\circ}$ & 31-40 \\
`mid' &5 & 0.01 & 0.05  & 30$^{\circ}$ & 41-50 \\
 &6 & 0.01 & 0.05 & 45$^{\circ}$ & 51-60 \\
\hline
 &7 & 0.005 & 0.01 & 0$^{\circ}$ & 61-70 \\
`low' &8 & 0.005 & 0.01  & 30$^{\circ}$ & 71-80 \\
 &9 & 0.005 & 0.01 & 45$^{\circ}$ & 81-90 \\
\hline
\end{tabular}
\end{center}
\end{table}
Ten survey tiles were then simulated for each image set described
above, each with a $3.53 \times 3.63$ arcmin$^2$ sky coverage area
(an ACS pointing) and at an output resolution of 0.03 arcsec
pixel$^{-1}$. These were created by:
\begin{enumerate}
\item Randomly selecting with replacement from the HUDF starter set
  described in Section \ref{sect:udf}.
\item Randomly perturbing, rotating and inverting each starter set
  galaxy model as described in Section \ref{sect:starter}.
\item Applying a lensing distortion as prescribed by Table
  \ref{tab:inputs} using the raytracing method presented in Section
  \ref{sect:raytrace} with an upsampling ratio of $r=2$.
\item Convolving each distorted image with the circularized GEMS PSF
  of Figure \ref{fig:gemspsf}. This convolution was performed at the
  image level using FFTs as described in Section \ref{sect:convol},
  again using $r=2$ for the upsampling ratio.
\item Placing each simulated galaxy image at a random position in the tile.
\end{enumerate}
In this way $3 \times 90$ simulated ACS tiles were created for each of
$\gamma$, $\fflex$ and $\gflex$.  We note that it was decided not to
include simulation tiles containing \emph{both} shear and flexion
input signals simultaneously.  The possibility of cross-contamination
between these signals is interesting (see \citealp*{violaetal12}), but
in this initial study we concentrate on examining each signal
individually.

In order to take advantage of the reduction in shape noise that can be
achieved by combining lensing measurements from appropriately rotated
galaxy images \citep{masseyetal07step}, we generate a further set of
`rotated' simulation images. These are identical to those described
above except for an additional rotation of 180$^{\circ}$, 90$^{\circ}$
and 60$^{\circ}$ (for $\fflex$, $\gamma$ and $\gflex$ respectively),
given to the starter set galaxies immediately after step (ii) above.

This allows the distortion measurements for matched pairs of rotated
images to be averaged, cancelling the leading order impact of noise
from the
intrinsic galaxy shape upon shear and flexion measurement. This also
allows the relative impact of galaxy shape noise and image pixel noise
to be compared, which is of particular interest for flexion.  The
total number of simulated ACS tiles in each suite is therefore $2
\times 90$, leading to a full suite of $6 \times 90 = 540$ simulated
ACS pointings ($\sim 100$ Gb of data in total).

\subsection{Correlated noise}\label{sect:noise}

The final ingredient in the creation of simulated images is the
addition of realistic measurement noise, due to diffuse background
light and finite photon number counts.  An important consideration for
shape measurement is the impact of spatially \emph{correlated}
noise. This is present due to the standard practice of combining
multiple, dithered ACS exposures to generate single `science' images
using software such as M\textsc{ultidrizzle}
\citep{fruchterhook02,koekemoeretal02}, and is generic even for more
carefully optimized linear combination schemes (e.g.,
\citealp*{roweetal11}).  Such dithered images were used as the basis
for weak lensing measurement in each of GEMS, COSMOS and STAGES
\citep{rixetal04,heymansetal05,leauthaudetal07,heymansetal08,caldwelletal08,grayetal09}.

We add realistic correlated noise to the simulation tiles described in
Section \ref{sect:strategy} in a novel manner, using a `noise mosaic'
image constructed as part of the imaging reduction of the GEMS survey
\citep{rixetal04}. This composite image, which is the size of a single
ACS pointing, is a mosaic of multiple blank sky regions in the GEMS
$V_{606}$ science images described by \citet{caldwelletal08}. Each
0.03 arcsec pixel$^{-1}$ GEMS science image was generated from a
dithered combination of three 0.05 arcsec pixel$^{-1}$ exposures, each
of total duration 2160s.  The composite noise image therefore reflects
precisely the noise-correlating effects of the GEMS dither and drizzle
strategy, which is typical of the reduction strategies employed for
high-resolution, space-based imaging data.  The galaxy image magnitude
zero-points were set to match the GEMS data, resulting therefore in
simulated survey images with a detection threshold around $V_{606}
\sim 27$, matching the GEMS survey characteristics quite closely
\citep{caldwelletal08}.  The fact that the images include many more
simulated objects, buried in the noise down to the HUDF starter set
cutoff magnitude of $V_{606} < 29$, is another realistic feature of
the simulations.

As galaxies are placed at random locations in each successive
simulation tile, it is possible to re-use this noise image each time.
Only a small fraction of galaxy images within tiles will consist of
galaxy models that, by chance, have been repeatedly placed into the
same location in the ACS pointing-sized noise mosaic ($\simeq 3.53
\times 3.63$ arcmin$^2$ in total area), so the level of unwanted pixel
noise repetition in the simulation images will be low.  In these
simulations we do not add an additional Poisson noise term to image
pixels to account for variation in variance due to varying flux.
However, as the overwhelming majority of simulated galaxies in the
catalogues are faint relative to the background this can be shown to
represent a very small correction to the images, arguing that the
effect can be safely neglected in light of other uncertainties in the
modelling overall.

One final concern in the use of the GEMS noise mosaic might be if
there were to be found to be some preferred direction in the noise
pattern.  Visual inspection of the noise map did not suggest any such
artefact, and it will be seen in Section~\ref{sect:flippipe} that
there is no evidence for an \emph{additive} bias in shear or flexion
results, or a variation in multiplicative `$m$' biases
\citep{heymansetal06step} with polar angle with respect to the pixel
grid. Global anisotropies or preferred directions in the noise map
might be expected to cause such effects in the presence of a
circularly-symmetric PSF such as that adopted
(Figure~\ref{fig:gemspsf}), and the fact that we are unable to detect
them within simulation uncertainties suggests that they exist at a
sufficiently low level as to not affect the conclusions of this study.

It should be noted that by reusing the GEMS noise mosaic for each tile
in this way we are applying the same noise field to both galaxy images
in the rotated and un-rotated galaxy pairs, as in these pairs the
simulated galaxies share the same location.  However, this offers an
opportunity to separate the effects of pixel noise in flexion
measurements from those of shape noise without diluting the
former. The price paid for this opportunity is a reduction in the
overall statistical power, by a factor of $\sim\sqrt{2}$ in the simplest
estimate, with which measurement bias parameters may be constrained.

\subsection{Summary and comparison to previous flexion simulations}
\begin{figure}
\begin{center}
  \psfig{figure=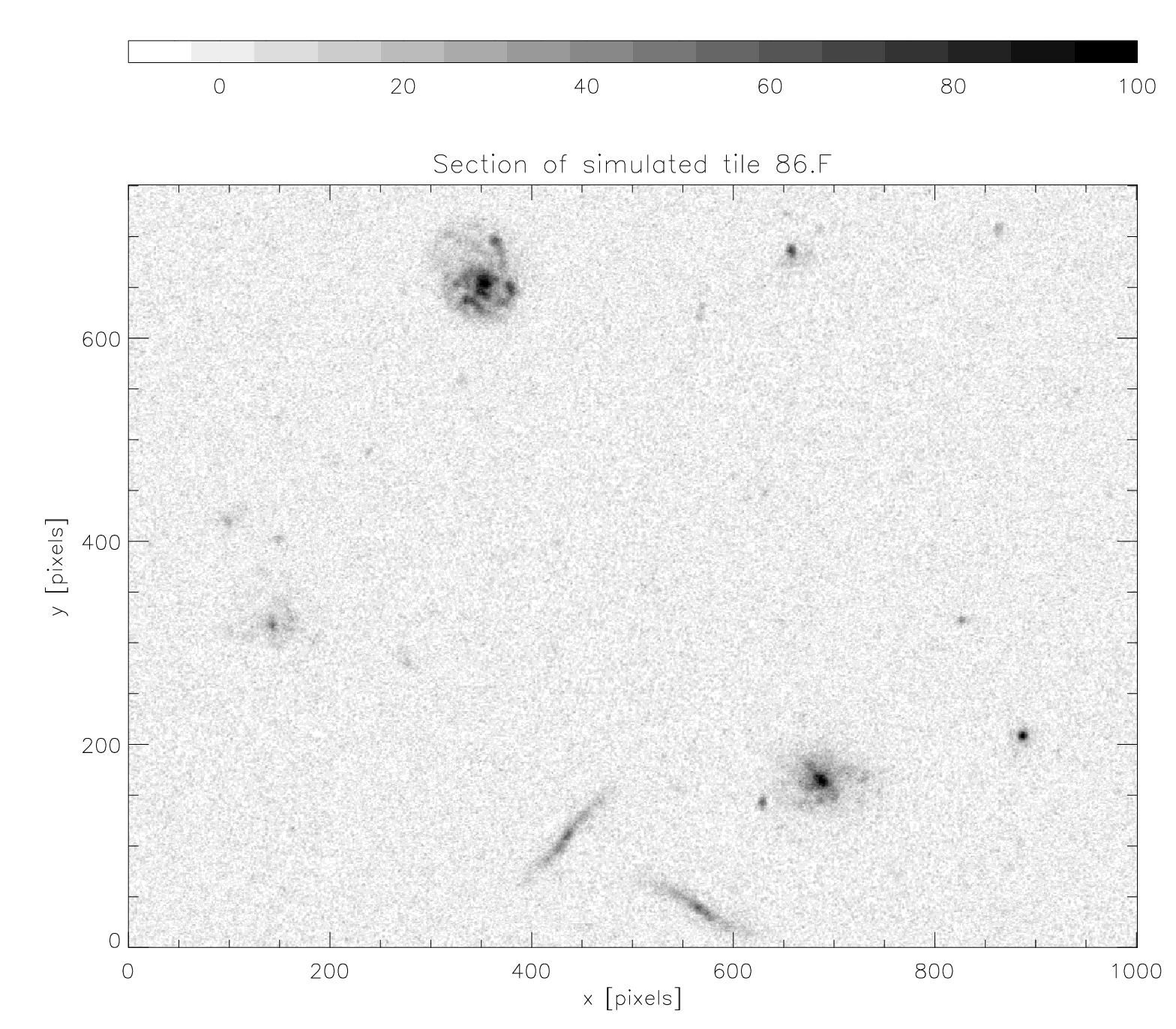, width=0.5\textwidth} {\caption[$30
    \times 22.5$ arcsec$^2$ section of a simulated galaxy
    image.]{Illustrative 30 arcsec $\times$ 22.5 arcsec section of the
      86th tile (chosen at random) in the simulated, convolved,
      $\fflex$-lensed images. The grey-scale is linear in
      flux.% for the $m_{606}$ $ab$-zeropoint of 34.648.
\label{fig:fliptile}}}
\end{center}
\end{figure}
After the addition of the noise image, the 540 ACS simulation tiles
are complete.  We now briefly summarize the differences between these
simulations and those of \citet{velanderetal11}, in case such a
comparison is of utility to the interested reader.  A primary
difference is that the simulations in this study employ complex
morphologies in the shapelet galaxy models, constructed from HUDF
imaging data, compared to the simpler parametric forms as used in
\citep{velanderetal11}.  Our simulations also employ a continuous
distribution of object sizes and signal-to-noise ratio (SNR), taken directly from
the UDF sample, rather than fixing galaxy sizes and SNR at
fixed values of interest.  Whereas the \citet{velanderetal11} PSF is a
Moffat profile motivated by ground-based PSFs (and made elliptical for
some sets of simulated images), we instead employ a circular PSF with
a radial profile taken from shapelet fits to GEMS stellar images.
Deblending, and realistic noise associated with deblending, is a
feature in our simulations, whereas \citet{velanderetal11} placed
galaxies on a regular grid.  Finally, our simulations employ
correlated noise taken from observations in the GEMS survey data
\citep{caldwelletal08}.

In Figure \ref{fig:fliptile} we show a small section
from one of these tiles, illustrating some of the realistic diversity
of galaxy morphology depicted in the simulation images.  All of these
simulation images are available by request from the authors.  We now
turn to a description of the measurement of shear and flexion in these
images using shapelets, allowing a calibration of the shear and
flexion signal recovery using this method.

\section{Testing a shapelet flexion pipeline}\label{sect:flippipe}

We now describe measurements of flexion and shear made from the
simulations described in Sections \ref{sect:udf} and \ref{sect:sims},
and investigate the recovery of flexion and shear as a function of
image properties such as SNR and galaxy size.
%Here, like \citet{velanderetal10}, we are testing a shapelet pipeline.

\subsection{Object detection and shapelet decomposition}
\label{sect:detect}
To test and calibrate shapelet measurements of shear and flexion, we
treat the 540 ACS simulation tiles as if they were new telescope data
(with properties such as galaxy positions and shapes unknown).  The
first step is therefore to detect objects in the images from peaks in
the surface brightness distribution, and we employ the same techniques
as described in Appendix~\ref{sect:selec} to create a catalogue of
galaxy objects in the HUDF.  We detect galaxy objects in the
simulations using the SE\textsc{xtractor} software
with the parameter choices given in
Table~\ref{tab:sexparams}.  An initial cut of $V_{606} < 27$ is
then applied to the catalogues, along with requiring
$\textsc{flux\_radius} > 2.4$ and $\textsc{flux\_auto} /
\textsc{fluxerr\_auto} > 10$ (see \citealp{bertinarnouts96} for
descriptions of these \textsc{SExtractor} output parameters).  These
cuts are motivated by the choices made in \citet{heymansetal05},
and result in catalogues containing approximately 64 galaxies
arcmin$^{-2}$.  This figure agrees well with galaxy densities found in surveys at
similar depth to that simulated here, such as GEMS, COSMOS, and STAGES 
(e.g., \citealp{heymansetal05,leauthaudetal07}).

Postage stamp images of each detected
galaxy object are then created as described in Appendix~\ref{sect:pstamp}
and shapelet models of the galaxies are created as described in
Appendix~\ref{sect:modudfgals}, except for two important differences
that we now describe.
The first and most important difference is in the PSF: for
this we use the shapelet model of the circularized GEMS PSF
(Figure~\ref{fig:gemspsf}) described in Appendix~\ref{sect:convol}.
Therefore our calibration of flexion and shear measurement does not
include the potential effects of a poorly-modelled PSF, such as might
be present working with real data.  While unrealistic, this
simplification will allows any measurement biases to be interpreted
cleanly rather than being subject to external factors such as poor PSF
modelling.  The problem of building accurate PSF models is important
enough to be addressed in its own right, and this is increasingly
reflected in the literature (e.g.\
\citealp{hoekstra04,jarvisjain04,jarvisetal08,paulinetal08,paulinetal09,
rowe10,kitchingetal11,kitchingetal13star}).

The second important difference is in the choice of the
\textsc{neighbour} input parameter to the shapelet software, setting
$\textsc{neighbour} = 1$.  As described in Appendix \ref{sect:pstamp},
this causes pixels in the masked areas of each postage stamp (i.e.\
those associated with nearby objects, or cosmic rays, bad pixels etc.\
in real data) to be given zero weighting at the shapelet modelling
stage. Having made these changes to the input settings, the modelling
provides a catalogue of shapelet coefficient values for each galaxy in
the rotated and unrotated simulations.

The galaxies in the rotated and unrotated simulations are then
matched, treating as pairs all galaxies with centroids separated by
less than 0.15 arcsec in centroid (five 0.03 arcsec pixels) and 0.25
in $V_{606}$ magnitude as estimated by SE\textsc{xtractor}. This was
found to produce a total of $\sim$$51 \, 000$ matched galaxy pairs for
each of the shear and flexion simulation suites, reducing the galaxy
density to approximately 45 galaxies arcmin$^{-2}$.  More stringent
pair exclusion criteria were seen to cause significant reductions in
the numbers of matched pairs. Opting for more tolerant criteria
produced slightly greater numbers of galaxies ($\sim$1-3\%), but not
in a manner that substantially altered final results. The magnitude
cut had greatest impact of the pair matching criteria, suggesting that
the allocation of objects and flux in the deblending process was
dominating over centroid errors in the contribution to the loss of
matching pairs.  The fraction of galaxies lost to catastrophic failures in shapelet
modelling was small (0.4\%).

%The lack of improvement in overall
%results as matching criteria were relaxed suggests that this
%ceiling is being approached with our adopted values. 

The difficulty of matching galaxies suggests that significant numbers
of objects are being affected by noise in the determination of their
properties at the SE\textsc{xtractor} detection and deblending stage;
we found $\sim$20-30\% of pairs could not be matched in this
simulation, although it should be stressed that, in general,
deblending errors will depend on PSF shape, noise correlation, other
aspects of data quality, galaxy population morphology and SNR.  In
these simulations the rate of
successful matching depended upon SNR in particular, with the brighter
galaxies being matched better than the faint.  However, deblending
does represent a realistic additional source of noise for shear and
flexion measurements that is not due to the intrinsic shape of
galaxies, and so it is an effect of interest for these simulations and
for the \citet{caldwelletal08} two-stage object detection strategy in
the adopted pipeline.

Unfortunately, this effect sets a ceiling on the number of objects
that can be successfully matched after simulating an end-to-end
pipeline in this way. While including deblending in the simulation
test contributes realistic noise to the end measurements, it also
reduces the constraining power of the simulations as a whole when
significant numbers of objects are lost.  To isolate the
effects, it may be preferable in future work to generate galaxies on
grids as done in GREAT08/GREAT10
\citep{bridleetal09,bridleetal10,kitchingetal11,kitchingetal12} and
compare results.  We will discuss the issue of deblending further in
Section \ref{sect:conc} in the context of the results of the pipeline
tests.

For the pairwise matched sample, flexion and shear
estimates were then generated for each galaxy from the catalogue of shapelet
models, and the estimators chosen will now be described.

\subsection{Estimating flexion and shear}\label{sect:estflexshear}
To estimate shear and flexion from shapelet models of galaxies we
adopt an approach similar to that described by M07, and compare the
relative values of the shapelet coefficients $f_{n,m}$ of best-fitting
shapelet models to derive estimators of lensing distortions. In
Appendix \ref{App:shapeflexest} we describe the generation of flexion
and shear estimators from catalogues of shapelet coefficient models
for a population of galaxies with a realistic distribution of fluxes
and sizes.  We note that the multiple-decade variation in these two
properties means that small modifications to the estimators proposed
in M07 must be adopted. Described in detail in Appendix
\ref{App:shapeflexest}, we label these estimators $\tilde{\gamma}$,
$\tilde{\fflex}$, and $\tilde{\gflex}$ for the shear, $\fflex$ flexion
and $\gflex$ flexion, respectively.  

We also note that in order to use these estimators,
it was made a condition that the shapelet model reach sufficient $n_{\rm max}$
as to contain non-zero values for all the shapelet coefficients required by equations
\eqref{eq:rawshear}, \eqref{eq:d11} and \eqref{eq:d33} when estimating
$\gamma$, $\fflex$ or $\gflex$, respectively.  This is to avoid bias
from artificially setting coefficient values to zero when this is not
the most appropriate prior expectation.

Overall, this approach differs from that of \citet{velanderetal11},
who instead take circular profiles and estimate the shear and flexion
required to distort these objects to match the data, in a manner
similar to the shear-only estimators of \citet{kuijken06} and
\citet{bernsteinjarvis02}.  Both methods are similar in that they rely
to some extent upon shapelet models being an accurate description of
the underlying surface brightness distribution of galaxies to avoid
what has been identified as `underfitting bias' in galaxy shape
estimation (e.g. \citealp{voigtbridle10,bernstein10}).  However, in
our simulation tests shapelet models have been used
to provide the underlying galaxy light profiles, and perfect knowledge
of the PSF is also available.  

In principle, this allows the direct probing of potential
biases due to the use of noisy data, such as imperfect deblending with
SE\textsc{xtractor}, or the biased response of non-linear estimators
under noise itself (so-called `noise bias', e.g.\
\citealp{bernsteinjarvis02,hirataetal04,refregieretal12,kacprzaketal12,melchiorviola12}).
However, in practice the shapelet method of \citet{masseyrefregier05}
truncates models once they become consistent with the noise in the
image, iterating the modelling parameters $\beta$ and $n_{\rm max}$
until this is achieved, while always seeking the model with the lowest
$n_{\rm max}$ that meets this criterion.  Galaxy models will therefore
in general still be truncated, leading to the distinct possibility of
underfitting biases still being present in lensing estimates from the
shapelet models constructed in Section \ref{sect:detect}.

Using the method described in Appendix \ref{App:shapeflexest},
estimators of shear and flexion are constructed for each of the
galaxies with successful detections and shapelet decompositions
described in Section \ref{sect:detect}.  In the following sections, we
compare these estimates to the input gravitational signal, and explore
how the the properties of these estimates vary with galaxy properties
such as SNR and size.

\subsection{Flexion and shear estimator results}\label{sect:calresults}
\begin{figure}
\begin{center}
  \psfig{figure=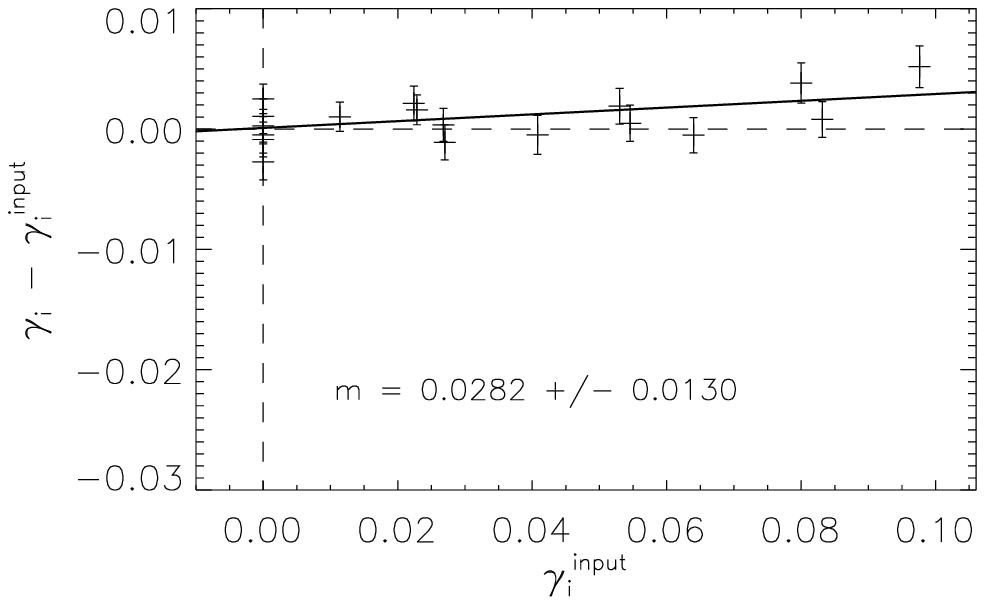, width=0.5\textwidth}
  \psfig{figure=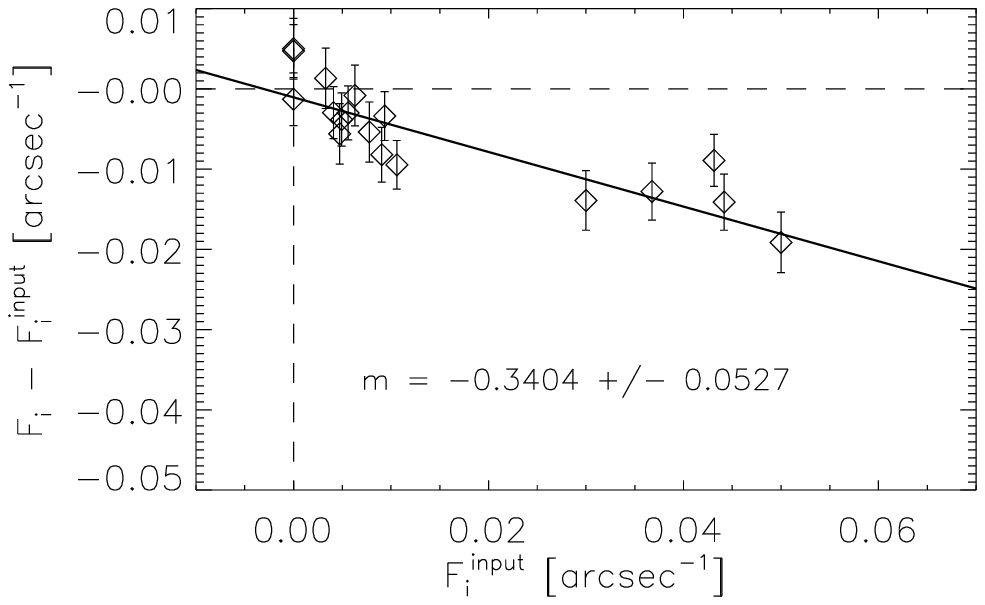, width=0.5\textwidth}
  \psfig{figure=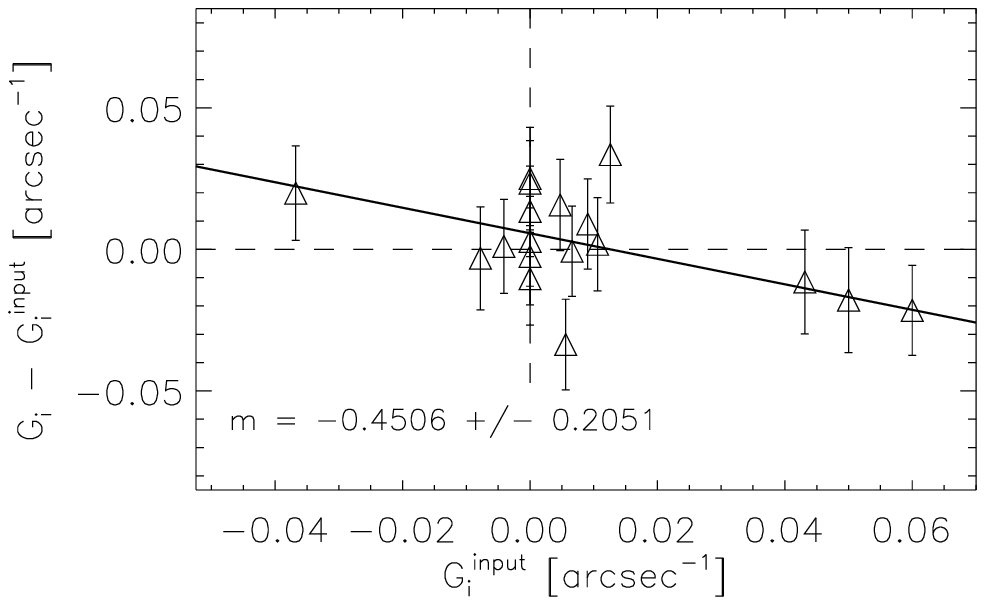, width=0.5\textwidth} {\caption{Lensing
      measurement calibration results from the full set of simulated
      galaxies in matched, rotated pairs, for $\tilde{\gamma}$ (top
      panel), $\tilde{\fflex}$ (central panel), and $\tilde{\gflex}$
      (bottom panel).\label{fig:calresults}}}
\end{center}
\end{figure}

In Figure \ref{fig:calresults} we plot $(\tilde{\gamma} -
\gamma_{\textrm{input}})$ versus $\gamma_{\textrm{input}}$ (top
panel), $(\tilde{\fflex} - \fflex_{\textrm{input}})$ versus
$\fflex_{\textrm{input}}$ (middle panel), and $(\tilde{\gflex} -
\gflex_{\textrm{input}})$ versus $\gflex_{\textrm{input}}$ (bottom
panel) for our simulated galaxies, with estimators for rotated and
unrotated pairs of galaxies being mean-averaged to reduce noise as
described by \citet{masseyetal07step}.  Results were not found to
differ between components (i.e.\ results for $\gamma_1$ were
consistent with those for $\gamma_2$, etc.) and so the plots show both
real and imaginary parts for each signal.  Results were binned
according to input signal via the image sets described in Section
\ref{sect:strategy}, and the points for each bin represent the
\emph{median} of the averaged rotated and unrotated estimates in each
case.  For shear, the median results were consistent with results
derived from the arithmetic mean; for flexion, the mean was found to
be very noisy due to the distribution of flexion estimators (see
Section \ref{sect:flexdist}), and so the median was preferred as the
comparison statistic in both cases.

The estimated median of each variate in a bivariate distribution,
taken along each of its two dimensions independently, is a good
estimator of the central tendency in both variates provided that
correlations between them are linear.  We found no evidence of
non-linear correlations between the estimates of $\fflex_1$ and
$\fflex_2$, indeed no evidence of correlations at all, and so do not
consider the use of the median (rather than, e.g., convex hull
stripping; \citealp{velanderetal11}) to be a source of bias in this
analysis.  The uncertainties plotted show the standard error on the
median in each bin \citep{lupton93}.

We fit a linear relation to the results of Figure
\ref{fig:calresults}, deriving best-fitting slope $m$ and offset $c$
bias parameters as used in the STEP project \citep{heymansetal06step}:
\begin{equation}\label{eq:mc}
\tilde{\gamma} - \gamma_{\textrm{input}}=  m \gamma_{\textrm{input}} + c,
\end{equation}
with similar expressions for the two flexion estimators. The
best-fitting values, and uncertainties, are given in Figure
\ref{fig:calresults}.  The real and imaginary parts were again found
to give consistent results, and so the best-fitting $m$ and $c$
describe input versus output for both components.  We found that the $c$
values were consistent with zero in all cases.  This is 
expected for the purely circular PSF chosen for these simulations, and
for the square grid which imparts no preferred sign for any of the shears or
flexion distortions applied.  However, it provides a useful null test
of the algorithms adopted.

The value of the multiplicative bias $m$ was found to be relatively
small in the case of shear, $m = 0.028 \pm 0.013$.  These results are
comparable to those obtained with a number of shear estimation methods
in the GREAT challenges \citep{bridleetal10,kitchingetal12}, although
there are a number of differences between these simulations and those
used in GREAT08 and GREAT10 (e.g.\ correlated noise; the distribution
of galaxy sizes and SNR; overlapping objects; a purely
circular PSF).  For flexion we detect stronger multiplicative biases,
finding $m = -0.340 \pm 0.053$ for the median of the $\fflex$
estimators, and $m = -0.45 \pm 0.21$ for the median of the $\gflex$
estimators.  As discussed above in Section \ref{sect:estflexshear},
the shapelet approach adopted may be affected by underfitting bias (or
model bias) caused by shapelet truncation, or bias due purely to
noise, or both.

The values of the biases seen are comparable to those
identified by \citet{velanderetal11}, for galaxies based on analytic
profiles and a different distribution of sizes and SNR values.  We now
discuss the variation of our measured estimator biases as a function
of these properties, and discuss some reasons for the possible
presence of this bias.

\subsection{Dependence of bias on noise and apparent galaxy size}\label{sect:biasvsSNRR}
\begin{figure}
\begin{center}
  \psfig{figure=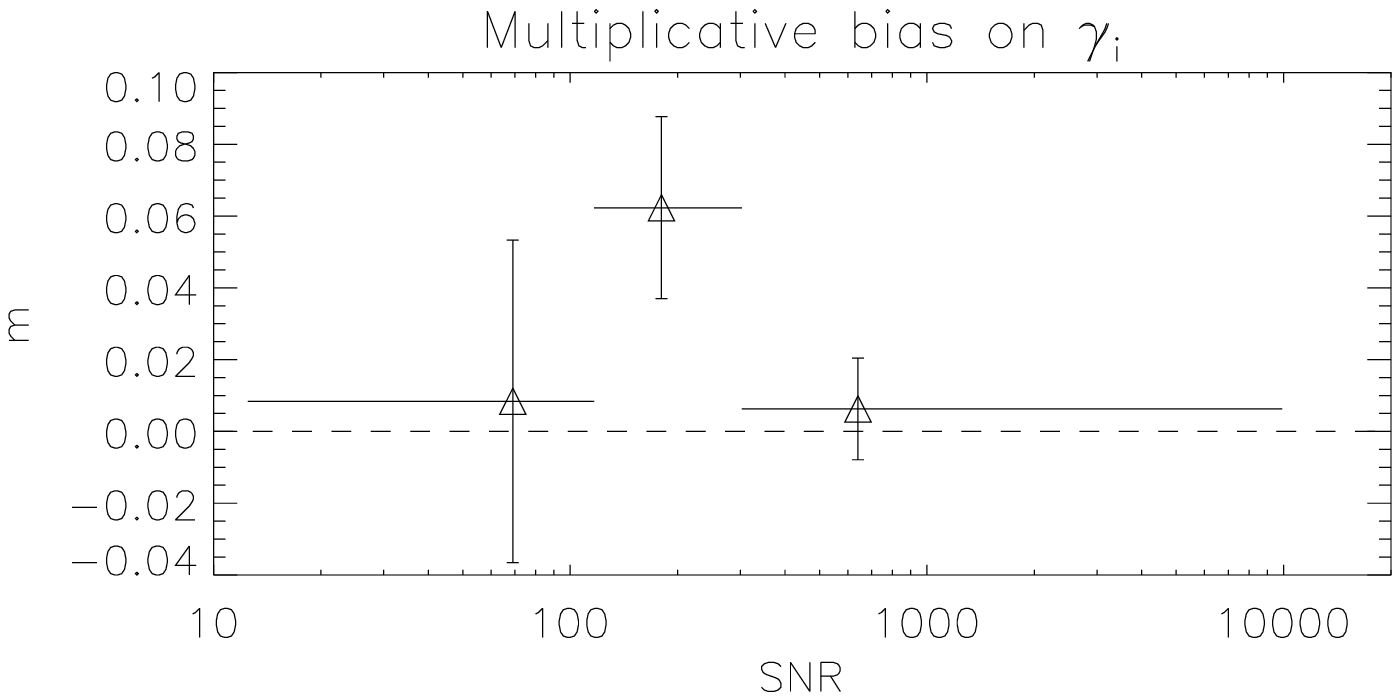,width=0.5\textwidth}
  \psfig{figure=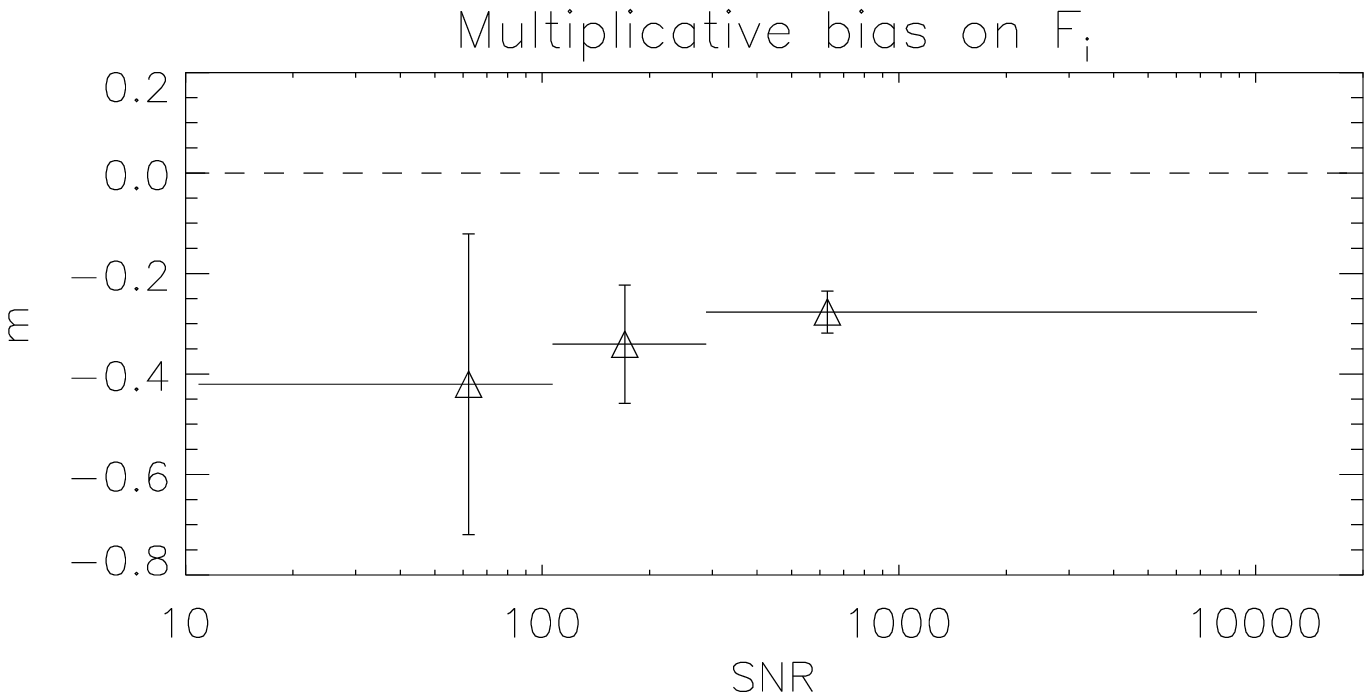,width=0.5\textwidth}
  \psfig{figure=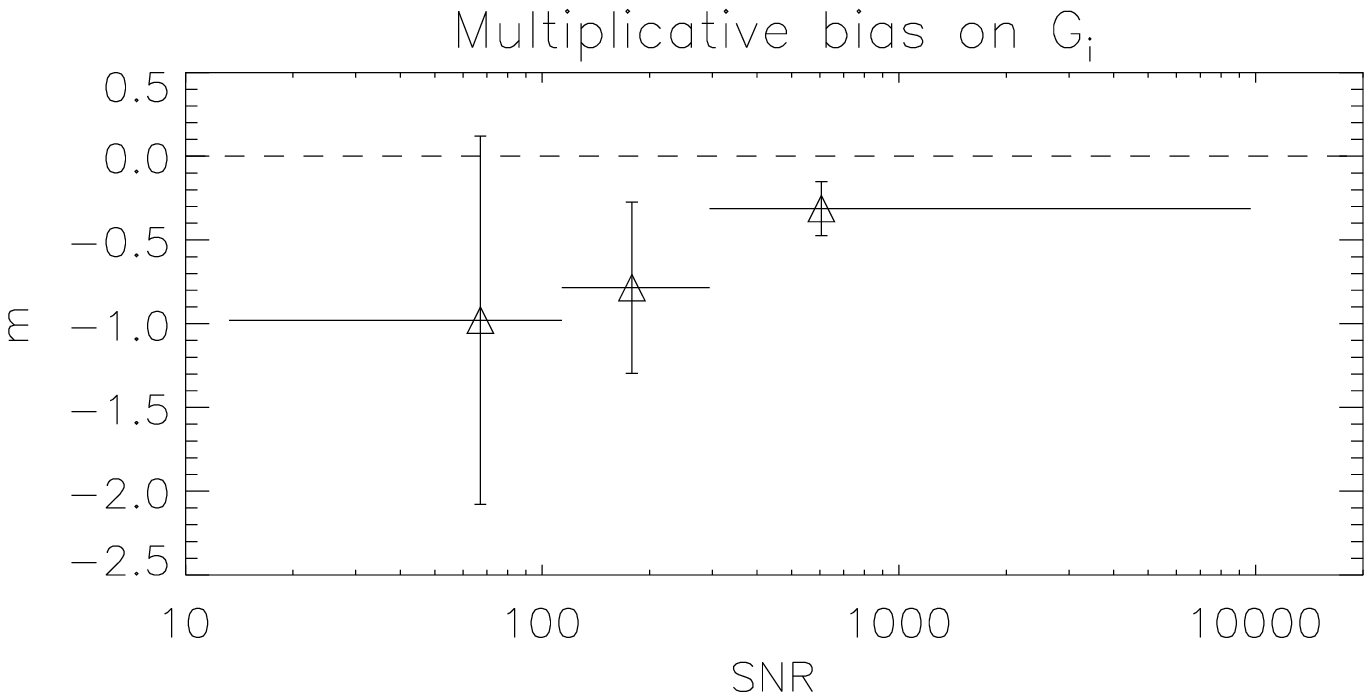,width=0.5\textwidth}
  {\caption{\label{fig:vsSNR} Variation of multiplicative bias $m$
      in
      shear and flexion estimation versus `observed' SNR for the
      simulation galaxies (see equation \ref{eq:mc}).  SNR bins were
      chosen to give equal numbers of galaxy in each bin: the increase
      in errors for low SNR objects is due to the increasing
      scatter in individual estimates. Solid, horizontal lines through points
      show the extent of each bin.}}
\end{center}
\end{figure}
In Figure~\ref{fig:vsSNR} we plot the dependence of the multiplicative
bias $m$ (see equation \ref{eq:mc}) upon
observed galaxy SNR for pair-matched shear and flexion
estimates from the simulations.  
Here, SNR is defined in terms of the \textsc{SExtractor} output
parameters \textsc{flux\_auto} and \textsc{fluxerr\_auto} as
\begin{equation}
{\rm SNR} = \frac{\textsc{flux\_auto}}{\textsc{fluxerr\_auto}} \times
1 / \sqrt{0.316},
\end{equation}
where the scaling factor $1/\sqrt{0.316}$ is taken from
\citet{leauthaudetal07} and adjusts SNR in drizzled \emph{HST} images
to account for correlated noise, adding the assumption that excess
Poisson variance due to object flux above the background is negligible
(it is not included in our simulations).  This is only an approximate
correction, based on a simplified model \citep{casertanoetal00} and an
assumption that the COSMOS drizzling approach closely resembles that
used in the GEMS noise mosaic (it does, although there is a small
difference in the kernel used for the Mk II GEMS reduction; see
\citealp{caldwelletal08}).  However, it does help take correlated
noise into account at a level of accuracy that is appropriate given
that differences between definitions of SNR can also introduce factor
$\sim$2 discrepancies.

The SNR ranges for the bins were chosen to
give equal numbers of galaxies in each bin; for reference, the bin
maxima and minima are given in Figure~\ref{fig:flexdist} and indicated
in Figure~\ref{fig:vsSNR} by solid horizontal lines.  It
is difficult to discern strong support for overall trends --- results for the lower SNR
bins are noisy compared
to those in higher SNR bins, particularly in the case of the flexion
--- although the higher SNR bin consistently gives arguably better results.
Overall, flexion results are broadly consistent with \citet{velanderetal11}, within
large errors, hinting
that the impact of more complex galaxy morphology (a key difference
between these simulations and those of \citealp{velanderetal11}) upon
flexion estimation bias is not great compared to other properties of
the observational data.  

\begin{figure}
\begin{center}
  \psfig{figure=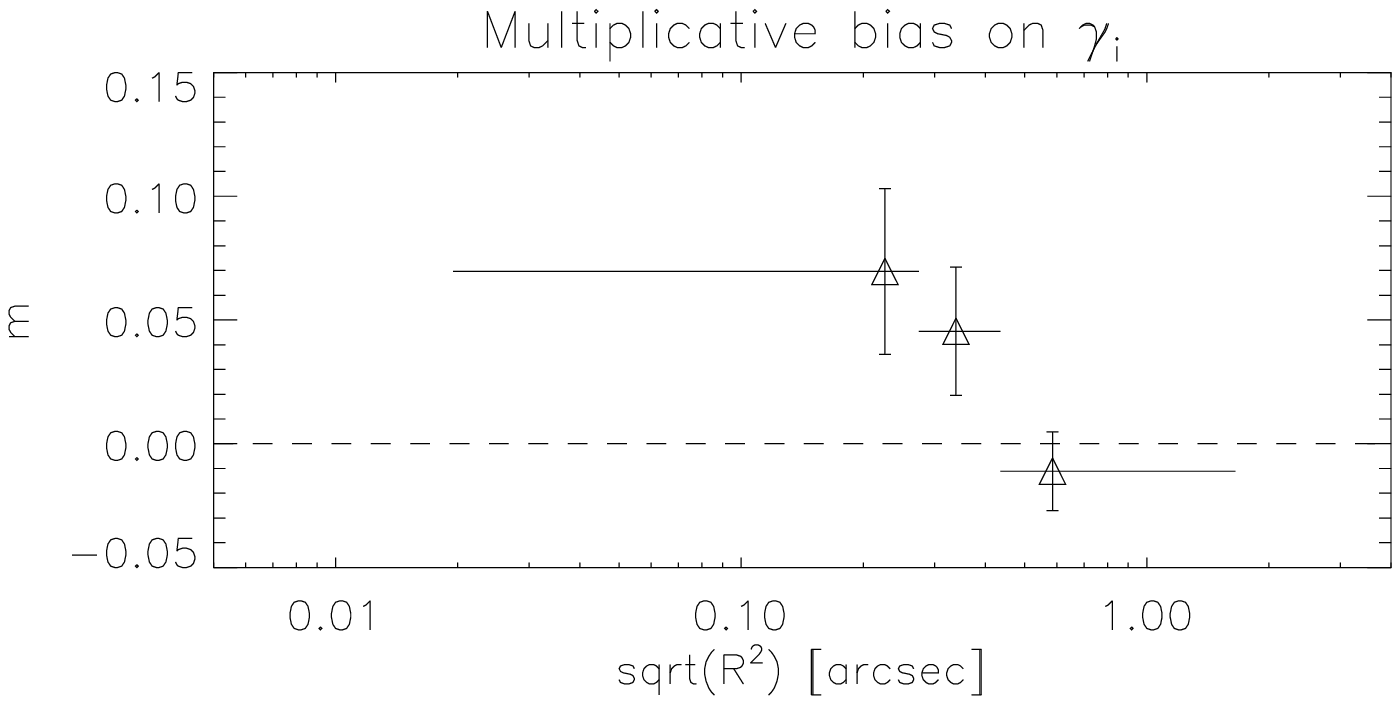,width=0.5\textwidth}
  \psfig{figure=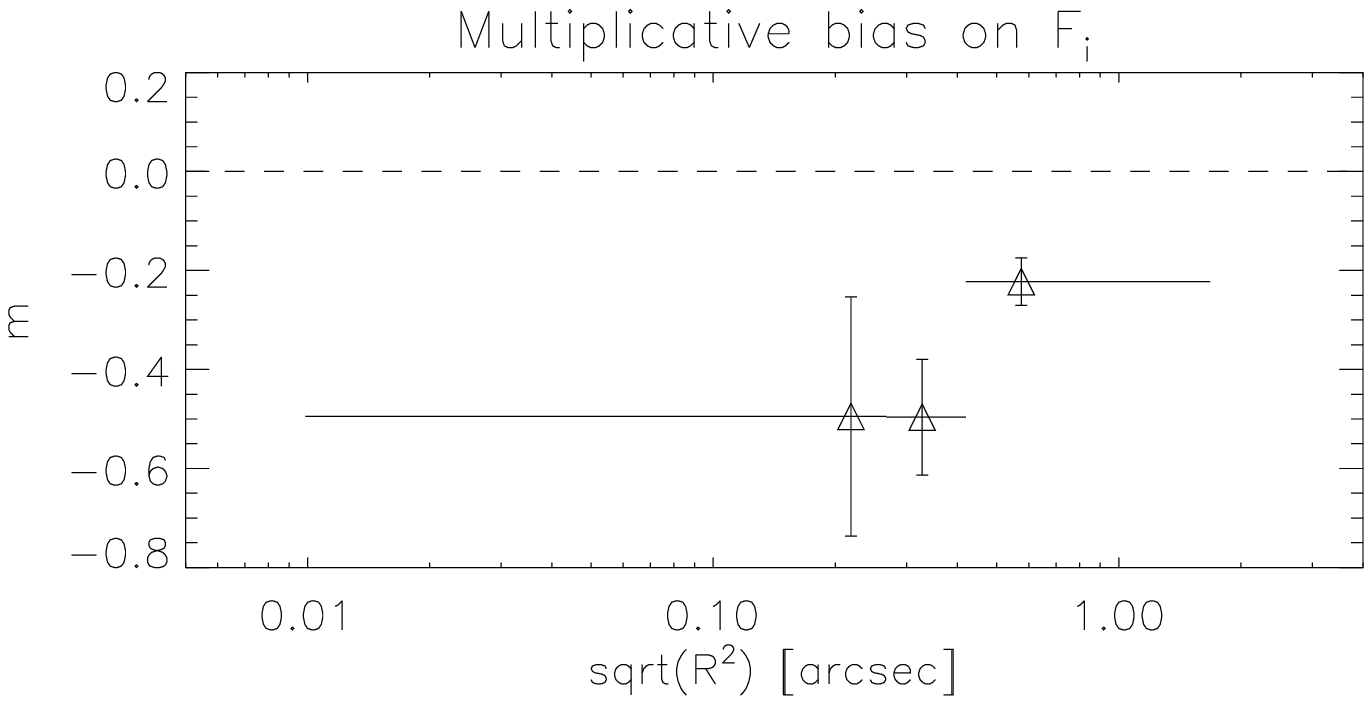,width=0.5\textwidth}
  \psfig{figure=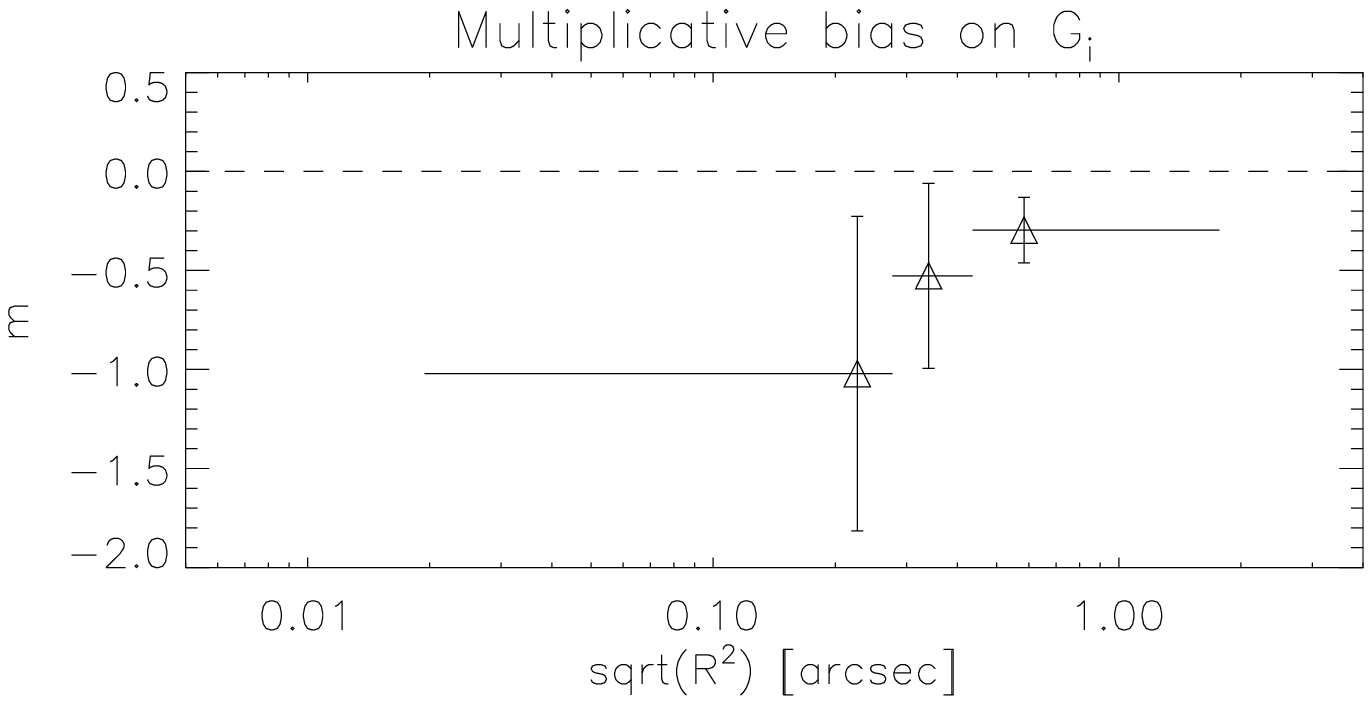,width=0.5\textwidth}
  {\caption{\label{fig:vsR}Variation of multiplicative bias $m$
      in
      shear and flexion estimation versus observed size for the
      simulation galaxies parametrized by $\sqrt{R^2}$
      \citep{masseyrefregier05}.  In a similar manner to the SNR
      binning, size bins were chosen to give equal
      numbers of galaxy in each bin: the increase in errors for
      smaller objects is therefore due to the increasing scatter in
      individual estimates.  Solid, horizontal lines through points
      show the extent of each bin.}}
\end{center}
\end{figure}
In addition to the SNR of galaxy objects, the size is an important
observable quantity for measuring flexion.  Unlike shear, flexion has
dimensions of inverse angle, and larger objects provide greater
leverage for measuring flexion.  In Figure \ref{fig:vsR} we plot $m$
for the simulated galaxy sample binned into three bins
reflecting angular size, chosen to given equal numbers of galaxies in
each bin.  The size estimate adopted for this binning is $\sqrt{R^2}$
for the best-fitting shapelet models, plotted in arcsec, where the
size measure $R^2$ derived from shapelet model coefficients is
described by \citet{masseyrefregier05} and given in equation
\eqref{eq:r2}.  It is equivalent to an unweighted, integrated
second-moment size over the shapelet model.

Results are qualitatively similar to those found for the SNR tests,
which is not altogether surprising as we expect a positive correlation
between size and SNR among the simulated galaxy population. There is
some evidence that estimators are better for larger apparent
sizes; the shear in particular now shows a marked trend towards improvement for
larger objects.  But the noise on estimates of $m$ for flexion
estimators become large for smaller objects, and so for flexion clear
trends are difficult to discern.

The large uncertainty on flexion estimates, despite the cancellation
of intrinsic shape noise at leading order, also provides a possible
explanation for the clear detection of significant $m$ biases.  Noise
bias (sometimes referred to as noise rectification bias, e.g.,
\citealp{bernsteinjarvis02,hirataetal04,melchiorviola12,refregieretal12,kacprzaketal12})
is an inherent property of statistical estimators derived from
non-linear combinations of random deviates, in this case the values of
noisy pixels.  For the shapelet technique used in this paper, the
linearity of the decomposition is broken by the presence of a
convolving PSF (see \citealp{masseyrefregier05}).  Flexion estimators,
being sensitive to higher order (and therefore noisier) shape moments
in galaxy images, are plausibly more sensitive to noise biases when
compared to shear estimators
derived from shapelets, as well as to underfitting biases
caused by shapelet model truncation (see Section
\ref{sect:estflexshear}). 
Another reason for a greater susceptibility to noise bias
might be the need to apply a centroid shift correction to in the
estimation of $\fflex$, although this would not be able to account for
the observed bias in $\gflex$ estimators.

\subsection{The distribution of flexion and shear measurements}
\label{sect:flexdist}
The results of Section \ref{sect:biasvsSNRR} (Figures \ref{fig:vsSNR}
and \ref{fig:vsR}) show a significant increase in the uncertainty of
estimates of $m$ as SNR (and the related property $\sqrt{R^2}$)
decrease, despite the number of objects being the same in each bin.
This effect is strongest in the measurements of $m$ for $\fflex$ and
$\gflex$.

\begin{figure*}
\begin{center}
\psfig{figure=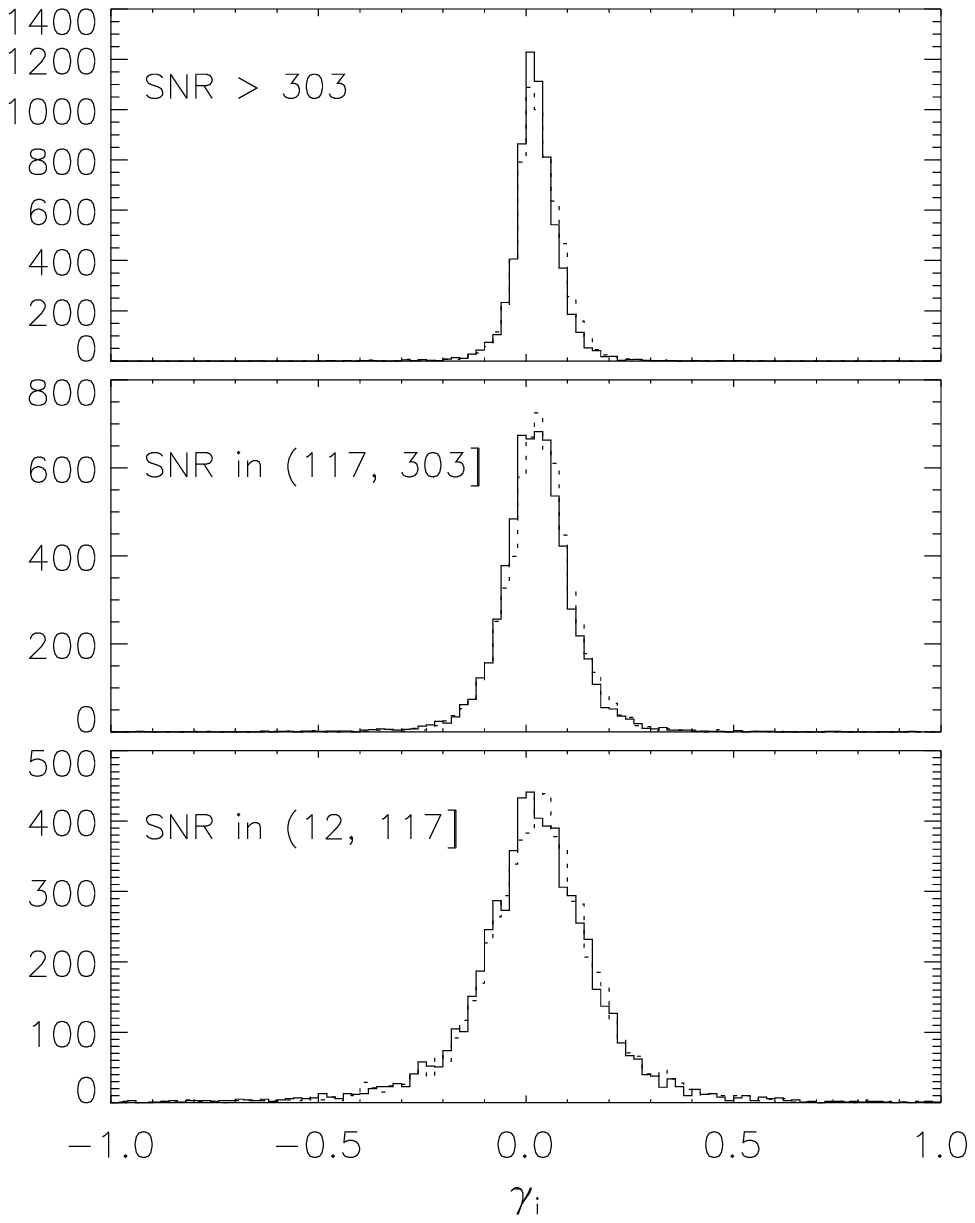,width=0.33\textwidth}
\psfig{figure=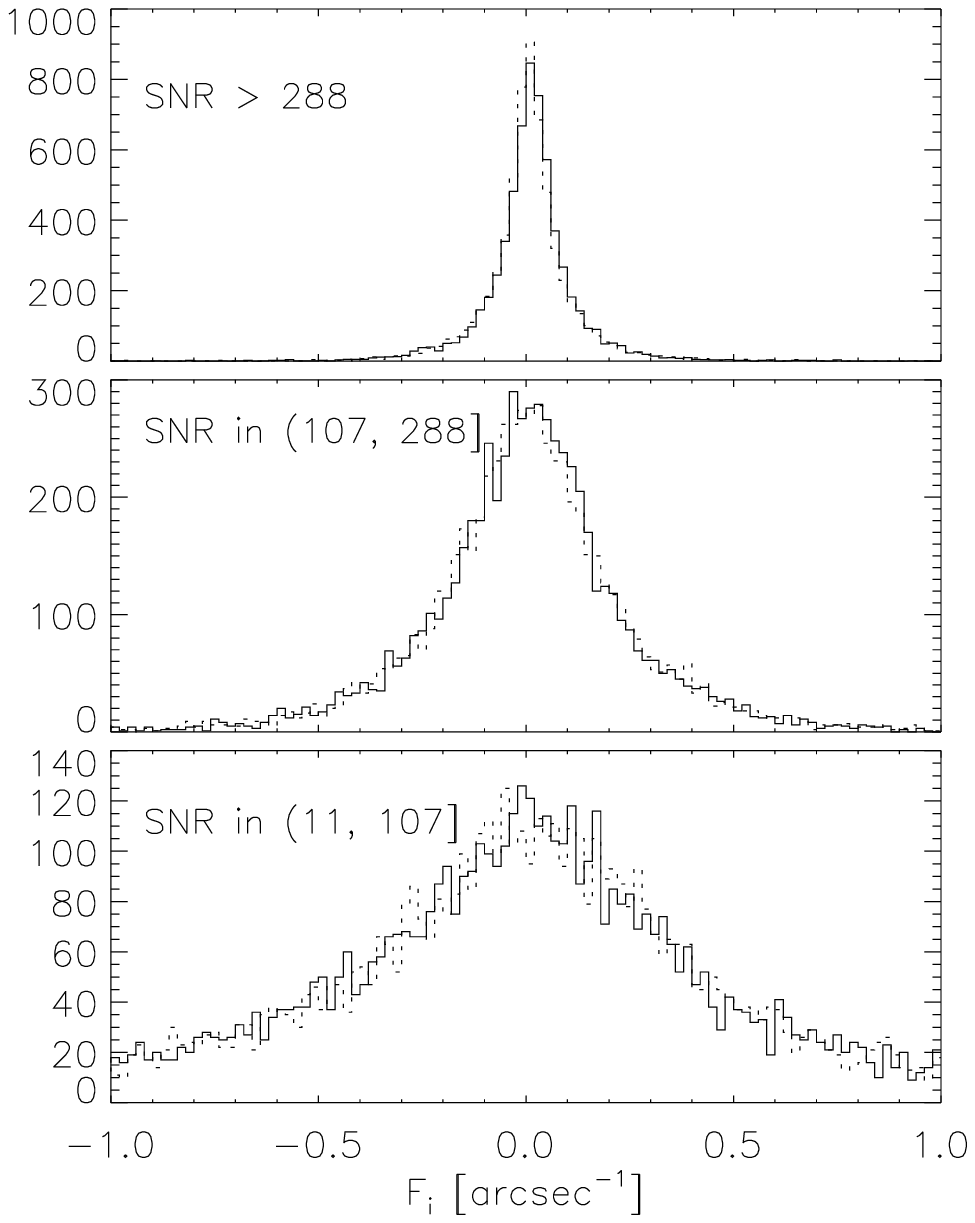,width=0.33\textwidth}
\psfig{figure=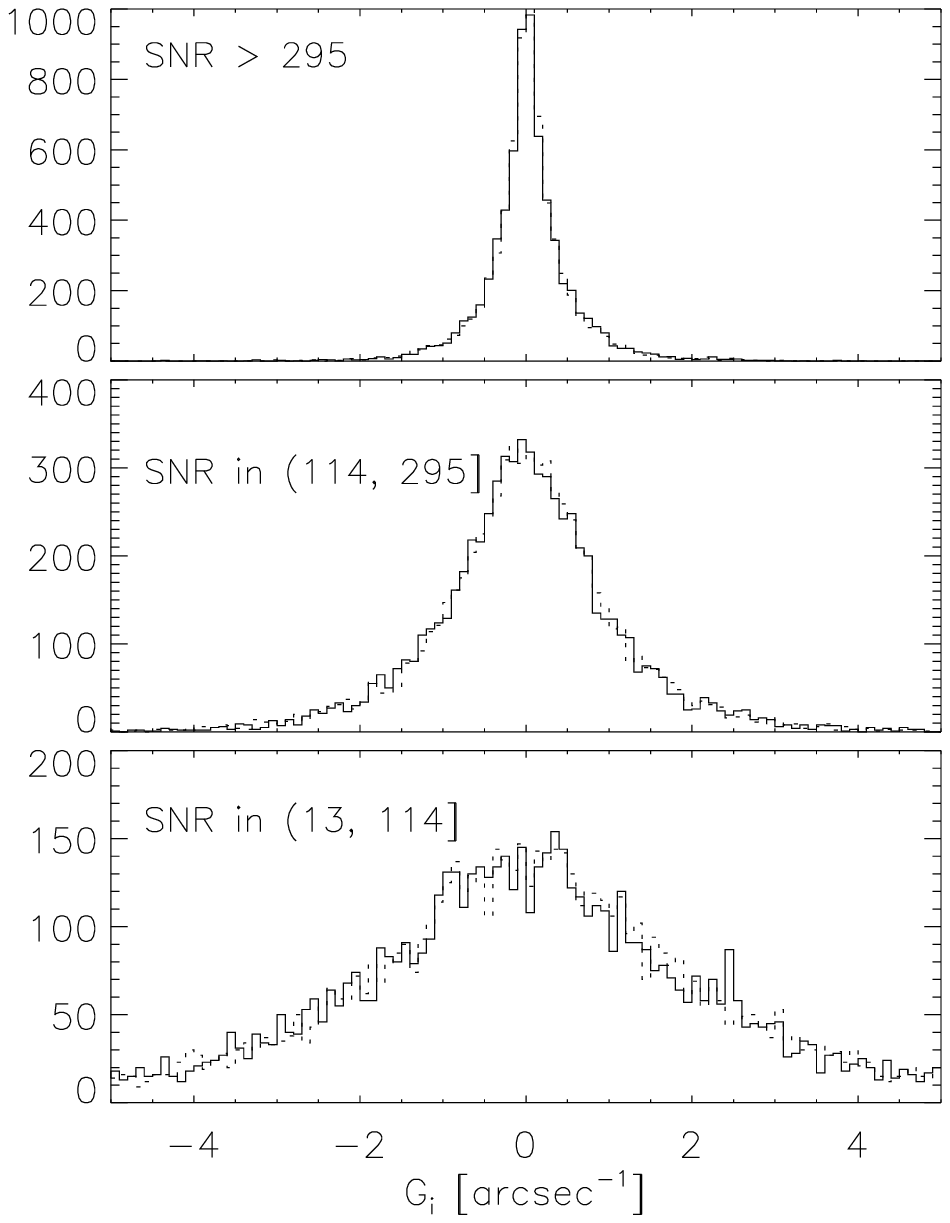,width=0.33\textwidth}
{\caption{\label{fig:flexdist}Distributions of matched measured
    $\tilde{\gamma}_1$ and $\tilde{\gamma}_2$
    (solid and dashed lines, respectively; left panel),
    $\tilde{\fflex}_1$ and $\tilde{\fflex}_2$
    (solid and dashed lines, respectively; centre panel), and
    $\tilde{\gflex}_1$ and $\tilde{\gflex}_2$
    (solid and dashed lines, respectively; right panel)
    from the mean of matched pairs of galaxies in the rotated and unrotated
    simulations (see Section \ref{sect:strategy}).  These paired
    combinations will cancel the leading order contribution of
    intrinsic galaxy shape to the estimation of each signal: what
    remains will be dominated by uncertainty due to pixel noise and deblending.}}
\end{center}
\end{figure*}
This can be explained by considering the two separable contributions
to measurement uncertainty: the scatter in the intrinsic shapes of
galaxies prior to lensing, and the noise in shape estimates due to
noise in pixels.  The latter will increase as SNR decreases, and may
also be a cause of bias as well as increased uncertainty in individual
shape estimates \citep{refregieretal12,melchiorviola12}.  In Figure
\ref{fig:flexdist} we plot the distributions of pair-matched shear and
flexion estimates as a function of SNR.  These paired combinations
cancel the leading order contribution of intrinsic galaxy shape to the
estimation of each signal, so that the remaining scatter is that
solely that due to the differing pixel noise applied to each simulated
galaxy.  The uncertainty of shear estimates increases as SNR
decreases, but the effect is far stronger for the estimators of
flexion.

Figure \ref{fig:flexdist} also illustrates the non-Gaussian
\emph{distribution} of the uncertainty in shapelet shear and flexion
estimators due to noise in image pixels.  For flexion in particular,
this distribution is highly non-Gaussian, but this is also noticeably
true for the shear estimates: the distributions show a sharp peak in
the region of central tendency, accompanied by broad wings.  However,
as the intrinsic ellipticity in galaxy images is typically $\sigma_e
\simeq 0.3$, the shear measurement uncertainty in the left-hand panel of
Figure \ref{fig:flexdist} only becomes a very significant additional
contribution for the lowest SNR galaxies.

For the first flexion $\fflex$ estimation of the uncertainty due to
intrinsic galaxy shapes has proved to be difficult.  A value of
$\sigma_{\mathcal{F}} \simeq 0.04$ arcsec$^{-1}$ was measured by
\citep{goldbergbacon05} for a population of galaxies that was
sbrighter, and mostly larger (within the uncertainties of PSF
correction for poorly resolved objects) than those in the current
simulation set. However, even incorporating scalings with galaxy size
and population it seems likely that measurement error will represent a
significant contribution to overall uncertainty for a broad range of
interest in SNR.  Indeed, the size of the measurement uncertainty itself
suggests that it may be quite challenging to determine reliable
estimates of $\sigma_{\mathcal{F}}$ for deeper galaxy populations than
those explored in \citet{goldbergbacon05}: it has proved to be so in
practice.   That there has been no detection of gravitational $\gflex$ (despite a number of
$\fflex$ detections) is also plausibly attributable to the extreme
measurement scatter seen in the right-hand panel of Figure
\ref{fig:flexdist}, and while the intrinsic $\sigma_{\mathcal{G}}$ is
yet to be reliably estimated for real galaxies this measurement
uncertainty is likely to be a significant contribution to noise in
future.  Flexion forecasts which fail to account for the extra
uncertainty due to pixel noise will provide overoptimistic estimates
of future prospects for flexion measurement.

\begin{figure}
\begin{center}
\psfig{figure=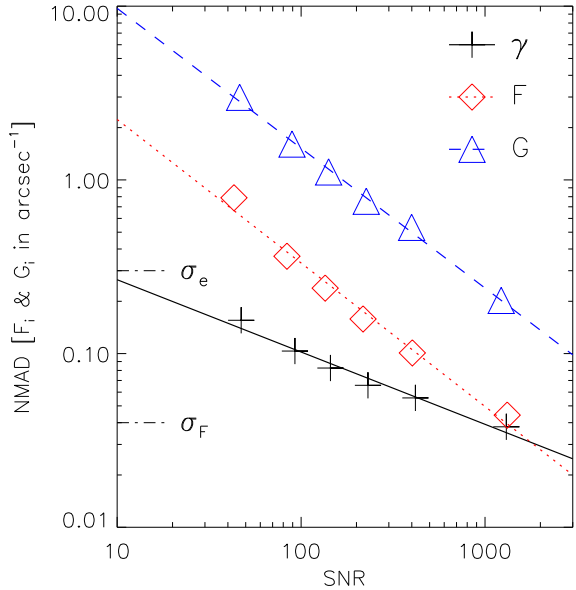,width=0.45\textwidth}
{\caption{Normalized median absolute deviation (NMAD) for the
    distributions of shear and flexion measurement error in Figure
    \ref{fig:flexdist}, plotted as a function of SNR for the binned
    galaxy sample.  The best-fitting power law models are
    plotted over the data points as straight lines (model
    parameters listed in Table \ref{tab:nmad}). \label{fig:nmad}}}
\end{center}
\end{figure}

To provide a simple description of the shear and flexion uncertainty
due to pixel noise as a function of SNR, we calculate the normalized
median absolute deviation (NMAD) for shear and flexion measurement
error distributions in six bins of SNR.  The NMAD is useful as a robust estimator of
dispersion; for distributions such as those in Figure
\ref{fig:flexdist} the population standard deviation itself may be poorly defined,
and the sample standard deviation may be extremely noisy and sensitive to
object cuts and outlier removal.
We plot the NMAD as
a function of SNR in Figure \ref{fig:nmad}, along with the
best-fitting power law description of these data
\begin{equation}\label{eq:nmad}
\textrm{NMAD} = A_{100} (\textrm{SNR} / 100)^b.
\end{equation}
Here SNR = 100 has been chosen as the reference scaling value since
this lies roughly in the middle range of simulated galaxies.  The
best-fitting parameters are given in Table \ref{tab:nmad}.  The
difference in the power law slope between the flexion and shear
results, and the similarity of the slopes for the $\fflex$ and
$\gflex$ estimators, is interesting: it suggests a common origin for
the increased flexion measurement noise in each case, despite
differences in the amplitude of the effect.  A definitive theoretical
explanation for differences between the shear and flexion slopes $b$
is unclear in the presence of multiple factors (noise on pixels,
deblending, centroid uncertainties), but the difference is a clear
effect in these simulated data, for these estimators.  One possibility
(suggested by anonymous referee) is that since fainter galaxies are
smaller, and flexion is a dimensional quantity with units of inverse
angle, the noise will necessarily increase more steeply than shear as
galaxy SNR decreases.  We discuss this scenario, along with
alternatives and possible tests, in Section \ref{sect:conc}.

These results highlight the importance of considering measurement
noise when discussing flexion estimation in practice.  We illustrate
the significance of this contribution to uncertainty by adding a
marker, showing the \citet{goldbergbacon05} intrinsic flexion
estimate, to Figure \ref{fig:nmad} (it should be stressed that the
galaxy population from which this value was derived is not that we are
simulating here, and so this point is merely illustrative; the
intrinsic $\gflex$ is still unmeasured, as has been mentioned
above).\footnote{\citet{goldbergleonard07} found $\sigma_{a |
    \mathcal{F} |} \simeq 0.03$ arcsec$^{-1}$ where $a$ is the size of
  the galaxy, indicating expected intrinsic flexion dispersions in the
  range 0.03-0.3 arcsec$^{-1}$ for the HUDF starter set described in
  Section \ref{sect:udf}. However, it is unclear how much of this may have
  been intrinsic shape, versus noise, dispersion.}  In these tests,
flexion estimation therefore appears to operate in a somewhat
different regime to that of shear estimation, where the measurement
contribution to uncertainty is dominated by the intrinsic ellipticity
$\sigma_e \simeq 0.3$ of galaxies.  This is true throughout the range
of SNR $\gtrsim 10$ where shear estimators have been shown have some
success in controlling bias \citep{bridleetal10,kitchingetal12}.  This
fact has allowed forecasts for shear surveys to proceed using the
intrinsic variance alone as a reasonable approximation to the
uncertainty in shear estimators from all sources.  The results of this
study suggest that this should not be done for flexion, where noise
due to the finite numbers of photons arriving at the detector is
significant.

\begin{table}
\caption{Best-fitting parameters for the power law model of equation
  \eqref{eq:nmad}, describing the NMAD of the shapelet estimator shear and
  flexion measurement error as a function of SNR in this study.  These models are
  plotted over the data in Figure~\ref{fig:nmad}. \label{tab:nmad}}
\begin{center}
\begin{tabular}{crr}
\hline
Lensing measurable & $A_{100}$ & $b$ \\ 
\hline
\hline
$\gamma$ & 0.10 &     -0.42 \\
$\fflex$     & 0.33 &     -0.83 \\
$\gflex$    & 1.52 &     -0.81 \\
\hline
\end{tabular}
\end{center}
\end{table}

\section{Discussion}\label{sect:conc}
We have undertaken a detailed investigation into the problem of
estimating flexion and shear from noisy galaxy images with substantial
variation in underlying galaxy morphology. The simulated galaxy models
also display a realistic distribution of sizes and apparent fluxes,
drawn directly from the galaxy sample in the HUDF, but with additional noise
realistic for a wider area (e.g., GEMS; COSMOS) survey.

We have found evidence of a qualitative difference between flexion and
shear measurement uncertainties.  Whereas noise in shear estimates is dominated by
intrinsic galaxy ellipticity at typical survey image depths, the
corollary appears not to be true for flexion.  Instead, there is large
uncertainty
due to a combination of: i) noise at the pixel level
(due to read noise and finite photon number counts); and ii) (related)
uncertainties in deblending. Furthermore this dispersion in flexion estimates
increases steeply for fainter galaxy images.  It will be
important to account for this fact when generating forecasts of the
flexion information content in future surveys: predictions based
solely on the scatter in intrinsic galaxy flexion
$\sigma_{\mathcal{F}}$ alone will be too optimistic. Existing
forecasts such as those of B06 that use $\sigma_{\mathcal{F}} \simeq
0.04$ arcsec$^{-1}$
\citep{goldbergbacon05} may only correspond to predictions in
the limit of large image SNR, and are only appropriate for the
population of galaxies for which the figure of $\sigma_{\mathcal{F}}$
was measured.  Other predictions, some of which have used
values as low as $0.03$ arcsec$^{-1}$ for the intrinsic flexion
dispersion, will be yet more optimistic.

To provide some aid to more realistic forecasts in future, we fit a simple power law
model to the dispersion of flexion measurements due to noise on pixels and
deblending, as a function of source galaxy image SNR
(Section~\ref{sect:flexdist}).  It was found that noise in flexion
estimates varies significantly more strongly with galaxy SNR than was
found for shear estimates, nearly a factor of 2 in power-law slope.
As the majority of galaxy images in any wide area survey are likely to
be faint, it will be important to consider this effect when
forecasting what may be learned from flexion in practice.

%As shear estimators can be expressed as combinations of second moments
%of galaxy images (e.g.\ \citealp{melchiorviola12}), similarly flexion
%estimators can be expressed as combinations of third moments (in the
%numerator of the estimator) and fourth moments (in the denominator) of
%galaxy images \citep{okuraetal07}.  

As well as considering the noise in flexion measurements, systematic
biases in shapelet shear and flexion estimators were also
investigated.  Such tests had only previously been performed with
galaxy simulations with less morphological richness in the underlying
models \citep{velanderetal11}.  The best performing flexion estimator
in our shapelet pipeline showed comparable performance to the shapelet
estimator tested by \citet{velanderetal11}.  Because of the direct use
of the HUDF galaxy size and SNR distribution in the simulations,
exploration of the dependence of biases on these properties is less
clean, requiring broad bins.  We found little strong evidence of clear
systematic trends within the errors for the flexion results, except
that flexion measurement appeared to perform somewhat better for
large, high-SNR galaxies.  Shear results for the shapelet pipeline
compare well alongside the better performing methods in the most
recent shear measurement simulation challenges, GREAT08 and GREAT10
\citep{bridleetal09,bridleetal10,kitchingetal11,kitchingetal12}.

In the future, theoretical tools from the increasing body of work
invested in understanding the effects of pixel noise in shear
measurement (e.g.\
\citealp{bernsteinjarvis02,hirataetal04,refregieretal12,kacprzaketal12,melchiorviola12,milleretal13})
will be useful in forming a deeper understanding of noise and bias in
flexion estimators.  In particular, it may be possible to understand
the apparent qualitative differences between flexion and shear estimators under
pixel noise.  

As mentioned previously in Section \ref{sect:biasvsSNRR}, one possible
source of difference between estimators of $\fflex$ and estimators of
$\gamma$ and $\gflex$ is the need to apply a simultaneous correction
for the induced centroid shift (see Appendix \ref{App:shapeflexest}).
This centroid term is itself noisy.  Furthermore, because it
consistently appears in the \emph{denominator} for flexion estimators
(e.g., \citealp{okuraetal08}) the shift correction could potentially increase
rectification biases, as well as introduce large wings to the
distribution of estimates of $\fflex$.  In future work it would be
interesting to explore the uncertainties in $\fflex$ flexion estimates
in synthetic test cases where galaxies model fits are forced to use
the fixed, true (i.e.\ pre-flexion) centroids.  An additional study of
interest in a related topic would be an analysis in which rotated
pairs of model galaxies were fitted about the same centroid: although
unrealistic in practice and untenable as a remedy to the issue, such a
study might give an insight into the nature of $\fflex$ flexion measurement
noise. 

However, the fact that the centroid shift affects only estimates of
$\fflex$ and not $\gflex$ suggests that it cannot provide a full
explanation for the steep power-law slope seen in the dispersion of
measurements of flexion as a function of SNR (see Figure
\ref{fig:nmad} and Table \ref{tab:nmad}).  The estimates of this slope
are similar for both $\fflex$ and $\gflex$, and markedly different
from those of shear, suggesting that there may be a common origin for
these noise properties.  One possibility for this difference, suggested by anonymous
referee, is the dimensionality of
flexion as briefly mentioned in Section \ref{sect:flexdist}.  Fainter
galaxies tend to be smaller (although the slope of this relationship
depends on the galaxy population), and flexion is a dimensional
quantity with units of inverse angle.  
An interesting experiment would be to return to tests such as those of
\citet{velanderetal11} or \citet{bridleetal10}, with simpler
parametric galaxy profiles and grid positions for galaxy centroids (to
remove noise due to deblending).  These would allow a controlled
exploration of noise in flexion measurement, for example as a function
of galaxy size at constant SNR, or as a function of SNR at constant
galaxy size. 

Recent examples of such tests in the case of shear
(GREAT08 and GREAT10: \citealp{bridleetal10,kitchingetal12}) provided a
compelling demonstration of the ubiquitous impact of noise biases
across many measurement methods.  This was something that had been
pointed out as potential issue for certain shear measurement methods
prior to the GREAT challenges (e.g.,
\citealp{bernsteinjarvis02,hirataetal04}).  However, the demonstration
of clear experimental dependences on SNR excited a recent surge in
interest in the problem, leading to a greatly improved understanding
of how noise biases are a consistent presence in all but the most
carefully constructed shear measurement methods
\citep{melchiorviola12,refregieretal12,kacprzaketal12,milleretal13}.
Similar work may provide an understanding of why the dispersion on
flexion measurements due to noise on image pixels increases so steeply
as a function of galaxy population.  It would also be interesting to
compare results for a HOLICS-type method, for which there are
indications of lesser uncertainty in flexion estimates relative to a
shapelet treatment
(\citealp{leonardetal11}, although we note that some details of the implementation
of shapelet estimates in this work differed markedly from this analysis).

Some fraction of the dispersion in the flexion estimates presented here is
likely to be due to deblending.
Deblending is an inevitably non-linear process, which introduces a
source of noise that is not intrinsic to the individual galaxy shape,
and will contribute most greatly at low SNR.  The simple gridded tests
pursued by \citet{bridleetal10}, \citet{velanderetal11} and
\citet{kitchingetal12}, while powerful for exploring many aspects of
shape measurement, are not able to elucidate noise due to deblending.
The random placement of galaxy models, and the presence of many HUDF
starter set galaxy models too faint to be detected versus the
GEMS-like noise levels, mean that these simulations are able to mimic
some important aspects of the deblending problem.

It was in order to capture a realistic contribution from this affect
that the adopted SE\textsc{xtractor} parameters used in the pipeline
(see Table \ref{tab:sexparams}) were based on the GEMS-optimized
choices made by \citet{caldwelletal08}.  In fact, the two-pass
strategy employed by these authors is a relatively sophisticated
attempt to tackle deblending: many more recent (e.g., COSMOS:
\citealp{leauthaudetal07}; CFHTLenS: \citealp{erbenetal13}) and
upcoming surveys rely on a single pass strategy alone, and will be
still more susceptible to the impact of deblending.

However, as discussed in Section \ref{sect:detect}, a fraction of
galaxies could not be matched with their pairwise rotated partner due
to noise on galaxy magnitudes and centroids from deblending.  Matching
criteria were made quite stringent to ensure a high purity of
legitimate pairs, at the cost of the loss of objects from the sample
(particularly at low SNR where matching was most difficult by far).
Despite these precautions, it is a cause for concern if deblending
causes the misidentification of paired sources and the breakdown of
the pairwise cancellation of the intrinsic galaxy shape contribution
to dispersion in flexion measurements.  This effect could,
potentially, lead to a contamination of the intrinsic shape dispersion
into the pairwise rotated pairs designed to be free from this
additional source of noise.  However, while these effects are
plausibly a contaminant for the lowest SNR results, the amplitude and
slope found at higher SNR are unlikely to be significantly
contaminated.  This can be seen from the shear results of the lower-left panel of Figure \ref{fig:flexdist}, for which the full width at
half-maximum (FWHM) of the
distribution is $\sim 0.3$ (corresponding to a standard deviation of
$\sigma \simeq 0.3/2.35 = 0.128$ in a Gaussian approximation to the
distribution).  If this were fully contaminated by miscancelling
pairs we would expect to see a Gaussian of $\sigma \simeq 0.3$, and
given that these additional shape noise terms add approximately in
quadrature it can be seen that the overall level of contamination
cannot be large even at low SNR.  At higher SNR, given the greater
rate of successful matching and the width of the shear distributions
in Figure \ref{fig:flexdist}, the contamination is negligible.

The results of this study, originally designed to merely calibrate a
flexion measurement pipeline for an analysis of ACS data (in a similar
manner to \citealp{velanderetal11}), warrant further investigation.
The dispersion of flexion measurements due to pixel noise, and the
related issue of deblending, has been demonstrated to be an extremely
important contribution to flexion measurement and to
depend steeply on galaxy SNR.
Understanding the fundamental source of this strong dependence upon
galaxy SNR (explanations include centroiding effects, noise
rectification or the additional angular dimensionality of flexion)
would be a fascinating topic for further investigation with
custom-designed tests rather than simulations seeking to represent
all aspects of real data. The clear
difficulties of flexion measurement in practice call for a better
understanding of such issues if the potential of flexion as a probe of
small scale power in matter structure is to be realized.

\section{Acknowledgements}
The authors would like to thank Dave Goldberg for interesting
discussions in the course of the work that led to this paper, and the
anonymous referee for many helpful suggestions and improvements. 
BR and CH acknowledge support from the European Research Council in the form of
a Starting Grant with numbers 240672 (BR) and 240185 (CH).  JR and BR were
supported by the Jet Propulsion Laboratory, which is run by California
Institute of Technology under a contract for NASA.  This work was
supported in part by the National Science Foundation under Grant No.\
PHY-1066293 and the hospitality of the Aspen Center for Physics.

\bsp

\bibliographystyle{mn2e}

\bibliography{btprmnras_letter}

\appendix

\section{Shapelet galaxy models from the HUDF}\label{app:udf}

We model the HUDF galaxies using version 2.2 of the shapelets software
package, presented by \citet{masseyrefregier05} and described in
practical detail by \citet{berge06}. The modelling of these galaxies,
including PSF deconvolution, is a multistage process: catalogue
creation, postage stamp image creation and PSF modelling all precede
the construction of the deconvolved shapelet catalogue for the starter
set.  These processes are now described in turn.

\subsection{Star and galaxy selection}\label{sect:selec}

The starting point for building accurate models of galaxy images is an
object catalogue.  A primary science goal of the
\citet{beckwithetal06} analysis was a catalogue of full, multicolour
photometry in the HUDF, but we require only reliable object detection in the
$V_{606}$ in order to build galaxy models in the same filter.
The \citet{masseyrefregier05} shapelet software
package requires certain input parameters for each object that were
not \emph{all} provided in the catalogues of \citet{beckwithetal06}.
%Examples are the image edge parameters \textsc{xmin\_image},
%\textsc{xmax\_image}, etc.  
We chose to construct our own catalogue from just the $V_{606}$ data,
using the SE\textsc{xtractor} software (\citealp{bertinarnouts96};
version 2.5.2) in single-image mode.  Using this software successfully
requires a number of choices regarding configuration parameters, and
we now describe the strategy adopted for object detection in the HUDF
$V_{606}$ data.  We note that this same strategy will also be used in
Section~\ref{sect:detect} as part of the lensing measurement pipeline
being tested, but for simulated imaging data at a much shallower depth
when compared to the HUDF data used to provide galaxy models.

\begin{table}
  \caption{\sextractor~ configuration parameters used to detect galaxy objects in both the HUDF (Section \ref{sect:udf}) and in the simulated ACS images (Section \ref{sect:flippipe});
    these values are the same as used for the GEMS survey galaxies in the two-pass strategy of
    \citet{rixetal04} and \citet{caldwelletal08}.} 
\label{tab:sexparams}
\begin{center}
\begin{tabular}{lrr}
\hline
Configuration Parameter & Cold Sample & Hot Sample \\ 
\hline \hline
 \textsc{detect\_thresh} & 2.30 & 1.4 \\
 \textsc{detect\_minarea} & 100 & 37  \\
 \textsc{deblend\_mincont} & 0.065 & 0.060 \\
 \textsc{deblend\_nthresh} & 64  & 32 \\
 \textsc{back\_size} & 214   & 214  \\
 \textsc{back\_filtersize} & 5  & 5 \\
\hline 
\end{tabular}
\end{center}
\end{table}
We adopt a two-pass SE\textsc{xtractor} deblending strategy when
constructing a source catalogue (see, e.g.,
\citealp{rixetal04,leauthaudetal07,caldwelletal08}). Two catalogues
are created, one with a low detection threshold so as to pick out as
many faint objects as possible, and one using a more conservative
detection strategy so as to limit the over-deblending of bright
objects; these catalogues will be referred to as `hot' and `cold',
respectively.  These are created using exactly the same input
parameter values as used by \citet{caldwelletal08} to create the hot
and cold samples of objects in the GEMS $V_{606}$ science tiles,
summarized in Table~\ref{tab:sexparams}.

All cold detections are then combined with non-overlapping objects in
the hot catalogue.  We define a hot object as overlapping if its
centroid lies within an ellipse of semimajor axis $5.5 \times a$ and
semi-minor axis $5.5 \times b$, where $a$ and $b$ are the
SE\textsc{xtractor}-output semimajor and semiminor axes,
respectively, and this larger ellipse is aligned with that defined by
SE\textsc{xtractor}.  This factor of 5.5 was found, by visual
inspection of the HUDF and segmentation maps output by
SE\textsc{xtractor}, to provide a suitable compromise between cold
object overdeblending and the erroneous removal of hot
objects. Finally, a mask is applied so as to exclude detections from
the boundary regions of the CCD image.  The combined hot/cold
catalogue then contains a total of 8203 objects, corresponding to
$\sim$900 detections arcmin$^{-2}$.

\begin{figure}
\begin{center}
\psfig{figure=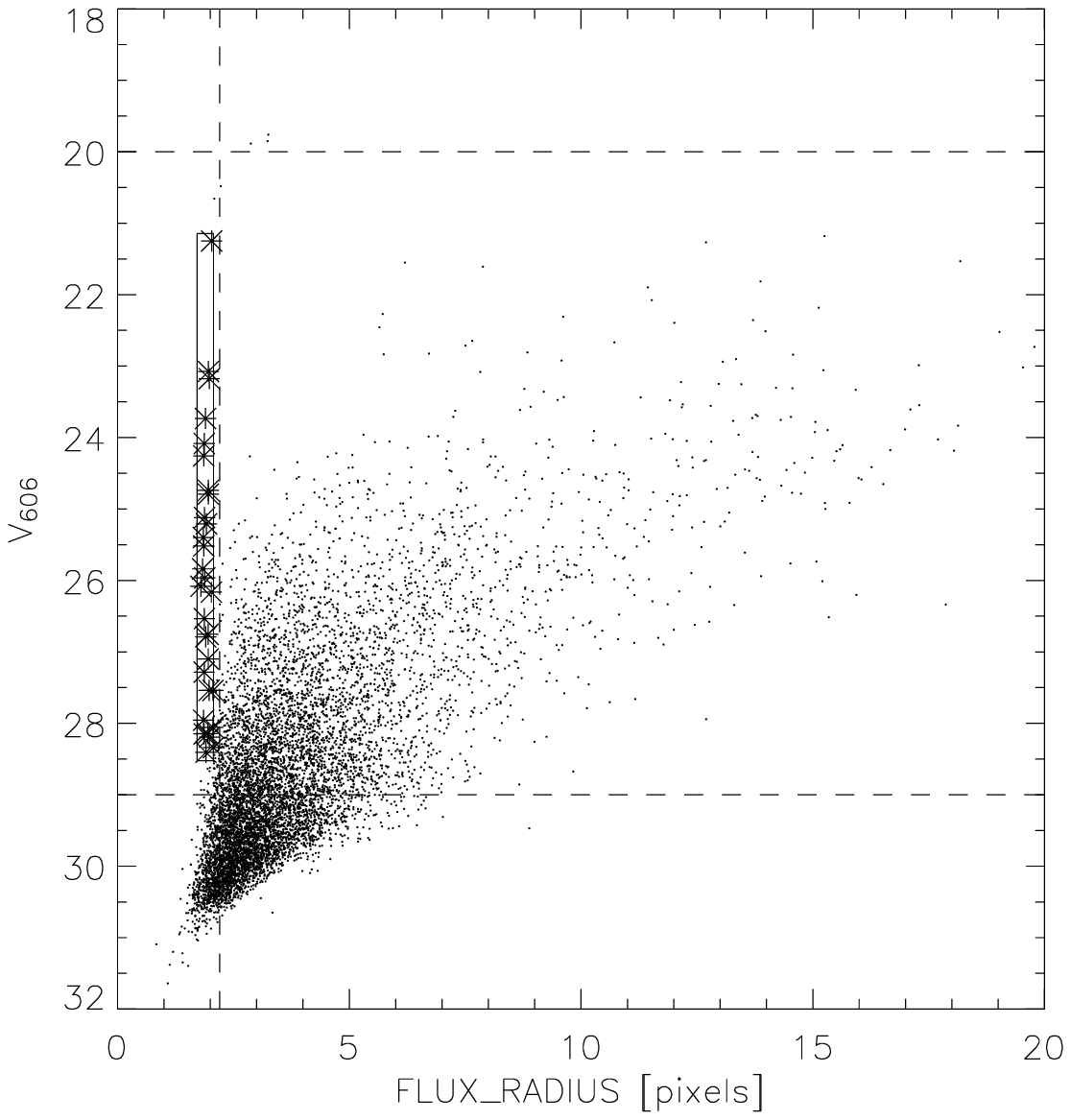,angle=0, width=0.5\textwidth}
\caption{Size-magnitude diagram for SE\textsc{extractor}-selected
  objects in the HUDF $V_{606}$ science image, showing the stellar
  locus and 30 selected stars (star-shaped points). The dashed lines
  show the size and magnitude limits used to define the galaxy sample
  chosen for shapelet modelling for inclusion in the simulation
  starter set. \label{fig:sizemag} }
\end{center}
\end{figure}
To select stars we use the fact that the \textsc{flux\_radius}
parameter found by SE\textsc{extractor} is typically constant,
irrespective of flux, for stellar images.  This radial profile of the
PSF will typically vary little across a given single ACS tile (the
same is therefore also true of the object Full-Width at Half Maximum:
FWHM).  This allows the stars to be easily identified via a straight
locus on a size-magnitude diagram, such as that showing all the masked
HUDF-selected objects in Figure ~\ref{fig:sizemag}.  The locus chosen
in the figure gives a total of 30 stellar images from which to build a
model of the HUDF PSF, avoiding the confused region at greater
magnitude and saturated images at lower magnitude. This low number of
stars is to be expected: the HUDF was specifically chosen as a
direction out of the plane of the Milky Way containing as few stars as
possible.  These 30 stars will be decomposed into shapelet models of
the HUDF PSF in Appendix~\ref{sect:modudfpsf}.

To isolate galaxies for shapelet modelling in Appendix
\ref{sect:modudfgals}, those to be included in the simulation starter
set, we cut for objects with $20.0 < V_{606} < 29.0$ and $2.2 <
\textsc{flux\_radius} < 150.0$ (in pixels). These cuts can also be
seen in Figure~\ref{fig:sizemag} (barring the large-radius cut at
$\textsc{flux\_radius} = 150.0$).  There are then 4128 galaxies from
the original masked sample that make these cuts, corresponding to
approximately 460 galaxies arcmin$^{-2}$.  It is this extremely
deep sample that will be used for generating simulated galaxy images, 
although of course many of these objects will be lost in
noise when simulating shallower data than the HUDF.

\subsection{Postage stamp image extraction}\label{sect:pstamp}

The decomposition of stars and galaxy images into shapelet models must
be preceded by the creation of `postage stamp' images of each object
in the catalogue, cropped around the object in question, and masked
for neighbours. A postage stamp is also made containing a map of the
noise and sky background in the same vicinity.

% These postage stamps may be made using noise and segmentation maps
% supplied by, for example, SE\textsc{xtractor}, or these may be
% created using the easily-modified routines supplied in the
% \citet{masseyrefregier05} software. The careful construction of
% postage stamps is an important consideration in shapelet modelling
% and so this second option is chosen, allowing the process to be
% fully customised in a way we we now describe in some detail (a
% summary of relevant parameters can also be seen in Table
% \ref{tab:shapeparams}).

For each object, the shapelet software draws circular postage stamps
centred on the \sextractor-measured centroid.  The radius of this
circular image is the integer number of pixels closest to a value
$r_{\rm PS}$, defined in terms of the user-specified shapelet input parameter
\textsc{nfwhm} as $r_{\rm PS} = \textsc{nfwhm} \times a + 4$,
where $a$ is the SE\textsc{xtractor}-output semimajor axis of the
object. The name of the
parameter \textsc{nfwhm} appears to be from an earlier incarnation of
the software in which $r_{\rm PS}$ was defined in terms of the FWHM.
% This definition is that adopted in the shapelet code itself: the
% reference to the FWHM in the name of the parameter \textsc{nfwhm} is
% likely a from an earlier incarnation of the software.
We choose $\textsc{nfwhm} = 6$ in this analysis, the default value of
5 being found to be often insufficiently large to allow the shapelet
modelling of the extended outer profiles of deep, space-based images.

% Furthermore, for the ACS data of the HUDF it was found necessary to
% make an algorithmic modification to the postage stamp construction
% routines in the publicly-available shapelet software.
The initial version of the shapelet software used in this analysis
flagged as a modelling failure any objects for which the effective
outer boundary of the shapelet model, defined as a locus of radius
$\theta_{\rm max} = \beta \sqrt{n_{\rm max} + 1} $ (see
\citealp{masseyrefregier05}) around the object centroid, extended
beyond the
edge of the postage stamp.  Small postage stamps led to an
unacceptable number of model failures due to this extension of light
profiles beyond the postage stamp boundaries, but it was found that
drawing overly large postage stamps ($\textsc{nfwhm} \ge 7$) around
every object was computationally prohibitive.  An algorithm for
iteratively redrawing the postage stamp in the event of such model
failure provided an efficient solution to this problem, and is now
part of the shapelet software, documented and available for download
online at the web location given in Section \ref{sect:shapelets}.  In
the iterative prescription used for this analysis, model failures due
to postage stamp outgrowth are resubmitted using a new postage stamp
that is increased in size by a factor $\textsc{redraw\_factor} =
1.3$. This process is repeated up to a maximum of
$\textsc{max\_n\_redraws} = 5$ times, after which a catastrophic
failure is flagged.  In tests, these parameter choices were found to
give a better compromise between modelling success rates and
computation time than other values tried: the number of floating point
operations required to generate images approximately varies $\propto
r^2_{\rm PS}$, so \textsc{redraw\_factor} and \textsc{max\_n\_redraws}
cannot be made arbitrarily large without computational cost.

After the drawing of a postage stamp around each object of interest, a
fundamental consideration in the shapelet approach is then the masking
of other, nearby objects: failure to do this well will often result in
the partial modelling of nearby objects as part of the object of
interest.  Unlike in the KSB approach, image pixels at a distance from
the object centroid are not explicitly down-weighted increasing the
importance of careful masking.  The shapelet software therefore
constructs over each object an elliptical mask defined with semi-major
axis \textsc{mask\_neigh}~$\times$~$a$ and semi-minor axis
\textsc{mask\_neigh}~$\times$~$b$, aligned with the
SE\textsc{xtractor}-defined object ellipse. The factor
$\textsc{mask\_neigh}=4$ was found to give better results than the
shapelet default value of 2.75: in the default setting there were
often portions of the outer galaxy light profile that were unmasked
and clearly visible. These caused an enhanced rate of catastrophic
modelling failure for the central galaxy of interest in such postage
stamps.  The larger value $\textsc{mask\_neigh} = 4$ provided a
significant reduction in such failures without generating unacceptable
numbers of cases where the nearby mask obscures the object of interest
in the postage stamp.

%The need to mask conservatively is perhaps heightened for deep space-based imaging by the more pronounced outer profiles of bright galaxies.
\begin{table}
  \caption{Shapelets software input parameters.  Upper section: input parameter settings that influence the construction of postage stamps.  Lower section: parameters that control the shapelet modelling and deconvolution.  No default value is indicated for parameters added as a modification to the publicly available software.} 
\label{tab:shapeparams}
\begin{center}
\begin{tabular}{lrr}
\hline
 Shapelets Input Parameter & Chosen Value & Default Value \\ 
\hline \hline
\textsc{nfwhm} & 6  &  5  \\
\textsc{redraw\_factor} & 1.3  & -- \\
 \textsc{max\_n\_redraws} & 5 & -- \\
\textsc{mask\_neigh} &  4.0 & 2.75 \\
\textsc{neighbour} & $1^{\dagger}, 0^{\star}$ &  0  \\
\textsc{sky} & 1 & 0 \\
%\textsc{g\_over\_t} & 1/2007 & -- \\
\hline
\textsc{n\_min} (minimum $n_{\mmax}$) & 0  & 2 \\
\textsc{n\_max} (maximum $n_{\mmax}$) & 20 & 20 \\
\textsc{theta\_min\_geom}  & $0.5^{\dagger}, 1.0^{\star}$ & 0.2       \\
\hline 
\end{tabular}
\end{center}
\medskip
$^{\dagger}$values chosen for modelling HUDF galaxies \\
$^{\star}$values chosen for modelling the ACS simulated galaxies
\end{table}

%\begin{figure}
%\begin{center}
%\psfig{figure=pstamp.ps,angle=0,width=8.4cm}
%{\caption[Example of a GEMS galaxy postage stamp image.]{Example of a GEMS %galaxy postage stamp image,
%showing a (severely-overlapping) masked neighbour. The small plotted
%ellipse is that defined by the $a$ and $b$ semi-major and semi-minor
%axes output by \sextractor}
%\label{fig:pstamp}}
%\end{center}
%\end{figure}

An inverse-variance noise weight map and an estimate of the sky
background level are then made via analysis of blank sky pixels in the
postage stamp, i.e.\ those unmasked by the central object or a
neighbour.  The sky background can be subtracted from the image
postage stamp by fitting a choice of surfaces to blank sky pixels
\citep{berge06}.  For the HUDF $V_{606}$ image only a very small
amount of residual sky background variation was found and the removal
of a simple constant sky level from each postage stamp was sufficient,
achieved by setting the shapelet input parameter $\textsc{sky}=1$.  In
order to make the noise map we estimate the root mean square (RMS)
blank sky pixel value to provide a constant, inverse-variance weight.

There is a further choice in how noise values are assigned for pixels
corresponding to masked neighbours, a choice controlled by the input
parameter \textsc{neighbour}.  For the default value
$\textsc{neighbour} = 0$ these pixels are assigned zero values in the 
inverse-variance weight map, and are therefore not considered in the
shapelet modelling. For $\textsc{neighbour} = 1$, the
pixels are assigned the same weight as elsewhere in the noise map and
set to the background level in the science image.  Although it is arguably better not
to include these masked pixels in any fit for shape inference purposes, for
the purposes of building a simulation galaxy starter set from HUDF
images it was found that $\textsc{neighbour} = 0$ sometimes caused large
negative flux patches in shapelet models for unconstrained
regions of the image.  So as to build as physically representative a
starter set as possible we therefore set $\textsc{neighbour} = 1$ for
the modelling of HUDF objects, but retain $\textsc{neighbour} = 0$
when later testing the shapelet lensing analysis on the derived, noisier,
simulated galaxies.

Finally, after these choices for the construction of the postage
stamps, the trimmed, masked, sky-subtracted science images and
inverse-variance noise weight maps are then ready to be supplied to
the shapelet decomposition and modelling routines as described in
\citet{masseyrefregier05}.  We now describe the use of this software
to model first the stars in the HUDF, and then the deconvolved,
high-resolution galaxy images that will be used in the flexion and
shear simulations.

\subsection{Modelling the HUDF PSF}\label{sect:modudfpsf}

Modelling the PSF in the HUDF is important.  A good PSF model allows
an approximate deconvolution of the PSF from galaxy images, desirable
for generating a starter set of models which accurately reflects real
galaxy properties for space-based data (see, e.g.,
\citealp{mandelbaumetal12} who
demonstrate a novel approach to simulating convolution-corrected
galaxy images).
An alternative approach is simply to model the ACS PSF-\emph{convolved} galaxy images of
the HUDF and use these as the starter set: such an approach would be
acceptable if the final use of the starter set was in simulating
observations with far larger PSF sizes, such as for ground-based data
(this was the approach taken in STEP2: \citealp{masseyetal07step}). However, when simulating ACS data
such an approach would lead to an unrealistic size distribution for
the small, fainter objects that are the most important carriers of
weak lensing information. Fainter galaxies, being not much larger than
the $\simeq 0.1$ arcsec PSF typical in ACS images, would be noticeably
oversmoothed, too large, and with an unrepresentative radial profile.
We therefore correct for this image blurring as much as possible.

\begin{figure}
\begin{center}
\psfig{figure=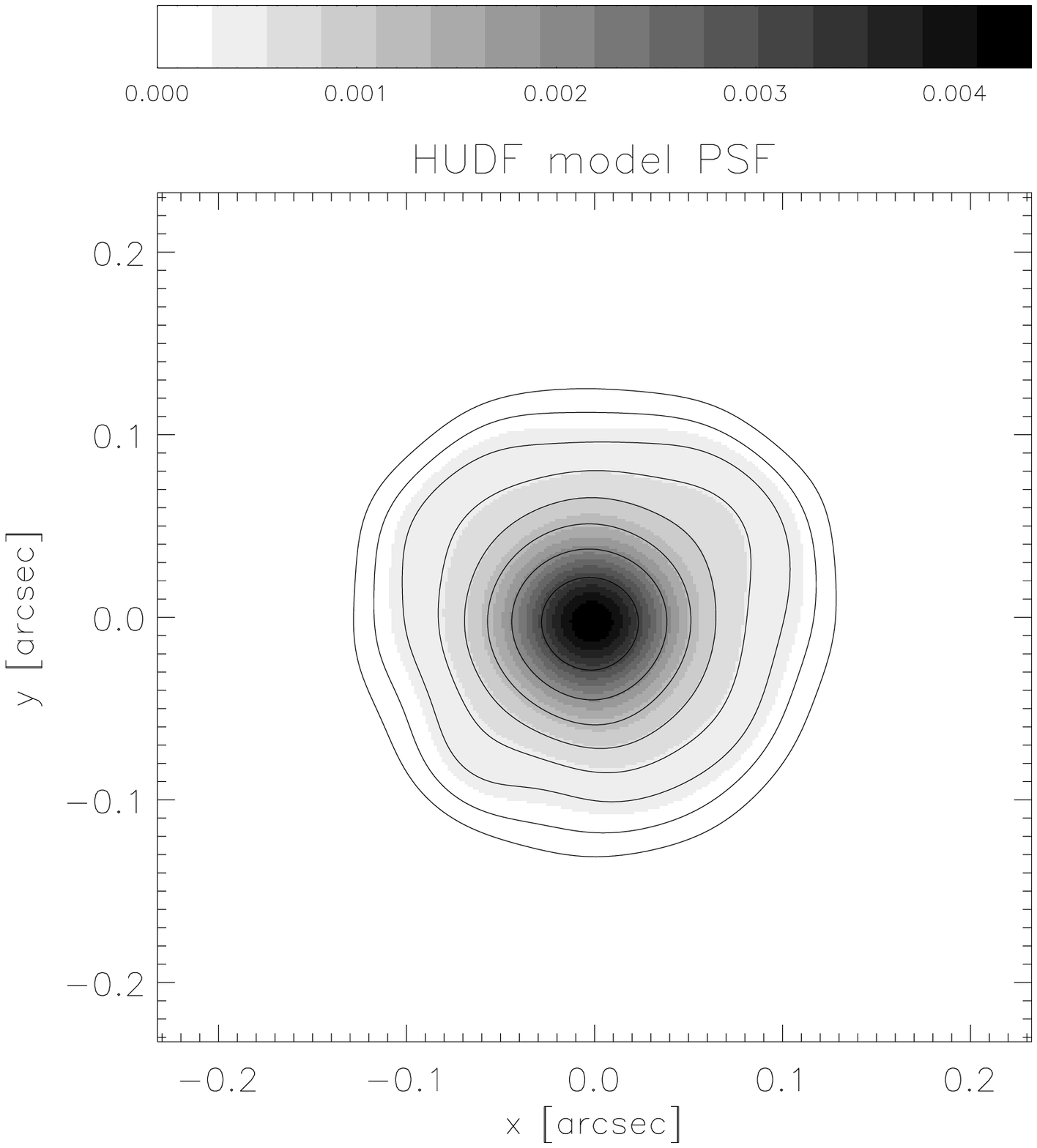,angle=0, width=0.4\textwidth}
\psfig{figure=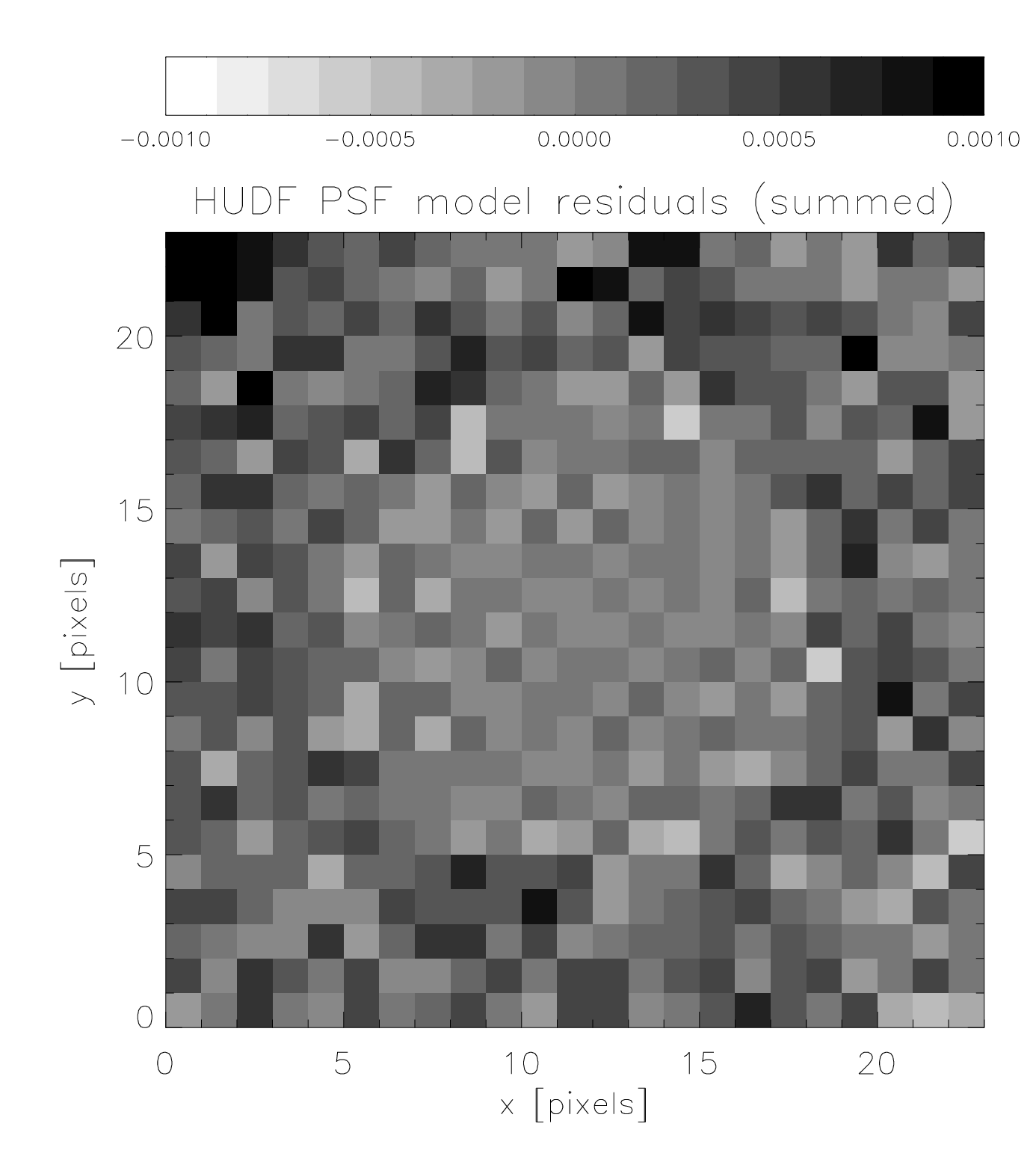, angle=0, width=0.4\textwidth}
\caption{Upper panel: PSF pattern created from the weighted average of
  30 shapelet models of selected stars in the HUDF $V_{606}$ science
  image.  The greyscale is linear in surface brightness whereas the
  contours are logarithmic. Lower panel: Mean residuals from shapelet
  model fits to the HUDF stars.  A flux excess beyond the central
  region of the postage stamp is visible.  There are two primary
  contributions to this excess in the residuals: the inability of the
  shapelet model to fully represent the extended wings of the HUDF
  PSF, and light from nearby objects not being fit in the
  modelling.\label{fig:psfudf}}
\end{center}
\end{figure}

Taking the 30 stellar objects selected as described in
Appendix~\ref{sect:selec} (see also Figure~\ref{fig:sizemag}) we first
create masked image and noise postage stamps as described in Appendix
\ref{sect:pstamp}. We then construct shapelet models of each star,
choosing fixed values of $\beta = 1.80$ and $n_{\mmax} = 20$.  The
fact that these values are fixed, and not allowed to vary as under the
amoeba-driven optimization described by \citet{masseyrefregier05} is
important: we wish to combine shapelet models to create an average PSF
for the field, so a fixed $n_{\mmax}$ and $\beta$ are practical as
they allow simple linear coaddition of models.  The shapelet software
also outputs the diagonal entries of the covariance matrix for all
shapelet coefficients, and so these are used to combined the models
for each star in an inverse-variance weighted average, giving the
resulting PSF model seen in Figure~\ref{fig:psfudf}.

That this model incorporates no spatial variation across the HUDF
field-of-view is not a great concern for the purposes of creating a
realistic starter set for lensing simulations.  Although the bulk
effect of image convolution will be largely corrected for, it does
mean that some residual ellipticity and flexion will remain in the
starter set images due to the residual anisotropy variation in the
HUDF PSF.  These faint distortions will therefore be retained in the
starter galaxy models, but as will be seen in Section~\ref{sect:sims}
these starter models are then randomly rotated, inverted and further
distorted before being used in simulations. Within these simulations
they are then lensed, reconvolved with a new PSF and significantly
noise-degraded.  Distortions due to variation in the original HUDF PSF
will not have a significant, coherent impact on the final galaxy
images at the level of measurement possible for simulations of this
size.

\subsection{Modelling the HUDF galaxies}\label{sect:modudfgals}

In constructing deconvolved (i.e.\ corrected for PSF convolution)
shapelet models of the HUDF galaxies, the \citet{masseyrefregier05}
software takes as its inputs the object catalogue constructed as
described in Appendix~\ref{sect:selec}, the accompanying postage stamp
images and noise maps, and the shapelet model of the PSF. The
best-fitting shapelet models output by the code will make up the
starter set that is used to create simulated galaxies.

However, as was necessary for the modelling of the HUDF PSF, choices
must be made for input parameter values that govern how these
best-fitting shapelet models are selected. The most important of these
are summarized in Table~\ref{tab:shapeparams}, and now described.

%There is therefore no
%need to model the simulation PSF using the methods described in Section
%\ref{sect:modelPSF}; in this
%way we are simulating shapelet shear and flexion measurement given
%%emph{perfect} knowledge of the distorting PSF, a luxury which is not
%ttainable for real data.  As discussed in Section \ref{sect:obsdist},
%no attempt is made to fit a time or spatially varying PSF
%model to the simulated images.  This is also the case in STEP, 
%but unlike these analyses we have additional access to the exact
%profile of the convolving PSF, rather than estimating it from the
%profiles of simulated stars in our images.
%Future work will be necessary to
%ascertain the extra uncertainty and possible biases 
%in shear and flexion measurements due
%to imperfect modelling of the PSF (such as may well be evident even
%for the $n_{\mmax} =20$ GEMS PSF model, see Section
%\ref{sect:modelPSF}), but it is nevertheless interesting to explore
%the undeniable limits of shape measurement accuracy even given an extremely
%well-modelled PSF.

Unlike when modelling the PSF, for constructing galaxy models we
allow $n_{\mmax}$ and $\beta$ to vary.  These quantities are
chosen by an amoeba-driven optimization process (described
by \citealp{masseyrefregier05}; we note that other implementations of the shapelet
method do not necessarily allow $n_{\max}$ to vary, e.g.,
\citealp{kuijken06}). However, the shapelet software does require the
choice of a limiting maximum value of $n_{\mmax} =$ \textsc{n\_max},
and a lower starting point \textsc{n\_min} from which to begin the
amoeba search.

%In modelling the HUDF galaxies we choose the input
%parameter values \textsc{n\_max} $= 20$ and \textsc{n\_min} $= 0$.

The choice of \textsc{n\_max} is largely motivated by computing
resource constraints.  In general, a significant fraction of overall
processing time is spent modelling a very small subset of large/bright
galaxies with complex structure.  The time taken and memory required
to model a given object increases roughly as $n^4_{\mmax}$
\citep{masseyrefregier05}. Above $n_{\mmax}=20$ the calculation of
pre-multipliers for the shapelet basis functions (see equation
\ref{eq:polarbasis}) causes numerical overflow in the unsigned, 64-bit
integers the shapelets software uses for the calculation and
tabulation of factorial terms in the numerator and denominator of
$P_{n,m}$.  For $n_{\mmax}>20$ the shapelets software therefore
calculates these numbers as double precision floating point numbers,
as and when they are required.  This leads to significant processing
overheads, albeit ones which might plausibly be averted with some
changes to the internal processing design of the shapelet software
itself.

However, the largest, brightest and best resolved galaxies are also
the most likely to be local and only \emph{very} weakly lensed in real
data, and this reduces the motivation for dedicating significant
amounts of development effort and processing time to producing
high-order shapelet models of this population.  We therefore choose
\textsc{n\_max}$=20$, which nonetheless allows a high degree of galaxy
substructural complexity to be realized (e.g.\ Figure
\ref{fig:fliptile}).  We choose \textsc{n\_min} $= 0$ so as to give
the shapelet code freedom to model galaxies with a simple circular
Gaussian if there is no strong statistical support for a more complex
model.

%Similarly, the GEMS galaxies were all modelled to a minimum
%$n_{\mmin}$ of 2.  This was found to cause overfitting, which we define as
%occuring for galaxies with shapelet model with 
%reduced $\chi^2 < 1$, in only a tiny minority of cases ($<200$).  

The other input parameter chosen to differ from the default choice was
$\textsc{theta\_min\_geom} = 0.5$.  This parameter sets a lower limit
on the quantity defined by \citet{masseyrefregier05} as
$\theta_{\mmin} = \beta / \sqrt{n_{\mmax} + 1}$, which is the minimum
geometric scale on which the shapelet model varies.  The requirement
upon the shapelet modelling amoeba of only accepting models with
$\theta_{\mmin} \ge 0.5$, also suggested by \citet{melchioretal07},
prevents excessive sub-pixel modelling that can lead to unphysical
`ringing' in shapelet models of galaxy images.

The best-fitting galaxy models resulting from these chosen  input parameters, postage
stamps, and HUDF PSF model, are described in Section \ref{sect:udf}.

%The full treatment of correlated pixel noise (such as present in 
%the drizzled GEMS images) requires the inversion of an extremely 
%large pixel noise
%covariance matrix.  The instability and processing time requirements
%of this process led to it being excluded from the shapelet software of
%\citet{masseyrefregier05}, which instead treats the pixel noise as
%uncorrelated, allowing the far simpler inversion of a purely diagonal
%noise matrix.  This currently represents a potential weakness in the 
%shapelet method, as overfitting correlated noise will result in the
%circularization of output galaxy models, possibly biasing lensing results.

\section{Testing image operations}

\subsection{Weak lensing distortions}\label{sect:upsample}

When performing real-space raytracing as described in Section
\ref{sect:raytrace}, we use the lens equation to map from the image
pixel positions back to a non-regular grid of positions in the source
plane.  At these points we sample the surface brightness from the
shapelet model of the source galaxy $I^{(s)}$.  Ignoring the overall
deflection of galaxy images (equivalent to setting $\thetab_c =
\thetab_c' = 0$), and using equation \eqref{eq:flextran}, we may write
\begin{equation}
I ( \thetab) = I^{(s)}(\thetab') = I^{(s)}(A_{i,j} \theta_j + D_{ijk}\theta_j \theta_k / 2).
\end{equation}
In practical terms the raytracing scheme is simple: we assign
$I(\thetab)$ for each desired image pixel position $\thetab$ by taking
the shapelet model value of $I^{(s)}$ at the position $\thetab' =
\thetab'(\thetab, \gamma, \fflex, \gflex)$.

Square pixels in the image plane do not map to square pixels in the
source plane, and when flexion is included with shear the mapped pixel
boundaries become curved in the source plane.  This means that it is
no longer possible to use the results of \citet{masseyrefregier05} to
perform the exact flux integral of $I(\thetab)$ across each pixel.  In
the simplest approximation one can adopt the pixel-centre surface
brightness to estimate flux but more accurate results can be achieved
by \emph{upsampling}, i.e., by creating a higher-resolution image and
then summing values at the high-resolution subpixel locations to
estimate the true integral. This allows approximation of the exact
shear and flexion transformation to an accuracy that depends on the
degree of upsampling adopted.

We can only upsample by a finite, preferably integral, factor.  In
order to understand what factor is necessary in our simulations we now
test its impact upon fundamental lensing measures (e.g.\ image
moments) for noise-free ACS galaxy models.

We first extracted a random sample of 1000 galaxies from the HUDF
starter set.  For each galaxy, a control postage stamp image was
constructed by integrating across ACS imaging survey-size,
post-drizzling pixels (0.03 arcsec, as used in the HUDF, COSMOS, GEMS and STAGES final science
images; see \citealp{beckwithetal06,leauthaudetal07,caldwelletal08,grayetal09}) using
the exact \citet{masseyrefregier05} analytic results. We refer to this
`true' pixelized control image as $I^{(t)}$. Simple, unweighted
moments for such images may be defined as follows
\begin{eqnarray}
S & = & \int \dif^2 \theta I(\thetab), \label{eq:momentsi} \\
q_{ij} & = & \frac{1}{S} \int \dif^2 \theta \Delta \theta_i \Delta \theta_j I(\thetab), \\
q_{ijk} & = & \frac{1}{S} \int \dif^2 \theta \Delta \theta_i \Delta \theta_j \Delta \theta_k I(\thetab), \\
q_{ijkl} & = & \frac{1}{S} \int \dif^2 \theta \Delta \theta_i \Delta \theta_j \Delta \theta_k  \Delta \theta_l I(\thetab),
\end{eqnarray}
where $\Delthetab = \thetab - \thetab_c$ as before.  In the noise-free
case the unweighted moments can be used to construct complex
polarization measures that provide estimators of shear and flexion
(e.g.\ \citealp{kaiseretal95,bartelmannschneider01,okuraetal07}).

%We use these moment measures to investigate the effects of upsampling upon the accuracy of shear and flexion measurement, in isolation from noise.  To explore shear 
We construct the unweighted polarization
\begin{equation}
e = e_1 + \mi e_2 = \frac{q_{11} - q_{22} + 2\mi q_{12}}{q_{11} + q_{22}},
\end{equation}
(see, e.g., \citealp{kaiseretal95,bartelmannschneider01}) and, following \citet{okuraetal07}, we define
\begin{eqnarray}
f & = & f_1 + \mi f_2 = \frac{q_{111} + q_{122} + \mi (q_{112} + q_{222})} {q_{1111} + 2q_{1122} + q_{2222}}, \\
g & = & g_1 + \mi g_2 = \frac{q_{111} - 3q_{122} + \mi (3q_{112} - q_{222})} {q_{1111} + 2q_{1122} + q_{2222}} \label{eq:momentsf}
\end{eqnarray}
as equivalent measures for $\fflex$ and $\gflex$ respectively. We
require that any finite degree of upsampling must cause fractional
errors in $e$, $f$ and $g$ that are significantly smaller than those
we expect due to noise in the final simulation results.

\begin{figure}
\begin{center}
\psfig{figure=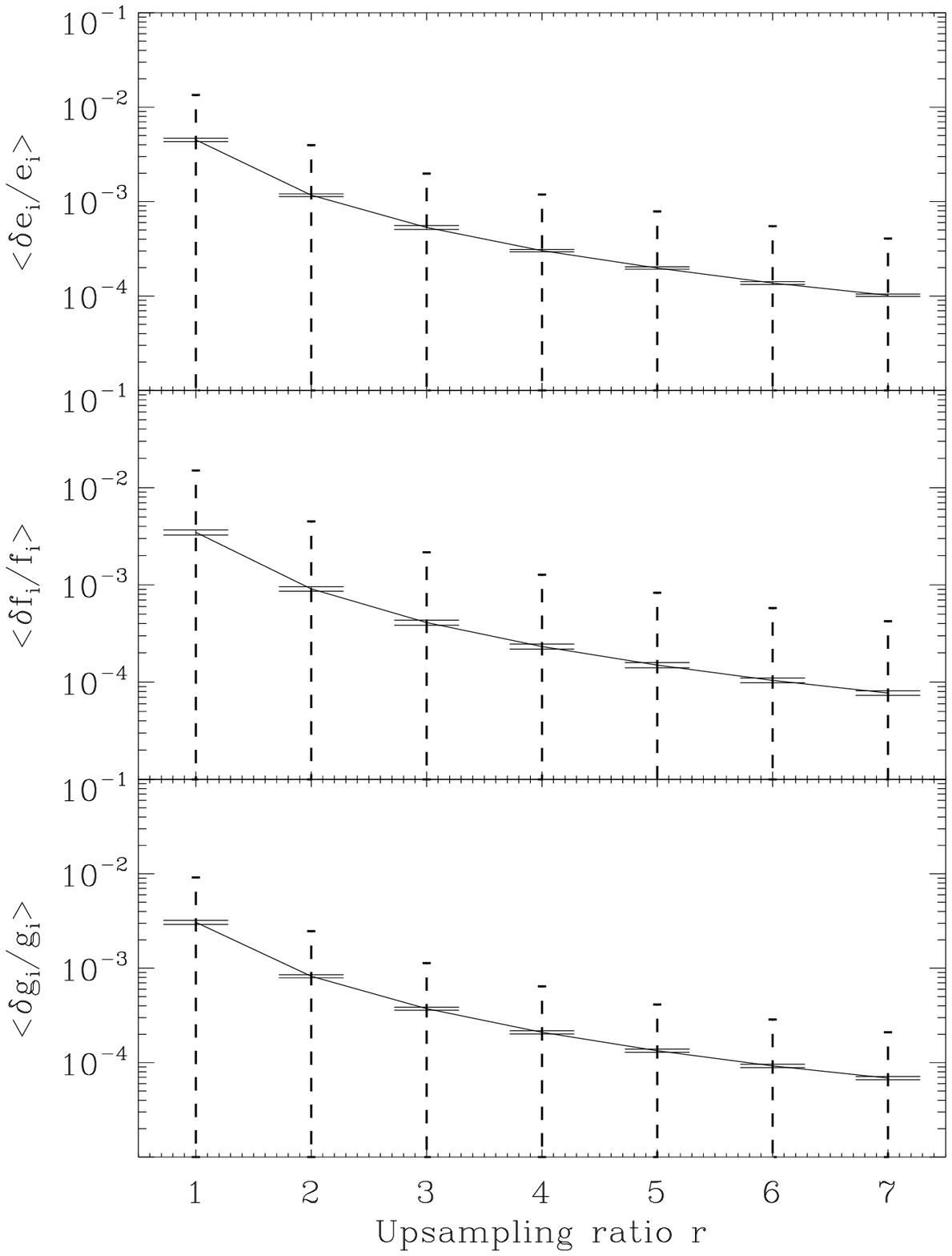, width=0.5\textwidth}
{\caption[]{Testing the impact of upsampling ratio for generating
    images of lensed (by shear and flexion) galaxies using multiple
    `ray-traced' image samples.  Plot of median fractional error in
    simple unweighted $e$ (top panel), $f$ (centre panel) and $g$
    (bottom panel) estimators, for a sample of 1000 galaxies randomly
    selected from the starter set, with increasing upsampling ratios
    $r$ as described in Appendix~\ref{sect:upsample}. Each $\delta
    e_i$, $\delta f_i$ etc.\ is calculated by comparing the
    upsampling-derived value relative to the exact (analytically
    integrated) value. The wide solid error bars on each point give
    the standard error on the median \citep{lupton93}, whereas the
    dashed error bars illustrate the typical range of the effect as
    described by the NMAD of the fractional
    error.}\label{fig:upsample}}
\end{center}
\end{figure}
The first step in the test is to calculate control values $e^{(t)}$,
$f^{(t)}$ and $g^{(t)}$ using the moments as defined in equations
\eqref{eq:momentsi}-\eqref{eq:momentsf} for each control image
$I^{(t)}$ from the sample of HUDF starter galaxies.  We then make
a series of raytraced estimates $I_r$ of $I^{(t)}$ by performing a
numerical integral of the flux across the upsampled pixels. We define
the upsampling ratio $r$ as the ratio between the linear scale of
output pixels and that of the sub-pixels in the upsampled image: $r=1$
is equivalent to simply taking the central pixel value; $r=2$ is
equivalent to four equally spaced sub-pixels; etc. For each galaxy
$I_r$ we measure a corresponding $e$, $f$, and $g$, and calculate the
fractional deviation in each component:
\begin{equation}
\frac{\delta e_i}{e_i}  =  \frac{e_i - e^{(t)}_i}{e_i},
\end{equation}
with equivalent expressions for $f$ and $g$, for $r=1,\dots,7$.  In
Figure~\ref{fig:upsample} we plot the median of these fractional
deviations across the sample of galaxies, as a function of $r$.

As a measure of the typical range for the error due to finite
upsampling, in Figure~\ref{fig:upsample} we also plot as dashed-line
error bars the NMAD of the fractional error.
The NMAD is a robust measure of the width of a distribution of some
variable $x$, defined as
\begin{eqnarray}
  \textrm{NMAD}(x) & \simeq & 1.4826 \times \textrm{MAD}(x) \\
                               & = & 1.4826 \times \textrm{Median}\left( \left | x -
    \textrm{Median}\left( x \right) \right| \right),
\end{eqnarray}
where the MAD is simply the median absolute deviation, and the
constant scale factor 1.4826 ensures that the NMAD of a Gaussian
distribution is approximately equal to its standard deviation,
$\textrm{NMAD} \simeq \sigma $.
Figure~\ref{fig:upsample} shows the NMAD for $\delta e_i / e_i$,
$\delta f_i / f_i$, and $\delta g_i / g_i$ at each value of $r$, to give an idea of the
typical range in fractional errors for these quantities due to the use
of finite element upsampling techniques.  It should be
noted that median statistical measures were used because the property
of interest, which in this case is specifically the fractional
uncertainty on an ellipticity or flexion measurement, leads to large
outlier values when there are small values of $e_i$, $f_i$ or $g_i$ in
the denominator.  Adopting median statistics downweights these noisy
outliers to reduce overall statistical uncertainty on the central
tendency.  As a check on the results, simple means were also used and
resulted in a consistent (but rather noisier) trend.

The results of Figure~\ref{fig:upsample} are encouraging and as
expected: there is clear convergence towards smaller and smaller
fractional errors with increasing $r$.  Practically, the results imply
that for the HUDF galaxies and an output pixel scale of 0.03 arcsec we
need only upsample galaxies by a factor of 2 in order to give
output images that are typically accurate in $e$, $f$ and $g$ to
within a factor of $\sim 10^{-3}$ of the correct value (a target value
motivated by the results of, e.g., \citealp{amararefregier08}).
Interestingly, Figure~\ref{fig:upsample} shows that good accuracy
may be achieved for $e$, $f$ and $g$, even when $r=1$, for which the
error in the idealized measures is often less than 1\% of the
underlying value.

\subsection{Convolution}\label{sect:convol}

To test the accuracy of numerical convolution we use the methods
developed inAppendix~\ref{sect:upsample}. We again select a random sample
of 1000 HUDF galaxies from the starter set.  For each galaxy we
construct a set of seven images, upsampled by a linear factor
$r=1,\dots,7$, and perform a convolution via FFT using an image of the
circularized GEMS PSF (Figure~\ref{fig:gemspsf}) upsampled by the same
factor. The convolved image is then summed back to the final
resolution of 0.03 arcsec pixel$^{-1}$, and then each of $e$, $f$ and
$g$ are measured.

Because there is no practical way of generating the `true' convolved
image in this manner as a control (as discussed before the exact
shapelet treatment is computationally unfeasible), we instead take the
high upsampling case of $r=11$ as the reference point.  The fractional
deviation from this $r=11$ is then calculated for each galaxy and each
$r$, and in Figure~\ref{fig:convol} we plot the resulting median
values and range as in Appendix~\ref{sect:upsample}.

\begin{figure}
\begin{center}
\psfig{figure=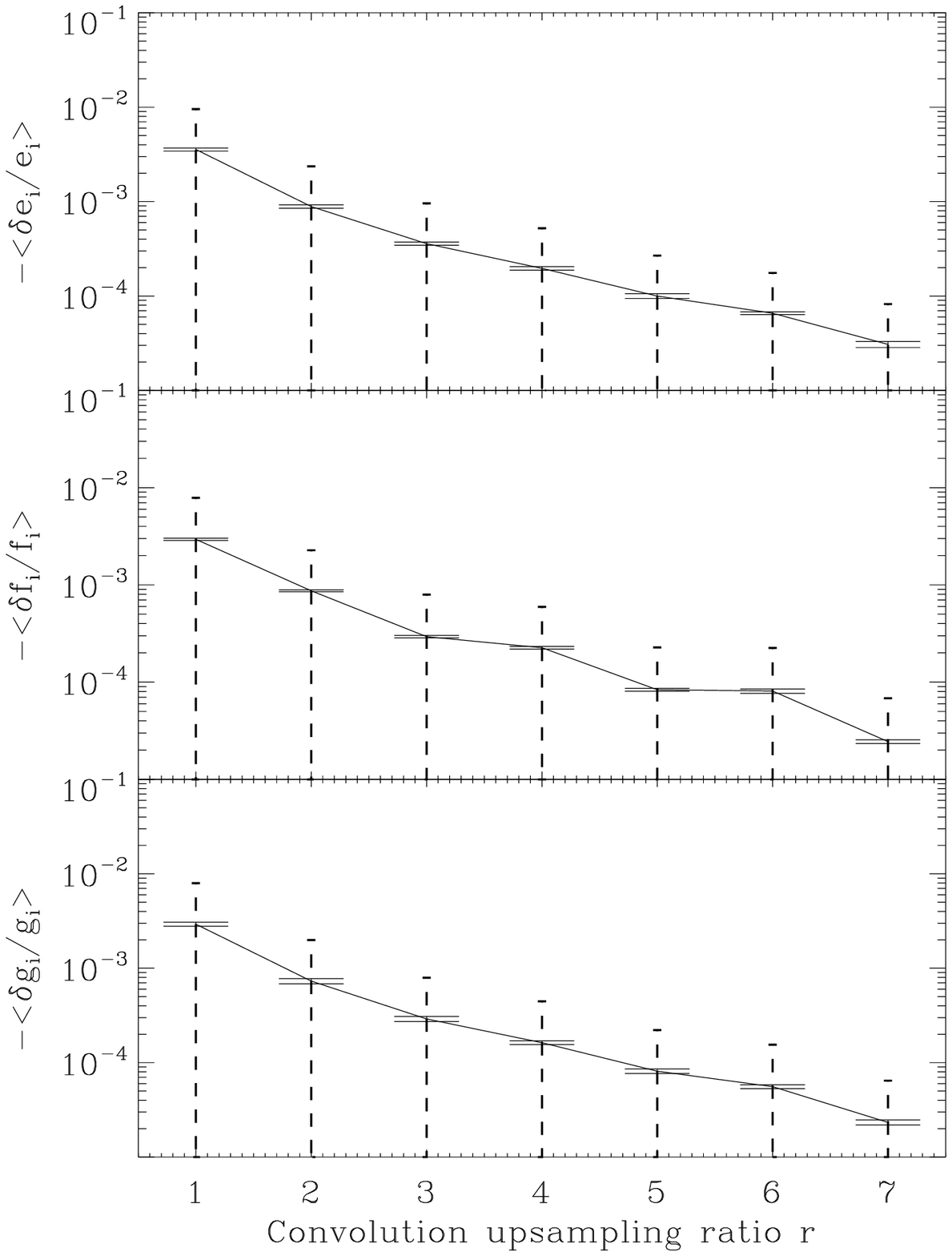, width=0.5\textwidth}
{\caption{{Testing the impact of upsampling ratio in numerical
      convolution via discrete Fourier transforms, using FFT.  Plot of
      median fractional error in simple unweighted $e$ (top panel),
      $f$ (centre panel) and $g$ (bottom panel) estimators, for a
      sample of 1000 galaxies randomly selected from the starter set,
      after performing a convolution using discrete Fourier transforms
      at an upsampling ratio $r$ as described in
      Appendix~\ref{sect:convol}. Each $\delta e_i$, $\delta f_i$
      etc.\ is calculated by comparing the ovesampling-derived value
      relative to the value found for $r=11$. The wide solid error
      bars on each point give the standard error on the median
      \citep{lupton93}, whereas the dashed error bars illustrate the
      typical range of the effect as described by the NMAD of the
      fractional error.}\label{fig:convol}}}
\end{center}
\end{figure}
As was found for the shear and flexion upsampling tests, the results
of Figure \ref{fig:convol} are encouraging: an upsampling ratio of
only $r=2$ allows a systematic fractional error in the estimates of
$e$, $f$ and $g$ that is typically 0.1\% or less.  Interestingly, the
effect is in the opposite direction to that induced when performing
raytracing lensing transformations.  The effect of a finite sampling
approximation to convolution is to cause a net reduction in these
moments of a galaxy image, whereas a finite sampling approximation to
the lens mapping artificially exaggerates these moments.  The overall
net effect of the two approximations will thus, on average, be
typically smaller than that described by either of
Figures~\ref{fig:upsample} or \ref{fig:convol} in isolation.

\section{Shapelet estimators of flexion and shear}
\label{App:shapeflexest}
A variety of polar shapelet estimators for shear and flexion are
described by M07.  In theory, the number of lensing estimators that
may be constructed using shapelets is only limited by the number of
shapelet modes available, $n_{\mmax}$.  However, the majority of
shapelet models will be highly truncated to somewhat low $n_{\mmax}$
in practice.  Estimators which make extensive use of higher order
information prove problematic for many galaxy images, particular those
which rely upon the convergence (in the sense of converging to a
limit) of sums over shapelet coefficients (see M07). 

In this paper we limit our investigations of shear and flexion measurement primarily to
the most simple, and therefore widely available in noisy galaxies, shapelet
estimators.  We now describe construction of useful estimators based
on the results of M07, taking into account important practical
realities such as the wide dynamic range of galaxy fluxes and sizes.

\subsection{Normalizing shapelet coefficients by galaxy size and flux}

The simplest shear estimator is that which employs the 
polar shapelet coefficient $f_{2,2}$ as the galaxy
polarization estimator (see M07, section 3, 
for the precise usage of this term).  This quantity can in fact be
straightforwardly shown to correspond to a Gaussian-weighted quadrupole moment of the shapelet
model, with a weighting radius of $\beta$.  

The raw form of the simplest well-motivated estimator that can be
constructed from $f_{2,2}$ is as follows:
\begin{equation}\label{eq:rawshear}
\tilde{\gamma}^{\mGaussian} = \sqrt{2} \frac{ f_{2,2}}{\expect{f_{0,0} - f_{4,0}}},
\end{equation}
where the expectation notation used in the denominator denotes an
estimate of the ensemble average over all galaxies in the source sample
(\citealp{refregierbacon03}; M07).  

However, we have found that this estimator
suffers an important disadvantage: both the
numerator and denominator share the
dimensionality of the shapelet coefficients themselves, scaling with
both object flux and inverse scale size $\beta$
\citep{masseyrefregier05}.  This means that estimators of shear (and
similarly constructed flexion estimators) vary strongly in magnitude
as a function of the object size and flux relative to those of the ensemble average.

It is therefore desirable to generalize the estimator of equation
\eqref{eq:rawshear} to create a new estimator we label $\tilde{\gamma}_{\nu}$, defined as
\begin{equation}\label{eq:shearnu}
\tilde{\gamma}_{\nu} = \sqrt{2} \frac{\nu f_{2,2}}{\expect{\nu(f_{0,0} - f_{4,0})}},
\end{equation}
where the parameter $\nu$ is suitably chosen to be a property of the
galaxy image that helps normalize terms in the numerator and
denominator, helping to lessen the impact of the wide dynamic ranges of flux and scale length in
galaxy samples.  This is analogous to the adoption of a normalizing
factor $1/S$ in the definition of the image moments $q_{ij}$, $q_{ijk}$
and $q_{ijkl}$ in Appendix~\ref{sect:upsample}.  It should be stated
from the outset that if $\nu$ is an
estimated property of the image itself it may add both uncertainty and
bias to the shear estimator.  We will discuss how this danger weighs
against the benefits of such a normalization later in Appendix~\ref{app:issues}.

%Another shear estimator, 
%also of interest due to its simplicity, is 
%closely related the shapelet-measured unweighted
%ellipticity of the galaxy; we label this estimator $\stild^{\munweighted}$ (see
%\citealp{masseyrefregier05}; M07).  It was shown in M07 that this estimator
%can be equivalently constructed by the requirement of a purely
%diagonal shear susceptibility matrix, calibrated by a ``shear
%responsivity factor'' $\mathcal{R} = 1 - \expect{\varepsilon^2}/2 $
%where $\expect{\varepsilon^2}$ is the measured ensemble variance of
%the unweighted ellipticity $\varepsilon$.  The unweighted ellipticity
%shear estimator for single galaxy is then simply $\stild^{\munweighted} =
%\varepsilon/(2\mathcal{R})$. 

The simplest possible polar shapelet flexion estimators can
be constructed from similar combinations of shapelet coefficients.
The first order $\fflex$ estimator can be expressed as
\begin{equation}\label{eq:fflexestgauss}
\tilde{\fflex}^{\mGaussian}=\frac{4}{3 \beta}
\frac{f_{1,1}}{ \expect{d_{1,1}}},
\end{equation}
where
\begin{equation}\label{eq:d11}
d_{1,1} = \left( 1 - \frac{R^2}{\beta^2} \right) f_{0,0} + \frac{R^2}{\beta^2} f_{2,0}  + \frac{5 \sqrt{2} R^2}{6 \beta^2} \varepsilon^* f_{2,2} - f_{4,0}.
\end{equation}
The shapelet model parameters $\varepsilon$ and $R^2$ are 
defined in \citet{masseyrefregier05} as
\begin{eqnarray}
R^2 & = & \frac{\sqrt{16 \pi} \beta^3}{F} \sum_n^{\rm even} (n + 1)
f_{n, 0}, \label{eq:r2} \\
\varepsilon & = & \frac{\sqrt{16 \pi} \beta^3}{R^2 F} \sum_n^{\rm
  even} \left[ n (n + 2) \right]^{1/2} f_{n, 2},
\end{eqnarray}
where $F$ denotes the shapelet model flux,
\begin{equation}
F = \beta \sqrt{4 \pi} \sum_{n}^{\rm even} f_{n,0},
\end{equation}
and where all sums are over all even $n$ coefficients $f_{n,0}$ and  
$f_{n, 2}$ in the model
($m=0$ and $m=2$ coefficients do not exist for odd $n$).
We note that $\varepsilon$ and $R^2$ appear only in those terms necessary
to correctly account for the flexion centroid shift at linear order in
the applied $\fflex$ (see \citealp{goldbergbacon05}; M07).
It should also be noted that $d_{1,1}$ in equation \eqref{eq:d11} contains
a term proportional to 
$\varepsilon^* \times f_{n,2}$, for $n=2$; in M07 these were argued to
vanish due to rotational symmetry in the source plane.  
We argue that this term should not be omitted since the definition of $\varepsilon$ above
shows the inclusion of $f_{2,2}$ as the first coefficient in the sum.  For most galaxies
lower-$n$ components of the shapelet model will dominate those with
higher-$n$, so that as a first approximation $\varepsilon^* \times
f_{2,2} \sim |f_{2,2}|^2$. This does not cancel due to rotational
symmetry across a population of galaxies..

For the first order $\gflex$ estimator we have the expression
\begin{equation}\label{eq:gflexestgauss}
\tilde{\gflex}^{\mGaussian}=\frac{4\sqrt{6}}{3 \beta }
\frac{f_{3,3}}{\expect{d_{3,3}}},
\end{equation}
where
\begin{equation}\label{eq:d33}
d_{3,3} = \left( f_{0,0} + f_{2,0} - f_{4,0} - f_{6,0} \right).
\end{equation}
Analogously to the case of $\tilde{\gamma}^{\mGaussian}$, these
estimators of both $\fflex$ and $\gflex$ are directly related to Gaussian-weighted
octupole moments (e.g., \citealp{okuraetal07}) of the shapelet model
galaxy.

As for shear, the large dynamic range in the shapelet coefficients due
to varying galaxy size and flux suggests it may be useful to rescale
numerator and denominator with a generalizing factor $\nu$, defining:
\begin{eqnarray}
\tilde{\fflex}_{\nu} & = & \frac{4}{3 \beta }
\frac{\nu f_{1,1}}{\expect{\nu d_{1,1}}}, \label{eq:fflexnu}\\
\tilde{\gflex}_{\nu} & = & \frac{4\sqrt{6}}{3 \beta }
\frac{\nu f_{3,3}}{\expect{ \nu d_{3,3}}} \label{eq:gflexnu}.
\end{eqnarray}
We now discuss options for a suitable choice of this normalization
parameter $\nu$.

\subsection{Choosing a suitable $\nu$}

\begin{figure}
\begin{center}
\psfig{figure=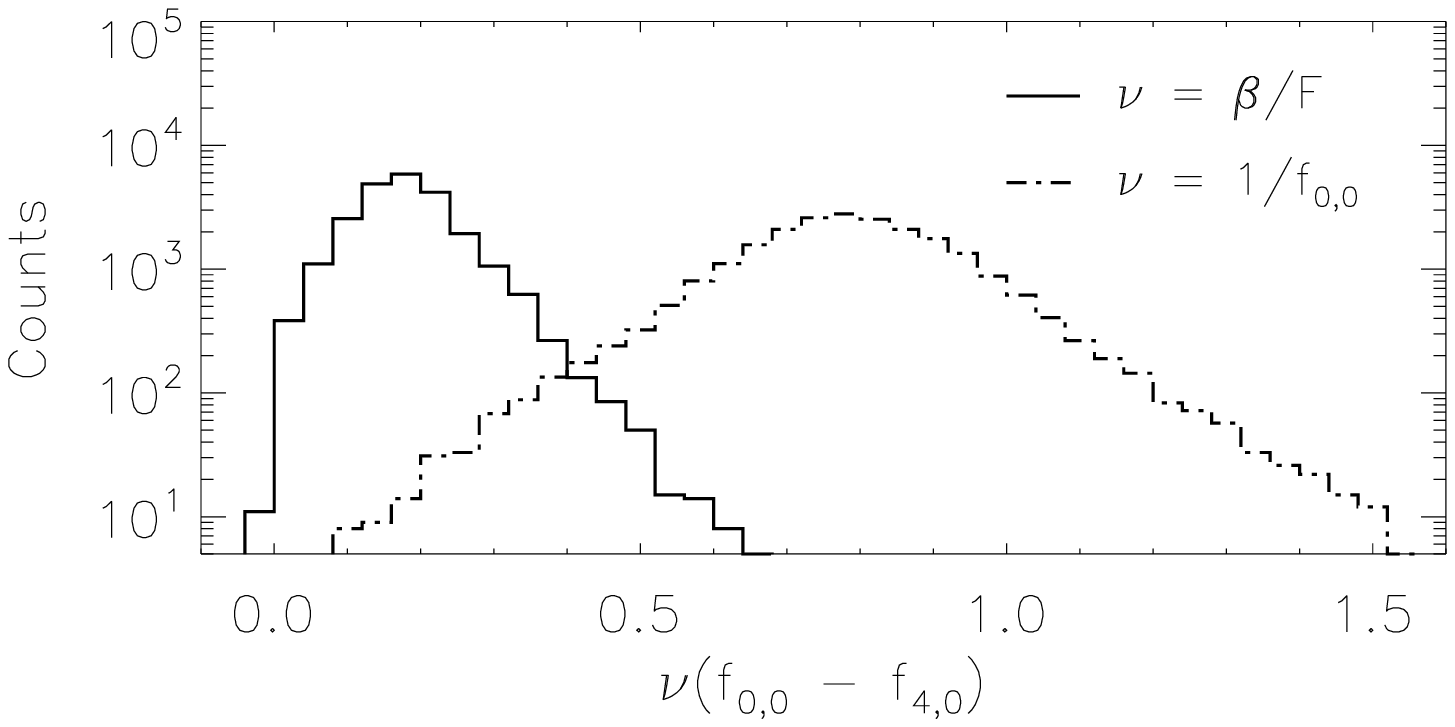,width=0.5\textwidth}
\psfig{figure=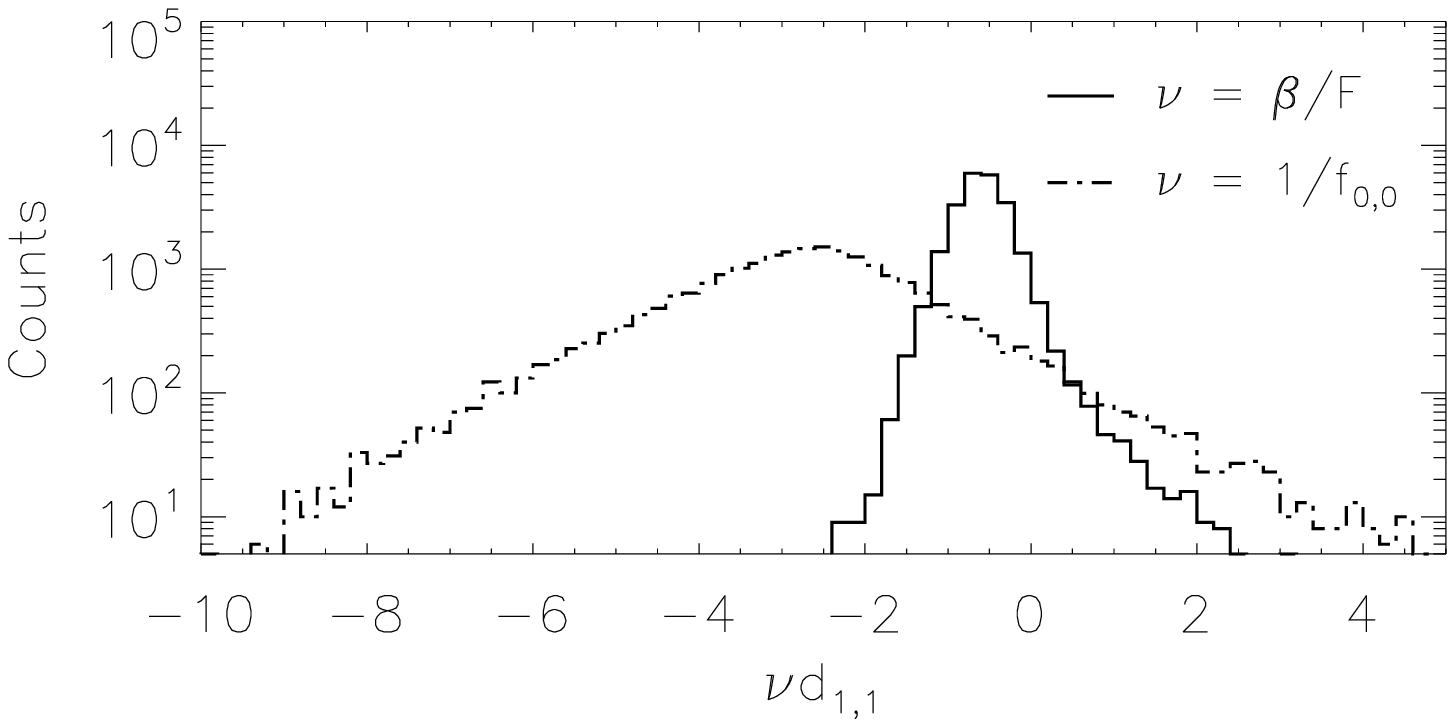,width=0.5\textwidth}
\psfig{figure=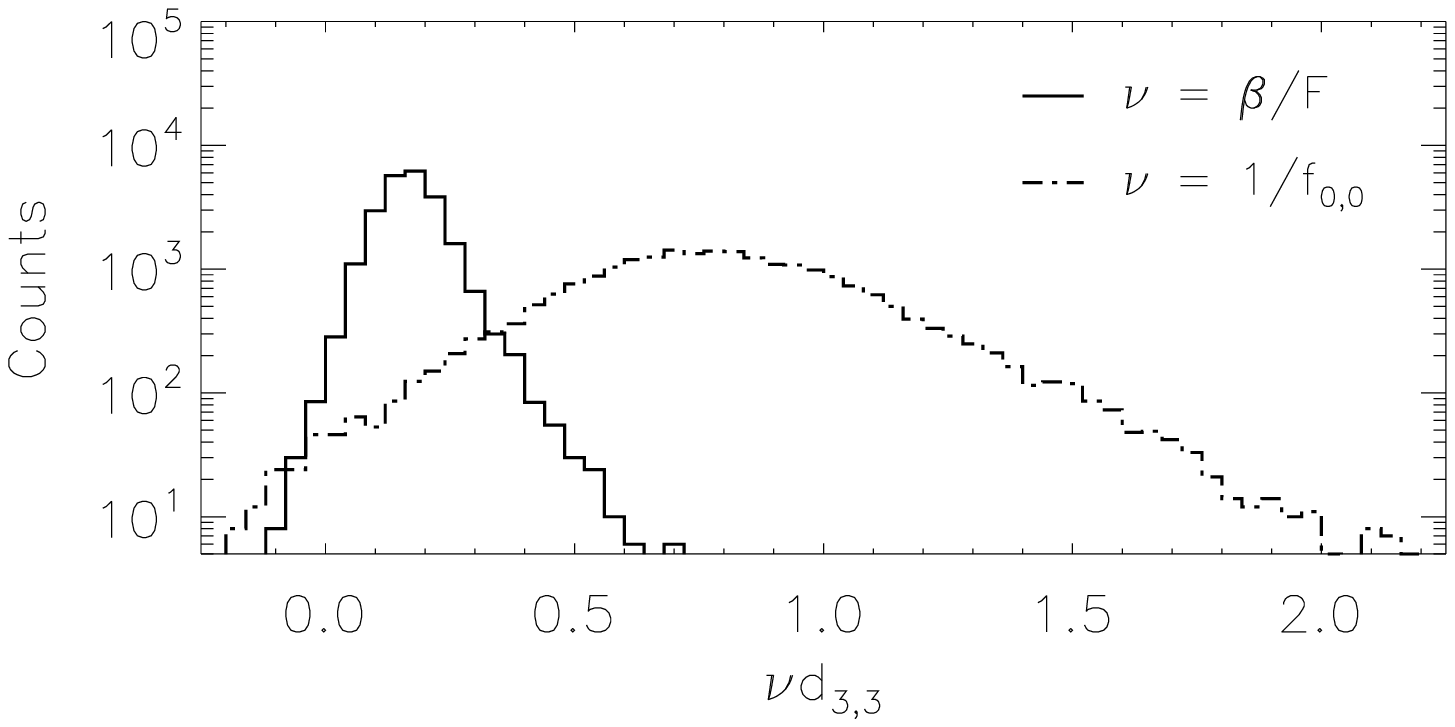,width=0.5\textwidth}
{\caption{Histograms showing distributions of the values averaged in
    the denominators of equations \eqref{eq:shearnu},
\eqref{eq:fflexnu} and \eqref{eq:gflexnu} --- upper, middle and lower
panels, respectively --- for two choices of
normalization parameter $\nu$. Solid line: choosing $\nu = \beta /
F$.  Dot-dashed line: choosing $\nu = 1/f_{0,0}$.\label{fig:denoms}}}
\end{center}
\end{figure}

Adopting $\nu = 1$ recovers the M07 $\tilde{\gamma}^{\mGaussian}$
estimator.  Natural alternative choices for $\nu$ are combinations of
shapelet model parameters that make the numerator and denominator of equation
\eqref{eq:shearnu} dimensionless.  Two of the simplest and most
easily-motivated potential combinations are
\begin{eqnarray}
\nu & = & \beta / F \label{eq:nubf}; \\
\nu & = & 1 / f_{0, 0} \label{eq:nuf00};
\end{eqnarray}
(c.f.\ the dimensionless shapelet basis functions introduced
in \citealp{refregier03}.) In Figure \ref{fig:denoms} we plot histograms of the values of the denominators
of the shear and flexion estimators defined by equations \eqref{eq:shearnu},
\eqref{eq:fflexnu} and \eqref{eq:gflexnu}, from shapelet fits to the
simulation galaxies in this study.  We plot results for the two
choices of $\nu$ expressed by equations \eqref{eq:nuf00} and
\eqref{eq:nubf}. The shape of these histograms appears to show a
consistent trend: using $\nu = 1/f_{0,0}$ results in
distributions of denominator values that have a more clearly-defined
central tendency than $\nu = \beta / F$.  This fact argues in favour
of the adoption of $1/f_{0,0}$ to normalize to a dimensionless
combination of shapelet coefficients.

To explore the distributions further we define the
following measure of skewness as a combination of the sample mean,
arithmetic median and standard deviation, commonly known as Pearson's second
skewness coefficient:
\begin{equation}
\Gamma(x) = 3\left[{\rm Mean}(x) - {\rm Median}(x) \right] / \sigma(x)  \label{eq:skewness}.
\end{equation}
Estimates of $\Gamma$ for the distributions of Figure
\ref{fig:denoms}, with uncertainties, are given in Table
\ref{tab:skewness}.  We see that the skewness in the $\nu = \beta / F$
distributions is consistently greater than that in the distribution of
denominator values when choosing $\nu = 1/f_{0,0}$.  Once again, this
argues in favour of the adoption of $\nu = 1/f_{0,0}$ as a
measurable quantity with which to make the numerators and denominators
of our shear and flexion estimators dimensionless.  No other
combination of observables was seen to provide better performance,
either in terms of showing a clear location for the central tendency of the denominator
distributions shown in Figure \ref{fig:denoms}, or in terms of 
reducing the required level of systematic bias calibration as shown in
Figure \ref{fig:calresults}.  To estimate shear and flexion from
shapelet fits to our simulated images we therefore adopt
$\tilde{\gamma} = \tilde{\gamma}_{\nu}$, $\tilde{\fflex} =
\tilde{\fflex}_{\nu}$ and $\tilde{\gflex} = \tilde{\gflex}_{\nu}$, where $\nu = 1/f_{0,0}$.

\begin{table}
\caption{Estimates of the skewness $\Gamma$ as defined by equation
  \eqref{eq:skewness} for the distributions of Figure \ref{fig:denoms}.} 
\label{tab:skewness}
\begin{center}
\begin{tabular}{cccc}
\hline
~ & ~ & Skewness $\Gamma(x)$ & ~ \\
$\nu$ ~ & $x = \nu (f_{0,0} - f_{0.4})$ & $x = \nu d_{1,1} $ & $x = \nu d_{3,3} $ \\
\hline \hline
 $\beta / F$ & 0.264 $\pm$ 0.002 & 0.173 $\pm$ 0.002 & 0.198 $\pm$ 0.001 \\
$1/f_{0,0}$ &  0.0275 $\pm$ 0.0002 & -0.0447 $\pm$ 0.0004 & 0.153 $\pm$ 0.001 \\
\hline 
\end{tabular}
\end{center}
\end{table}

%We can also make a third order $f_{3,1}$ estimator for $\fflex$:
%\begin{equation}
%\tilde{\fflex}^3 = \frac{4 \sqrt{2}}{3}\frac{f_{3,1}}{\expect{d_{3,1}}}
%\end{equation}
%where
%\begin{eqnarray}\label{eq:d31}
%d_{3,1} & = & \beta f_{0,0} + \left( 3 \beta - \frac{2R^2}{\beta} \right) f_{2,0}  - \left( \beta - \frac{2R^2}{\beta} \right) f_{4,0} \nn 
%& - & 3 \beta f_{6,0} + \frac{5 \sqrt{2} R^2}{6 \beta} \varepsilon^* \left( f_{4,2} - f_{2,2} \right) 
%\end{eqnarray}
%It should be noted that $d_{1,1}$ and $d_{3,1}$ in equations
%\eqref{eq:d11} and \eqref{eq:d31} both contain t rms proportional to $\varepsilon^* \times f_{n,2}$, for $n=2$ and $n=4$; in M07 these were argued to vanish due to rotational symmetry in the source plane.  These should not be omitted as $f_{2,2}$ and $f_{4,2}$ both in fact scale with the image ellipticity \citep{masseyrefregier05}.  These terms are therefore $\sim |\varepsilon|^2$, which does not cancel due to rotational symmetry.

\subsection{Potential issues}\label{app:issues}
The quantity $\nu$, which is estimated for each object individually,
is a function of random variables and therefore a random variable
itself.  Its inclusion in the estimators of equations \eqref{eq:shearnu},
\eqref{eq:fflexnu} and \eqref{eq:gflexnu} is therefore a cause of both
additional uncertainty in estimators of galaxy shape, and potential
bias (see e.g.\ \citealp{melchiorviola12}).  

These undesirable properties must be weighed against the
practical advantage of the technique in providing dimensionless combinations of
shapelet coefficients for shape estimation.  Given the large dynamic range in both apparent
galaxy sizes and fluxes in extragalactic data, and thus in the raw
values of best-fitting shapelet coefficients $f_{n,m}$, this advantage is
considerable. Figure \ref{fig:denoms} shows that such combinations can
provide distributions with a desirably compact support, and the same
is true for the denominators of the shear and flexion estimators of
equations \eqref{eq:shearnu}, \eqref{eq:fflexnu} and
\eqref{eq:gflexnu}.

To provide robust estimators of shear and flexion in the presence of
such a large dynamic range in $f_{n,m}$ values would probably require
the splitting of the galaxy sample into cells of apparent size and
flux in a manner similar to that described by \citet{kaiser00}.  Such
a scheme carries its own biases, due to the action of shear and
flexion to carry galaxies between adjacent cells.  Given the degree of
statistical uncertainty in both shear and flexion measurements from
space-based data sets in the near and medium term, the motivation for
constructing such dimensional cell-based estimators was not strong.
However, such a scheme, if correctly implemented, might help reduce the
calibration factors found in Section \ref{sect:calresults}.

\end{document}